\documentclass[11pt]{article}

\usepackage{epubtk}    
\usepackage{booktabs}  
\usepackage{graphicx}  
\usepackage{subfigure}
\usepackage{dsfont}
\usepackage{cancel}
\usepackage{placeins}
\usepackage{color}
\usepackage{enumerate}

\usepackage{amssymb,amsmath,amsthm,amsfonts,epsfig}
\usepackage{aas_macros}
\usepackage{epsf}
\usepackage{longtable}

\usepackage{epstopdf,pifont}
\usepackage{cite}

\usepackage{bm}
\usepackage{dcolumn}
\usepackage[latin1]{inputenc}
\usepackage{latexsym}
\usepackage{rotating}
\usepackage{color}
\usepackage{longtable}
\usepackage{enumerate}
\usepackage{tensor}
\usepackage{multirow}
\usepackage{url}
\usepackage{authblk}
\usepackage{gensymb}
\newcommand{\yes}{\checkmark}
\newcommand{\no}{\ding{55}}

\newcommand{\tn}{\textnormal}

\newcommand{\ben}{\begin{enumerate}}
\newcommand{\een}{\end{enumerate}}

\def\be{\begin{equation}}
\def\ee{\end{equation}}
\def\bea{\begin{eqnarray}}
\def\eea{\end{eqnarray}}

\newcommand{\beq}{\begin{eqnarray}}
\newcommand{\eeq}{\end{eqnarray}} 
\newcommand{\ba}{\begin{align}}
\newcommand{\ea}{\end{align}}

\begin{document}

\title{Testing the nature of dark compact objects: a status report}

\author{
\epubtkAuthorData{Vitor Cardoso}{%
CENTRA, Departamento de F\'{\i}sica, Instituto Superior T\'ecnico, Universidade de Lisboa, Avenida~Rovisco Pais 1, 1049 Lisboa, Portugal\\
CERN 1 Esplanade des Particules, Geneva 23, CH-1211, Switzerland}{vitor.cardoso@ist.utl.pt}{https://centra.tecnico.ulisboa.pt/network/grit/}
\and
\epubtkAuthorData{Paolo Pani}{Dipartimento di Fisica, ``Sapienza'' Universit\`a di Roma \& Sezione INFN Roma1, Piazzale Aldo Moro 5, 00185, Roma, Italy}{
paolo.pani@uniroma1.it}{https://www.roma1.infn.it/\~pani}
}

\date{\today}
\maketitle

\begin{abstract}
Very compact objects probe extreme gravitational fields and may be the key to understand 
outstanding puzzles in fundamental physics. These include the nature of dark matter, 
the fate of spacetime singularities, or the loss of unitarity in Hawking evaporation.
The standard astrophysical description of collapsing objects tells us that massive, dark 
and compact objects are black holes. Any observation suggesting otherwise would be an indication of beyond-the-standard-model physics. 
Null results strengthen and quantify the Kerr black hole paradigm.
The advent of gravitational-wave astronomy and precise measurements with very long baseline interferometry
allow one to finally probe into such foundational issues. We overview the physics of 
exotic dark compact objects and their observational status, including the observational 
evidence for black holes with current and future experiments.
\end{abstract}

\epubtkKeywords{Black Holes, Event horizon, Gravitational Waves, Quantum gravity, Singularities}

\pagebreak
\tableofcontents



\newpage
\hspace{1.8cm}
\parbox{0.8\textwidth}{{\small 
\noindent {\it ``The crushing of matter to infinite density by infinite tidal gravitation forces is a phenomenon with which one cannot live comfortably.
From a purely philosophical standpoint it is difficult to believe that physical singularities are a fundamental and unavoidable feature of our universe
[...] one is inclined to discard or modify that theory rather than accept the suggestion that the singularity actually occurs in nature.''}
\begin{flushright}
Kip Thorne, Relativistic Stellar Structure and Dynamics (1966) 
\end{flushright}
}
}
\vskip 1cm
\hspace{1.8cm}
\parbox{0.8\textwidth}{{\small 
\noindent {\it ``No testimony is sufficient to establish a miracle, unless the testimony be of such a kind, that its falsehood would be more miraculous than the fact which it endeavors to establish.''}
\begin{flushright}
David Hume, An Enquiry concerning Human Understanding (1748) 
\end{flushright}
}
}

\section{Introduction}

The discovery of the electron and the known neutrality of matter led in 1904 to 
J. J.~Thomson's ``plum-pudding'' atomic model. Data
from new scattering experiments was soon found to be in tension with this model, which 
was eventually superseeded by Rutherford's, featuring an atomic nucleus.
The point-like character of elementary particles opened up new questions. 
How to explain the apparent stability of the atom? How to handle the singular behavior of
the electric field close to the source? What is the structure of elementary particles? 
Some of these questions were elucidated with quantum mechanics and quantum field theory. 
Invariably, the path to the answer led to the discovery of hitherto unknown phenomena 
and to a deeper understanding of the fundamental laws of Nature. The history of elementary particles is a timeline of the understanding of the electromagnetic~(EM) interaction, and is pegged to its characteristic $1/r^2$ behavior 
(which necessarily implies that other structure {\it has} to exist on small scales within 
any sound theory).

Arguably, the elementary particle of the gravitational interaction are 
black holes~(BHs). Within General Relativity~(GR), BHs are indivisible and the simplest macroscopic objects that one can conceive.
The uniqueness results --~establishing that the two-parameter Kerr family of BHs describes any vacuum, stationary 
and asymptotically flat, regular solution to GR~--have turned BHs into somewhat of a miracle elementary particle~\cite{Chandra}. 

Even though the first nontrivial regular, asymptotically flat, 
vacuum solution to the field equations describing BHs were written already in 
1916~\cite{Schwarzschild:1916uq,Droste:1916uq}, several decades would elapse until such 
solutions became accepted and understood. The dissension between Eddington and Chandrasekhar over 
gravitational collapse to BHs is famous --~Eddington firmly believed that nature would 
find its way to prevent full collapse~-- and it took decades for the community to overcome 
individual prejudices. Ironically, after that BHs quickly became the {\it only} acceptable solution. So 
much so, that currently an informal definition of a BH might well be ``any dark, compact 
object with mass above roughly three solar masses.''

\subsection{Black holes: kings of the cosmos?}
There are various reasons why BHs were quickly adopted as the only possible dark and compact sources
triggering high-energy, violent phenomena in the Universe. 
The BH interior is causally disconnected from the exterior by an event horizon. Unlike 
the classical description of atoms, the GR description of the BH exterior is 
self-consistent and free of pathologies. 
The ``inverse-square law problem'' --~the GR counterpart of which is the appearance of 
pathological curvature singularities~-- is swept to inside the horizon and therefore 
harmless for the external world. There are strong indications that classical BHs are stable against small fluctuations~\cite{Klainerman:2017nrb}, and attempts to produce naked 
singularities, starting from BH spacetimes, have failed. 
In addition, BHs in GR can be shown to satisfy remarkable uniqueness properties~\cite{Chrusciel:2012jk}. These features promote
BHs to important solutions of the field equations and ideal testbeds for new physics. But BHs are not only curious 
mathematical solutions to Einstein's equations: their {\it formation} process is sound and well understood. At the 
classical level, there is nothing spectacular with the presence or formation of an event horizon. The 
equivalence principle dictates that an infalling observer crossing this region (which, by 
definition, is a \emph{global} concept) feels nothing extraordinary: in the case of 
macroscopic BHs all of the local physics at the horizon is rather unremarkable. 
Together with observations of phenomena so powerful that could only be 
explained via massive compact objects, the theoretical understanding of BHs turned them 
into undisputed kings of the cosmos.

There is, so far, no evidence for objects other than BHs that can explain all observations.
Nonetheless, given the special nature of BHs, one must {\it question and quantify} their existence.
Can BHs, as envisioned in vacuum GR, hold the same surprises that the electron and the 
hydrogen atom did when they started to be experimentally probed?
This overview will dwell on the existence of BHs, and signatures of possible alternatives.
There are a number of important reasons to do so, starting from the obvious: we {\it can} do it.
The landmark detection of gravitational waves~(GWs) showed that we are now able to analyze and understand the details 
of the signal produced when two compact objects merge~\cite{Abbott:2016blz,TheLIGOScientific:2016src}. 
An increase in sensitivity of current detectors and the advent of next-generation interferometers on ground and in 
space will open the frontier of \emph{precision} GW astrophysics.
GWs are produced by the coherent motion of the sources 
as a whole: they are ideal probes of strong gravity, and play the role that EM waves did to test the Rutherford model. 
In parallel, novel techniques such as radio and deep infrared 
interferometry~\cite{Doeleman:2008qh,Antoniadis:2013pzd} are now providing direct {\it images} of the center of ours and 
others galaxies, where a dark, massive and compact object is 
lurking~\cite{Genzel:2010zy,Falcke:2013ola,Johannsen:2015hib,Abuter:2018drb,Akiyama:2019cqa}.

The wealth of data from GW and EM observations has the potential to inform us on the following 
outstanding issues:

\subsection{Problems on the horizon}

Classically, spacetime singularities seem to be always cloaked by horizons and hence
inaccessible to distant observers; this is in essence the content of the weak cosmic 
censorship conjecture~\cite{Penrose:1969pc,Wald:1997wa}. However,
there is as yet no proof that the field equations always evolve regular initial data 
towards regular final states.

Classically, the BH exterior is pathology-free, but the interior is not. The Kerr family of BHs harbors singularities 
and closed timelike curves in its interior, and more generically it features a Cauchy horizon signaling the breakdown of 
predictability of the theory~\cite{Penrose_CCC,Reall_CCC,Dafermos:2003wr,Cardoso:2017soq}. The geometry describing the 
interior of an astrophysical spinning BH is currently unknown. A resolution of this problem most likely requires 
accounting for quantum effects. It is conceivable that these quantum effects are of no consequence whatsoever on physics 
outside the horizon. Nevertheless, it is conceivable as well that the resolution of such inconsistency leads to new 
physics that resolves singularities and does away with horizons, at least in the way we understand them currently. Such 
possibility is not too dissimilar from what happened with the atomic model after the advent of quantum electrodynamics.

Black holes have a tremendously large entropy, which is hard to explain from microscopic states of the progenitor star. 
Classical results regarding for example the area (and therefore entropy) 
increase~\cite{Hawking:1971tu} and the number of microstates can be tested using GW measurements~\cite{Lai:2018pmi,Brustein:2018ixz}, but assume classical matter. 
Indeed, semi-classical quantum effects around BHs are far from being under control. Quantum field theory
on BH backgrounds leads to loss of unitarity, a self-consistency requirement that any predictive theory ought to 
fulfill. The resolution of such conundrum may involve non-local effects changing the 
near-horizon structure, or doing away with horizons 
completely~\cite{Giddings:1992hh,Mazur:2004fk,Mathur:2005zp,Mathur:2008nj,Giddings:2009ae,Mathur:2009hf,Giddings:2012bm,Barcelo:2015noa, 
Giddings:2016tla,Giddings:2017mym,Almheiri:2012rt,Unruh:2017uaw,Giddings:2017jts,Bianchi:2018mml,Giddings:2019}. 

As a matter of fact, there is no tested nor fully satisfactory theory of quantum gravity, in much the same way 
that one did not have a quantum theory of point charges at the beginning of the 20th century.

GR is a purely classical theory. One expects quantum physics to become important 
beyond some energy scale. It is tacitly assumed that such ``quantum gravity effects'' are relevant only near the 
Planck scale: at lengths $\ell_P\sim\sqrt{G\hbar/c^3}\sim 10^{-35}\,{\rm meters}$, the 
Schwarzschild radius is of the order of the Compton wavelength of the BH and the notion of 
a classical system is lost. However, it has been argued that, in the orders of magnitude 
standing between the Planck scale and those accessible by current experiments, new physics 
can hide. To give but one example, if gravity is fundamentally a higher-dimensional 
interaction, then the fundamental Planck length can be substantially {\it 
larger}~\cite{ArkaniHamed:1998rs,Randall:1999vf}. In addition, some physics related to
compact objects have a logarithmic dependence on the (reasonably-defined) Planck length~\cite{Cardoso:2017cqb} (as also 
discussed below). Curiously, some attempts to quantize the area of BHs predict sizable effects even at a classical 
level, resulting in precisely the same phenomenology as that discussed in the rest of this 
review~\cite{Bekenstein:1995ju,Saravani:2012is,Foit:2016uxn,Cardoso:2019apo,Chakraborty:2017opo}. Thus, quantum-gravity 
effects may be within reach.

\subsection{Quantifying the evidence for black holes} \label{subsec:quantifying}

Horizons are not only a rather generic prediction of GR, but their existence is in fact 
\emph{necessary} for the consistency of the theory at the classical level. This is the 
root of Penrose's (weak) Cosmic Censorship 
Conjecture~\cite{Penrose:1969pc,Wald:1997wa}, which remains one of the most urgent 
open problems in fundamental physics. In this sense, the statement that there is a horizon 
in any spacetime harboring a singularity in its interior is such a remarkable claim, that 
(in an informal description of Hume's statement above) it requires similar remarkable 
evidence.

It is in the nature of science that paradigms have to be constantly questioned and 
subjected to experimental and observational scrutiny. Most specially because if the answer 
turns out to be that BHs do not exist, the consequences are so extreme and profound, that 
it is worth all the possible burden of actually testing it. 
As we will argue, the question is not just whether the strong-field gravity region near 
compact objects is consistent with the BH geometry of GR, but rather to \emph{quantify} 
the limits of observations in testing event horizons. This approach is common practice in other contexts.
Decades of efforts in testing the pillars of GR resulted in formalisms (such as the parametrized post-Newtonian 
approach~\cite{Will:2014kxa}) which quantify the constraints of putative deviations from GR. For example, we know that 
the weak equivalence principle is valid to at least within one part in $10^{15}$~\cite{Berge:2017ovy}. On the other 
hand, 
no such solid framework is currently available to quantify deviations from the standard BH paradigm.
In fact, as we advocate 
in this work, the question to be asked is not whether there is a horizon in the 
spacetime, but how close to it do experiments or observations go. It is important to 
highlight that some of the most important tests of theories or paradigms --~and GR
and its BH solutions are no exception~-- arise from entertaining the existence of 
alternatives. It is by allowing a large space of solutions that one can begin to exclude 
--~with observational and experimental data~-- some of the alternatives, thereby
producing a stronger paradigm.

\subsection{The dark matter connection}
{\it Known} physics all but exclude BH alternatives as explanations for the dark, massive 
and compact objects out there. Nonetheless, the Standard Model of fundamental interactions 
is not sufficient to describe the cosmos --~at least on the largest scales. The nature of dark matter (DM) is one of the 
longest-standing puzzles in physics~\cite{Bertone:2018krk,Barack:2018yly}. Given that the 
evidence for DM is --~so far~-- purely gravitational, further clues may well be hidden 
in strong-gravity regions or GW signals generated by dynamical compact objects.

As an example, new fundamental fields (such as axions, axion-like particles, 
etc~\cite{Marsh:2015xka,Clifton:2011jh}), either minimally or non-minimally coupled to 
gravity, are essential for cosmological models, and are able to explain all known 
observations concerning DM. Even the simplest possible theory of minimally 
coupled, massive scalar fields give rise to self-gravitating {\it compact} objects, which 
are dark if their interaction with Standard Model particles is weak. These are called 
boson stars or oscillatons, depending on whether the field if complex or real, 
respectively. Such dark objects have a maximum mass\footnote{A crucial property of BHs in GR is that --~owing to 
the scale-free nature of vacuum Einstein's equation~-- their mass is a free parameter. This is why the same Kerr metric 
can describe any type of BH in the universe, from stellar-mass (or even possibly primordial) to supermassive. It is 
extremely challenging to reproduce this property with a material body, since matter fields introduce a scale.} which 
is regulated by the mass of the 
fundamental boson itself and by possible self-interaction terms; they form naturally 
through gravitational collapse and may cluster around an ultracompact configuration 
through ``gravitational cooling''~\cite{Seidel:1993zk,Liebling:2012fv,Brito:2015yga,DiGiovanni:2018bvo}.

Furthermore, DM could be composed of entirely different fields or particles, and many of these are expect to lead to 
new classes of dark compact objects~\cite{Narain:2006kx,Raidal:2018eoo,Deliyergiyev:2019vti}.

\subsection{Taxonomy of compact objects: a lesson from particle physics}

From a phenomenological standpoint, BHs and neutron stars could be just two ``species'' of a larger family of 
astrophysical compact objects, which might co-exist with BHs rather than replacing them. These objects are theoretically 
predicted in extended theories of gravity but also in other scenarios in the context of GR, such as 
beyond-the-Standard-Model fundamental fields minimally coupled to gravity, or of exotic states of matter.

In this context, it is tempting to draw another parallel with particle physics. After the Thomson discovery of the 
electron in 1897, the zoo of elementary particles remained almost unpopulated for decades: the proton was discovered 
only in the 1920s, the neutron and the positron only in 1932, few years before the muon~(1936). Larger and more 
sensitive particle accelerators had been instrumental to discover dozens of new species of elementary particles during 
the second half of the XXth century, and nowadays the Standard Model of particle physics accounts for hundreds of 
particles, either elementary or composite.
Compared to the timeline of particle physics, the discovery of BHs, neutron stars, and binary thereof is much more 
recent; it is therefore natural to expect that the latest advance in GW astronomy and very long baseline interferometry 
can unveil new species in the zoo of astrophysical compact objects. Of course, this requires an understanding of the 
properties of new families of hypothetical compact objects and of their signatures.

\subsection{The small $\epsilon$-limit}
In addition to the above phenomenological motivations, dark compact objects are also interesting from a mathematical 
point of view. For instance, given the unique properties of a BH, it is interesting to study how a dark compact object 
approaches the ``BH limit'' (if the latter exists!) as its compactness increases. Continuity arguments would suggest 
that any deviation from a BH should vanish in this limit, but this might occur in a highly nontrivial way, as we shall 
discuss.
The first issue in this context is how to parametrize ``how close'' a self-gravitating object is to a BH in a rigorous 
way, by introducing a ``closeness'' parameter $\epsilon$, such that $\epsilon\to0$ corresponds to the BH limit. As we 
shall discuss, there are several choices for $\epsilon$, for example the tidal deformability, the inverse of the maximum 
redshift in the spacetime, or a quantity related to the compactness $M/R$ such as $\epsilon=1-2M/R$, where $M$ is the 
object mass in the static case and $R$ is its radius. 

In the context of DM self-gravitating objects $\epsilon$ is expected to be of order unity. However, when quantifying 
the evidence for horizons or in the context of quantum corrected spacetimes, one is usually interested in the $\epsilon 
\ll 1$ limit. The physics of such hypothetical objects is interesting on its own:
these objects are by construction regular everywhere and causality arguments imply that all known BH physics must be 
recovered in the $\epsilon \to 0$ limit. Thus, the small $\epsilon$-limit may prove useful in the understanding of BH 
themselves, or to help cast a new light in old murky aspects of objects with a teleological nature. Moreover, as we will 
see, such limit is amenable to many analytical simplifications and describes reasonably well even finite $\epsilon$ 
spacetimes. In this regard, the $\epsilon\to 0$ limit can be compared to large spacetime-dimensionality limit in 
Einstein field equations~\cite{Emparan:2013xia}, or even the large ${\cal N}$ limit in QCD~\cite{tHooft:1973alw}.
Here, we will focus exclusively on four-dimensional spacetimes.

\newpage

\section{Structure of stationary compact objects\label{sec:stage}}

\hskip 0.2\textwidth
\parbox{0.8\textwidth}{
\begin{flushright}
{\small 
\noindent {\it ``Mumbo Jumbo is a noun and is the name of a grotesque idol said to have been worshipped by some tribes. In its figurative sense, Mumbo Jumbo is an object of senseless veneration or a meaningless ritual.''}\\
Concise Oxford English Dictionary
}
\end{flushright}
}

\vskip 1cm

The precise understanding of the nature of dark, massive and compact objects can follow different routes,

\begin{itemize}
 \item[i.] a pragmatic approach of testing the spacetime close to compact, dark objects, 
irrespective of their nature, by devising model-independent observations that yield 
unambiguous answers; this often requires consistency checks and null-hypothesis tests of 
the Kerr metric.
 \item[ii.] a less ambitious and more theoretically-driven approach, which starts by 
constructing objects that are very compact, yet horizonless, within some framework. It 
proceeds to study their formation mechanisms and stability properties, and then discarding 
solutions which either do not form or are unstable on short timescales; finally, 
understand the observational imprints of the remaining objects, and how they differ from 
BHs.
\end{itemize}

In practice, when dealing with outstanding problems where our ignorance is extreme, 
pursuing both approaches simultaneously is preferable.
Indeed, using concrete models can sometimes be a useful guide to learn about broad, model-independent signatures.
As it will become clear, one can design exotic horizonless models which mimic all 
observational properties of a BH with arbitrary accuracy. 
While the statement ``BHs exist in our Universe'' is \emph{fundamentally unfalsifiable}, 
alternatives can be ruled out with a single observation, just like 
Popper's black swans~\cite{Popper}.

Henceforth we shall refer to horizonless compact objects other than a neutron star as \emph{Exotic Compact 
Objects~(ECOs)}. The aim of this section is to contrast the properties of BHs with those of ECOs and to find 
a classification for different models.

\subsection{Anatomy of compact objects}
%
\begin{figure}[ht!]
\begin{center}
\includegraphics[width=0.8\textwidth]{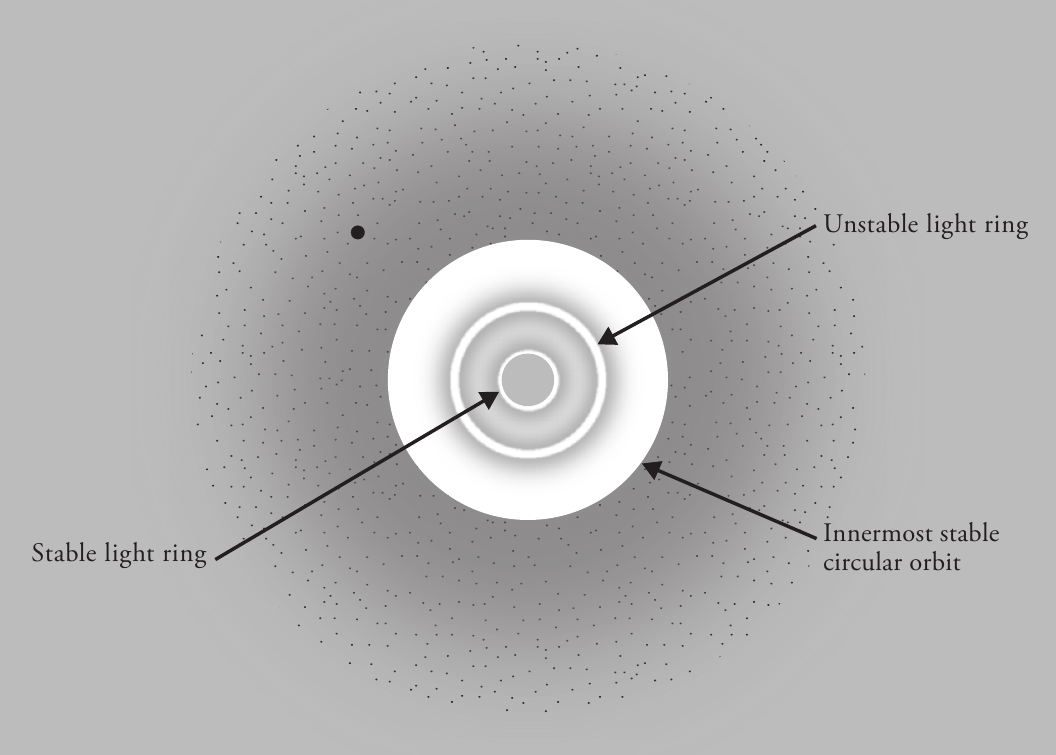}
  \caption{An equatorial slice of a very compact object, together with the most 
significant (from a geodesic perspective) locations.
At large distances away from the central region, physics is nearly Newtonian: planets 
--~such as the small dot on the figure~-- can orbit on stable orbits. The external gray 
area is the entire region where stable circular motion is possible. At the innermost 
stable circular orbit ($r=6M$), timelike circular motion is marginally stable, and 
unstable as one moves further within.  High-frequency EM waves or GWs can be on circular 
orbit in one very special location: the light ring ($r=3M$). Such motion is unstable, and 
can also be associated with the ``ringdown'' excited during mergers. For horizonless 
objects, as one approaches the geometric center another significant region may appear: a 
second, {\it stable} light ring. Once rotation is turned on, regions of negative energy (``ergoregions'') are possible. 
The astrophysical properties of a dark compact object depends on where in this diagram 
its surface is located.
\label{fig:diagram2}}
\end{center}
\end{figure} 
For simplicity, let us start with a four-dimensional spherically symmetric object and 
assume that it exterior is described by vacuum GR.
Static, spherically symmetric spacetimes are described (in standard coordinates with $r$ 
being the areal radius) by the line element
\begin{equation}
ds^2=-f(r) dt^2+g(r)^{-1}dr^2+r^2d\Omega^2\,. \label{metric}
\end{equation}
with $d\Omega^2=d\theta^2+\sin^2\theta d\phi^2$. Birkhoff's theorem guarantees that any 
vacuum, spherically-symmetric spacetime (in particular, the exterior an isolated compact, 
spherically-symmetric object) is described by the Schwarzschild geometry, for which
\begin{equation}
f(r)=g(r)=1-\frac{2M}{r}\,,
\end{equation}
and $M$ is the total mass of the spacetime (we use geometrical $G=c=1$ units, except if 
otherwise stated).

\subsubsection{Event horizons, trapped surfaces, apparent horizons}

A BH owns its name~\cite{Herdeiro:2018ldf} to the fact that nothing --~not even light~-- can escape 
from the region enclosed by its \emph{horizon}. Since the latter is the real defining 
quantity of a BH, it is important to define it rigorously.
In fact, there are several inequivalent concepts of horizon~\cite{Hawking:1973uf,Curiel:2019nna}.
In asymptotically-flat spacetime, a BH is the set of events from which no future-pointing 
null geodesic can reach future null infinity. The \emph{event 
horizon} is the (null) boundary of this region.
The event horizon is a \emph{global} property of an entire spacetime: on a given 
spacelike slice, the event horizon cannot be computed without knowing the entire future 
of the slice.
Strictly speaking, an event horizon does not ``form'' at a certain time, but it is a 
nonlocal property; as such, it is of limited practical use in dynamical situations.

On the other hand, in a $3+1$ splitting of spacetime, a \emph{trapped surface} is defined 
as a smooth closed $2$-surface on the slice whose future-pointing outgoing null geodesics 
have negative expansion~\cite{Hawking:1973uf,Thornburg:2006zb,Wald:1991zz}.
Roughly speaking, on a trapped surface light rays are all directed inside the trapped surface \emph{at that 
given time}. The \emph{trapped region} is the 
union of all trapped surfaces, and the outer boundary of 
the trapped region is called the \emph{apparent horizon}.
At variance with the event horizon, the apparent horizon is defined locally in time, but 
it is a property that depends on the choice of the slice.
Under certain hypothesis --~including the assumption that matter fields satisfy the 
energy conditions~-- the existence of a trapped surface (and hence of an apparent 
horizon) implies that the corresponding slice contains a 
BH~\cite{Hawking:1973uf}. The converse is instead not true: an arbitrary (spacelike) 
slice of a BH spacetime might not contain any apparent horizon. If an apparent horizon 
exists, it is necessarily contained within an event horizon, or it coincides with it. In 
a stationary spacetime, the event and apparent horizons always coincide at a classical level (see Refs.~\cite{Bardeen:1981zz,York:1983zb,Arzano:2016twc} for possible quantum effects).

In practice, we will be dealing mostly with quasi-stationary solutions, when the 
distinction between event and apparent horizon is negligible. For the sake of 
brevity, we shall often refer simply to a ``horizon'', having in mind the apparent 
horizon of a quasi-stationary solution. Notwithstanding, there is no direct observable associated to the horizon~\cite{Abramowicz:2002vt,Cardoso:2017cqb,Nakao:2018knn}.
There are signatures which can be directly associated to timelike surfaces, and whose presence would signal new physics.
The absence of such signatures strengthens and quantifies the BH paradigm.

\subsubsection{Quantifying the shades of dark objects: the closeness parameter 
$\epsilon$}

\begin{flushright}
{\small 
\noindent {\it ``Alas, I abhor informality.''}\\
That Mitchell and Webb Look, Episode 2
}
\end{flushright}

\vskip 0.5cm

Since we will mostly be discussing objects which look like BHs on many scales, it is 
useful to introduce a ``closeness''
parameter $\epsilon$ that indicates how close one is to a BH spacetime. There is an 
infinity of possible choices for such parameter (and in fact, different choices have been made in the literature, 
e.g., Refs.~\cite{Giddings:2014ova,Giddings:2016tla}).
At least in the case of spherical symmetric, Birkhoff's theorem provides a natural choice 
for the closeness parameter: if the object has a surface at $r_0$, then $\epsilon$ is 
defined as
\begin{equation}
r_0=2M(1+\epsilon)\,.\label{eq_epsilon_def}
\end{equation}
We are thus guaranteed that when $\epsilon \to 0$, a BH spacetime is recovered. For 
spherical objects the above definition is coordinate-independent ($2\pi r_0$ is the proper equatorial circumference of the object). Furthermore, one can also define the proper 
distance between the surface and $r_0$, $\int_{2M}^{r_0}dr f^{-1/2}\sim 
4M\sqrt{\epsilon}$, which is directly related to $\epsilon$. Some of the observables 
discussed below show a dependence on $\log \epsilon$, making the distinction 
between radial and proper distance irrelevant.

We should highlight that this choice of closeness parameter is made for convenience. None 
of the final results depend on such an arbitrary choice. In fact, there are objects 
--~such as boson stars~-- without a well defined surface, since the matter fields are 
smooth everywhere. In such case $r_0$ can be taken to be an effective radius beyond which 
the density drops sharply to zero. 
In some cases it is possible that the effective radius depends on the type of 
perturbations or on its frequency.
It sometimes proves more useful, and of direct significance, to use instead the 
coordinate time $\tau$ (measurable by our detectors) that a radial-directed light signal 
takes to travel from the light ring to the surface of the object. For 
spherically symmetric spacetimes, there is a one-to-one correspondence with the 
$\epsilon$ parameter, $\tau=M(1-2\epsilon-\log (4\epsilon^2))\sim -2M\log\epsilon$, where 
the last step is valid when $\epsilon\to0$.
In the rest, when convenient, we shall refer to this time scale rather than to $r_0$.

Overall, we shall use the magnitude of $\epsilon$ to classify different models of dark 
objects. 
A neutron star has $\epsilon\sim{\cal O}(1)$ and models with such value of the closeness 
parameter (e.g., boson stars, stars made of DM, see below) are expected to have 
dynamical properties which resemble those of a stellar object rather than a BH. For 
example, they are characterized by observables that display 
${\cal O}(1)$ corrections relative to the BH case and are therefore easier to distinguish. On 
the other hand, to test the BH paradigm in an agnostic way, or 
for testing the effects of quantum gravity, one often has in mind $\epsilon \ll 1$. For 
instance, in certain models $r_0-2M=2M\epsilon$ or the proper distance $\sim 
M\sqrt{\epsilon}$ are of the order of the Planck 
length $\ell_P$; in such case $\epsilon \sim 10^{-40}$ or even smaller. These models are 
more challenging to rule out.

Finally, in dynamical situations $\epsilon$ might be effectively time dependent. Even when $\epsilon\sim \ell_P/M$ 
at equilibrium, off-equilibrium configurations might have significantly large $\epsilon$ (see, e.g., 
Refs.~\cite{Brustein:2017nis,Brustein:2017kcj,Wang:2018mlp,Wang2019}).

\subsubsection{Quantifying the softness of dark objects: the curvature parameter} 
\label{sec:softness}
In addition the closeness parameter $\epsilon$, another important property of a dark object is its 
curvature scale. The horizon introduces 
a cut-off which limits the curvature that can be probed by an external observer. For a 
BH the largest curvature (as measured by the Kretschmann scalar ${\cal K}$) occurs at the 
horizon and reads
\begin{equation}
  {\cal K}^{1/2}\sim \frac{1}{M^2}\approx 4.6\times 
10^{-13}\left(\frac{10M_\odot}{M}\right)^2\,{\rm cm}^{-2}\,. \label{KBH}
\end{equation}
For astrophysical BHs the curvature at the horizon is always rather small, and it might 
be large only if sub-$M_\odot$ primordial BHs exist in the universe. As a reference, the 
curvature at the center of an ordinary neutron star is ${\cal K}^{1/2}\sim 
10^{-14}\,{\rm cm}^{-2}$.

By comparison with the BH case, one can introduce two classes of 
models~\cite{Raposo:2018xkf}: (i)~\emph{``soft'' ECOs}, for which the maximum
curvature is comparable to that at the horizon of the 
corresponding BH; and (ii)~\emph{``hard'' ECOs}, for which the curvature is much larger.
In the first class, the near-surface geometry smoothly approaches that at the horizon in the BH limit~(hence their 
``softness''), whereas in the second class the ECO can support large curvatures on its surface without collapsing, 
presumably because the 
underlying theory involves a new length scale, ${\cal L}$, such that ${\cal 
L}\ll {M}$.  In these models high-energy effects drastically modify the near-surface geometry (hence their 
``hardness''). An example are certain classes of wormholes (see Section~\ref{sec:taxonomy}).

An interesting question is whether the maximum curvature ${\cal K}_{\rm max}$ depends on $\epsilon$.
Indeed, an ECO with a surface just above the BH limit $(\epsilon\to0$) may always require large \emph{internal} 
stresses in order to prevent its collapse, so that the curvature in the interior is very large, even if the exterior is 
exactly the Schwarzschild geometry. 
In other words, an ECO can be soft in the exterior but hard in the interior.
Examples of this case are thin-shell gravastars and strongly anisotropic stars (see 
Section~\ref{sec:taxonomy}). 
Thus, according to this classification all ECOs might be ``hard'' in the $\epsilon\to0$ limit.
Likewise, the exterior of hard ECOs might be described by soft ECO solutions far from the 
surface, where the curvature is perturbatively close to that of a BH.

\subsubsection{Geodesic motion and associated scales}

The most salient geodesic features of a compact object are depicted in 
Fig.~\ref{fig:diagram2}, representing the equatorial slice of a spherically-symmetric 
spacetime. 

The geodesic motion of timelike or null particles in the geometry~\eqref{metric} can be 
described with the help of two conserved quantities, the specific energy $E=f\,\dot{t}$ 
and angular momentum $L=r^2\dot{\varphi}$, where a dot stands for a derivative with 
respect to proper time~\cite{MTB}. The radial motion can be computed via a 
normalization condition,
\begin{equation}
\dot{r}^2=g\left(\frac{E^2}{f}-\frac{L^2}{r^2}-\delta_1\right)\equiv E^2-V_{\rm geo}\,, \label{radial_motion}
\end{equation}
where $\delta_{1}=1,0$ for timelike or null geodesics, respectively. The null limit can be 
approached letting $E,L\to \infty$ and rescaling all quantities appropriately.
Circular trajectories are stable only when $r\geq6M$, and unstable for smaller 
radii. The $r=6M$ surface defines the innermost stable circular orbit~(ISCO), and has an 
important role in controlling the inner part of the accretion flow onto compact objects. It corresponds 
to the orbital distance at which a geometrically thin accretion disk is typically truncated~\cite{Novikov:1973kta} and 
it sets the highest characteristic frequency for compact emission region (``hotspots'') orbiting 
around accreting compact objects~\cite{2005MNRAS.363..353B,Broderick:2005jj}.
%
%
\begin{figure}[th]
\begin{center}
\includegraphics[width=0.9\textwidth]{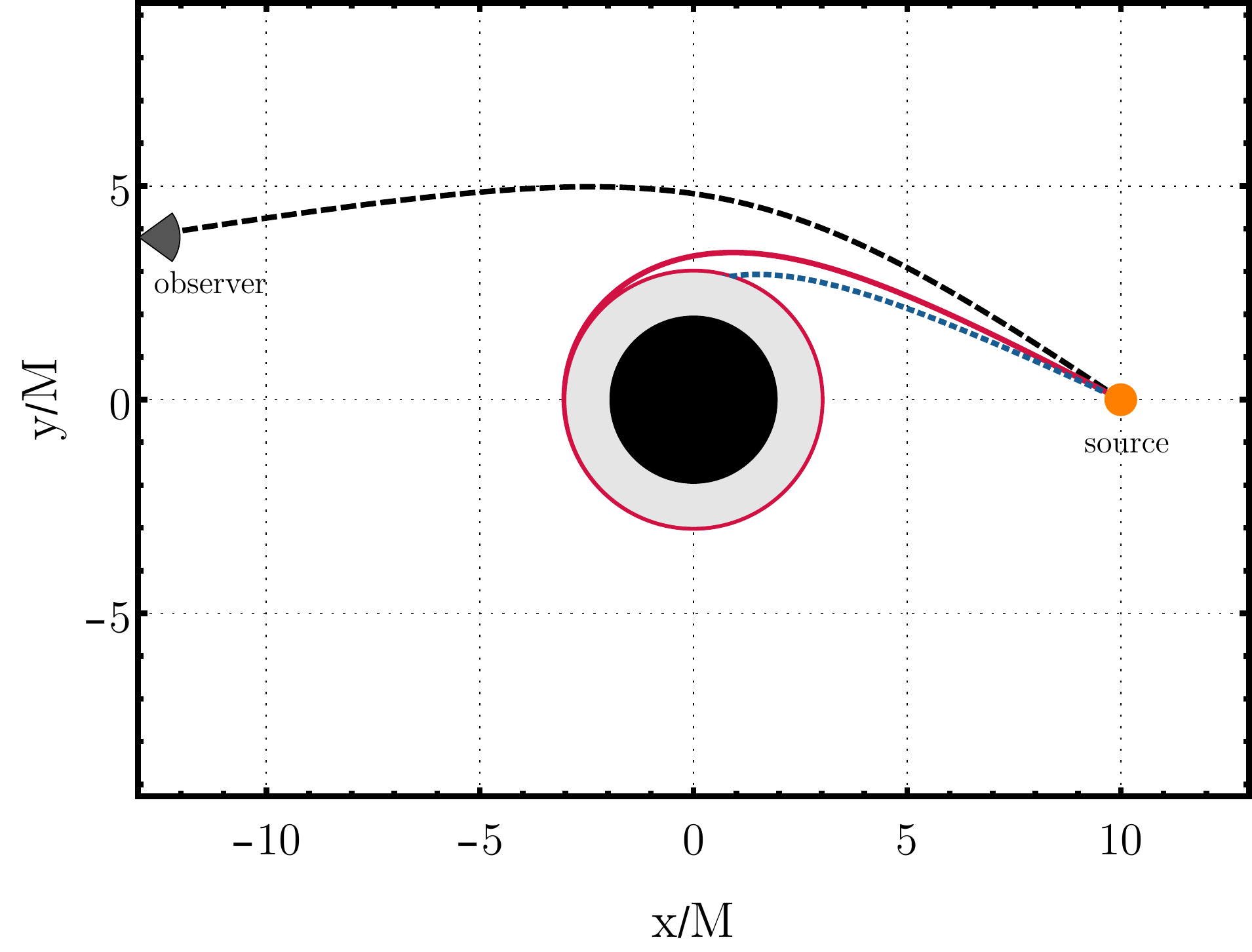}
\caption{A source (for example, a star) emits photons in all directions in a region of 
spacetime where a compact object exists (black circle). Photons with high impact 
parameter are weakly bent (dashed, black curve), while those with small impact parameter 
(short-dashed blue) are absorbed and hit the object. The separatrix corresponds to photons 
that travel an infinite amount of time around the light ring (solid red curve) before 
being scattered or absorbed. Such critical photons have an impact parameter 
$b=3\sqrt{3}M$~\cite{MTB}. The gray shaded area is the photon sphere.
\label{fig:scattering_capture}}
\end{center}
\end{figure}
Another truly relativistic feature is the existence of circular null geodesics, i.e., of 
circular motion for high-frequency EM waves or GWs. 
In the Schwarzschild geometry, a circular null geodesic is possible only
at $r=3M$~\cite{MTB}. This location defines a surface called the 
\emph{photon sphere}, or, on an equatorial slice, a \emph{light ring}. The 
photon sphere has a number of interesting properties, and is useful to understand certain 
features of compact spacetimes. 

For example, assume that an experimenter far away throws (high-frequency) photons in all 
directions and somewhere a compact object is sitting, as in 
Fig.~\ref{fig:scattering_capture}. Photons that have a very large impact parameter (or 
large angular momentum), never get close to the object. Photons with a smaller impact 
parameter start feeling the gravitational pull of the object and may be slightly 
deflected, as the ray in the figure. Below a critical impact parameter all photons 
``hit'' the compact object. It is a curious mathematical property that the critical 
impact parameter corresponds to photons that circle the light ring an infinite number of 
orbits, before being either absorbed or scattered. Thus, the light ring is fundamental for 
the description of how compact objects and BHs ``look'' like when illuminated by 
accretion disks or stars, thus defining their so-called \emph{shadow}, see 
Sec.~\ref{sec:shadow} below. 

The photon sphere also has a bearing on the 
spacetime response to any type of high-frequency waves, and therefore describes how 
high-frequency GWs linger close to the horizon. 
At the photon sphere, $V_{\rm geo}''=-2 E^2/(3 M^2)<0$. Thus, circular null geodesics are 
unstable: a displacement $\delta$ of null particles grows 
exponentially~\cite{Ferrari:1984zz,Cardoso:2008bp}
\begin{equation}
\delta(t) \sim \delta_0 e^{\lambda t}\,,\qquad \lambda=\sqrt{\frac{-f^2 
V_r''}{2E^2}}=\frac{1}{3\sqrt{3}M}\,.\label{LR_instability}
\end{equation}
A geodesic description anticipates that light or GWs may persist at or close to the 
photon sphere on timescales 
$3\sqrt{3}M \sim 5M$. Because the geodesic calculation is local, these conclusions hold 
irrespectively of the spacetime being vacuum all the way to the horizon or not.

For any regular body, the metric functions $f,g$ are well behaved at the center, never 
change sign and asymptote to unity at large distances. Thus, the effective potential 
$V_{\rm geo}$ is negative at large distances, vanishes with zero derivative at the light 
ring, and is positive close to the center of the object. This implies that there must be 
a second light ring in the spacetime, and that it is {\rm 
stable}~\cite{Cardoso:2014sna,Macedo:2013jja,Cunha:2017qtt}.  
Inside this region, there is stable timelike circular motion 
everywhere\footnote{Incidentally, this also means that the circular timelike geodesic 
at $6M$ is not really the ``innermost stable circular orbit''. We use this description to 
keep up with the tradition in BH physics.}.

\subsubsection{Photon spheres}

An ultracompact object with surface at $r_0=2M(1+\epsilon)$, with $\epsilon \ll 1$, 
features exactly the same geodesics and properties close to its photon sphere as BHs. 
From Eq.~\eqref{LR_instability}, we immediately realize that after a (say) three $e$-fold 
timescale, $t \sim 15M$, the amplitude of the original signal is only $5\%$ of its 
original value.
On these timescales one can say that the signal died away.
If on such timescales the ingoing part of the signal did not have time to bounce off the 
surface of the object and return to the light ring, then for an external observer the 
relaxation is identical to that of a BH. This amounts to requiring that $\tau\equiv \int_{2M(1+\epsilon)}^{3M}\gtrsim 15M$, or
\begin{equation}
\epsilon\lesssim\epsilon_{\rm crit}\sim 0.019\,. \label{eps_crit}
\end{equation}
Thus, the horizon plays no special role in the response of high frequency waves, nor 
could it: it takes an infinite (coordinate) time for a light ray to reach the horizon.
The above threshold on $\epsilon$ is a natural sifter between two classes of compact, dark 
objects. For objects characterized by $\epsilon \gtrsim 0.019$, light or GWs can make the 
roundtrip from the photon sphere to the object's surface and back, before dissipation of 
the photon sphere modes occurs. For objects satisfying \eqref{eps_crit}, the waves trapped 
at the photon sphere relax away by the time that the waves from the surface hit it back.

We can thus use the properties of the ISCO and photon sphere
to distinguish between different classes of models:
\begin{itemize}
 \item \emph{Compact object}: if it features an ISCO, or in other words if its surface 
satisfies $r_0<6M$ ($\epsilon<2$). Accretion disks around compact objects of the same 
mass should have similar characteristics;
 \item \emph{Ultracompact object~(UCO)~\cite{1985CQGra...2..219I}}: a compact object that features a 
photon sphere, $r_0<3M$~($\epsilon<1/2$). For these objects, the phenomenology related to 
the photon sphere might be very similar to that of a BH;
 \item \emph{Clean-photon sphere object~(ClePhO)}: an ultracompact object which satisfies
condition~\eqref{eps_crit} and therefore has a ``clean'' photon sphere, 
$r_0<2.038M$~($\epsilon\lesssim0.019$). The early-time dynamics of ClePhOs is expected 
to be the same as that of BHs. At late times, ClePhOs should display unique signatures of 
their surface.
\end{itemize}

An ECO can belong to any of the above categories. There are indications that the photon sphere is a fragile concept and that it suffers radical changes in the presence of small environmental disturbances~\cite{Shoom:2017ril}. The impact of such result on the dynamics on compact objects is unknown.

\subsection{Escape trajectories and shadows\label{sec:shadow}}
%
\begin{figure}[th]
\begin{tabular}{m{0.6\linewidth}m{0.4\linewidth}}
 \hspace{-0.4cm}\includegraphics[width=0.6\textwidth]{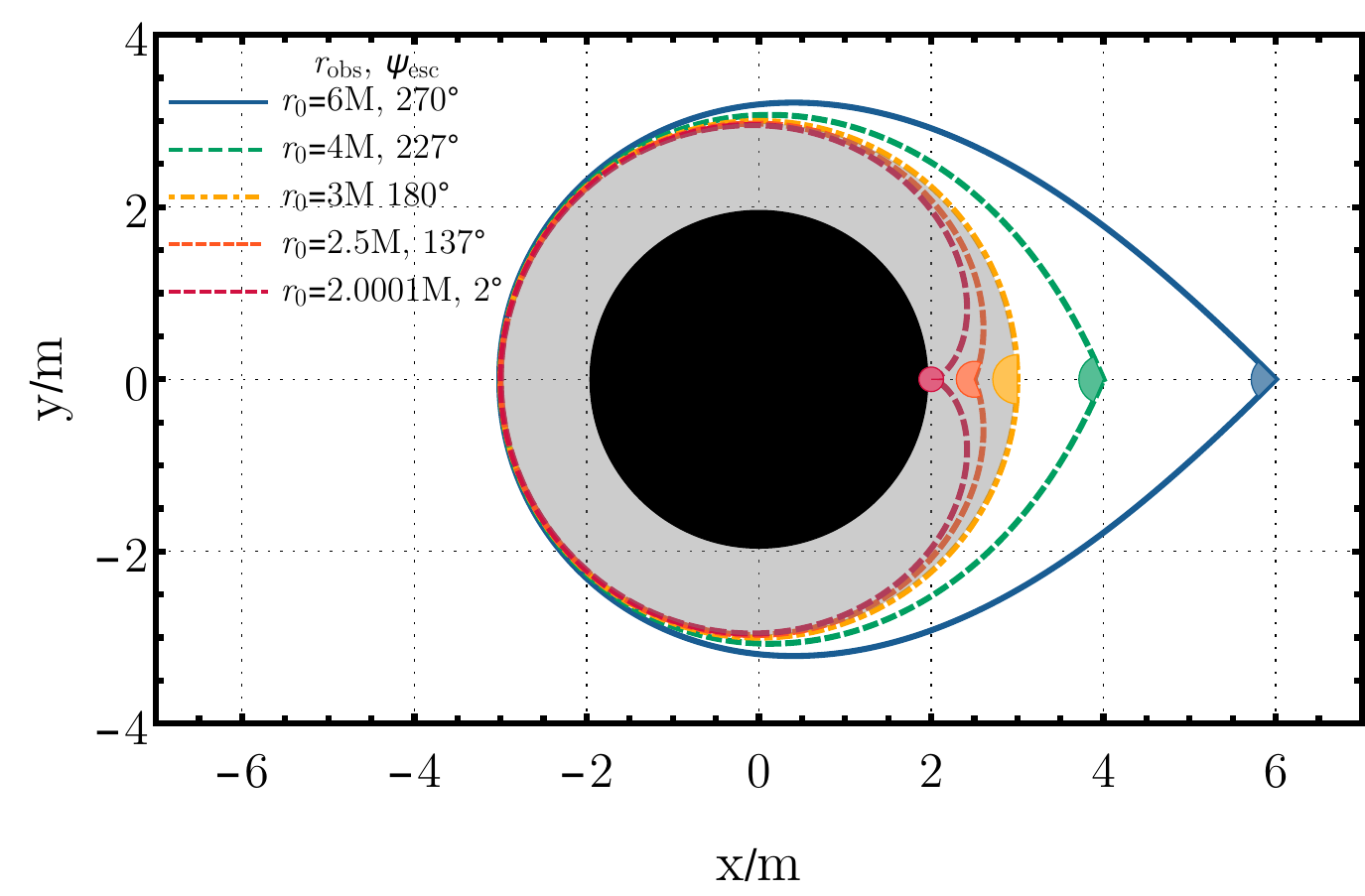} &
\hspace{-1.2cm} \includegraphics[width=0.392\textwidth]{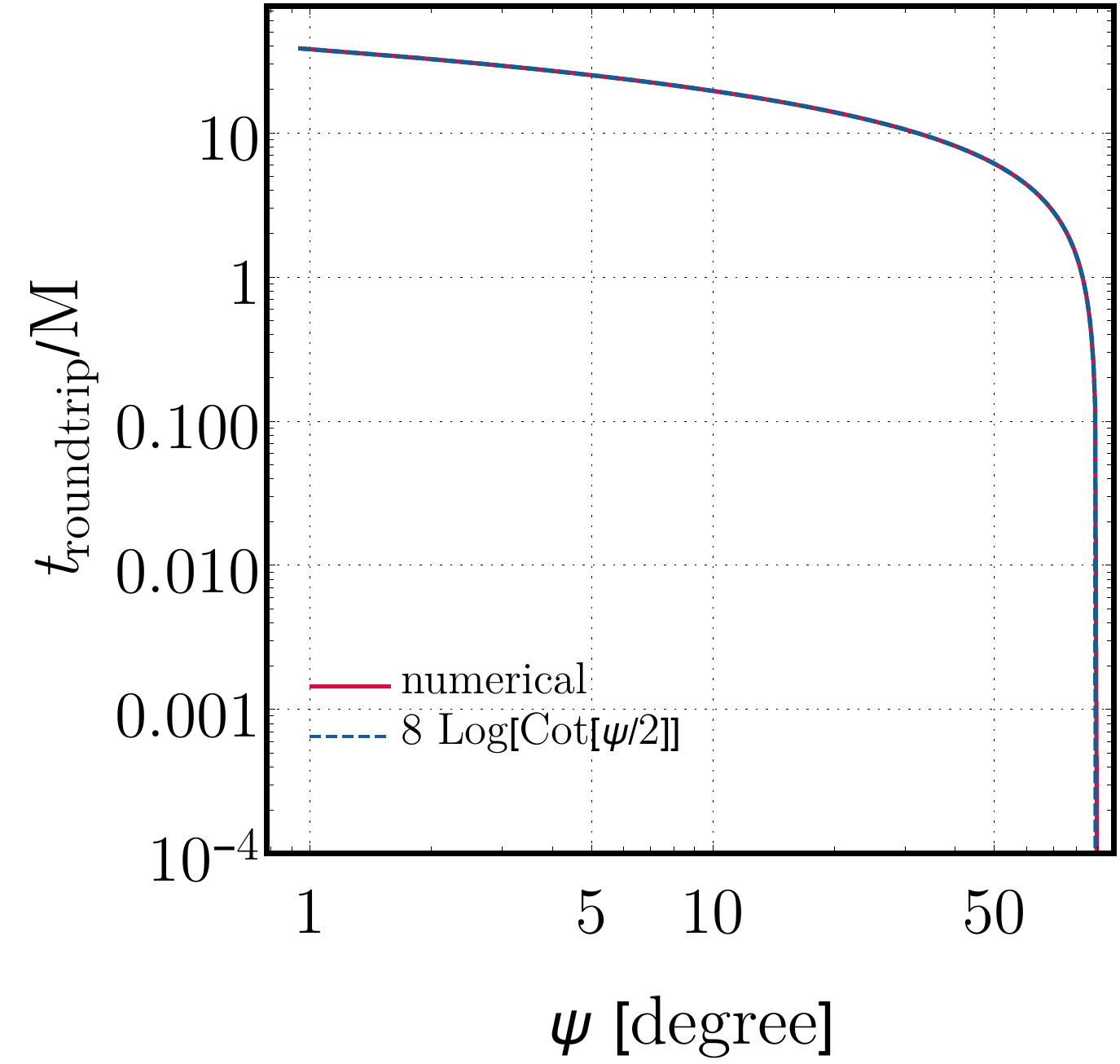}
\end{tabular}
\caption{
Left: Critical escape trajectories of radiation in the Schwarzschild geometry. A locally 
static observer (located at $r=r_{\rm obs}$) emits photons isotropically, but those 
emitted within the colored conical sectors will not reach infinity. The gray shaded area is the photon sphere.
\noindent Right: Coordinate roundtrip time of photons as a function of the emission angle 
$\psi>\psi_{\rm esc}$ and for $\epsilon \ll 1$.
\label{fig:capture}}
\end{figure}

An isolated BH would 
appear truly as a ``hole'' in the sky, since we observe objects by 
receiving the light they either emit or reflect. 
The boundary of this hole, i.e. the ``silhouette'' of a BH, is called 
the \emph{shadow} and is actually larger than the BH horizon and intimately related with the existence of a photon sphere. 

Indeed, according to Eq.~\eqref{radial_motion}, there exists a critical value of the angular 
momentum $L\equiv KME$ for a light ray to be able to escape to infinity. By requiring that 
a light ray emitted at a given point 
will not find turning points in its motion, Eq.~\eqref{radial_motion} yields $K_{\rm esc}=3\sqrt{3}$~\cite{MTB}.
This corresponds to the dimensionless critical impact parameter of a photon at very large distances.
Suppose now that the light ray is emitted by a locally static observer at $r=r_0$.
In the local rest frame, the velocity components of the photon are~\cite{Shapiro:1983du}
\begin{eqnarray}
v_{\varphi}^{\rm local}&=&\frac{MK}{r_0}\sqrt{f_0}\,,\qquad v_{r}^{\rm local}=\sqrt{1-K^2M^2\frac{f_0}{r_0^2}}\,,
\end{eqnarray}
where $f_0\equiv f(r_0)=1-2M/r_0$. 
With this, one can easily compute the escape angle, $\sin\psi_{\rm esc}=3M \sqrt{3f_0}/r_0$.
In other words, the solid angle for escape is
\begin{equation}
 \Delta \Omega_{\rm esc}= 2\pi\left(1-\sqrt{1-\frac{27 M^2(r_0-2M)}{r_0^3}}\right)\sim 27\pi\left(\frac{r_0-2M}{8M}\right)\,, \label{solidangle}
\end{equation}
where the last step is valid for $\epsilon\ll 1$.
For angles larger than these, the light ray falls back and either hits the surface of the object, if there is one, or will be absorbed by the horizon.
The escape angle is depicted in Fig.~\ref{fig:capture} for different emission points $r_0$. The rays that are not able to escape reach a maximum coordinate distance,
\begin{equation}
r_{\rm max} \sim 
2M \left(1+\frac{4f_0M^2}{r_0^2\sin^2\psi}\right)\,.
\end{equation}
This result is accurate away from $\psi_{\rm esc}$, whereas for $\psi\to\psi_{\rm esc}$ 
the photon approaches the photon sphere ($r=3M$).
The coordinate time that it takes for photons that travel initially outward, but eventually turn back and hit the surface of the object,
is shown in Fig.~\ref{fig:capture} as a function of the locally measured angle $\psi$, and is 
of order $\sim M$ for most of the angles $\psi$, for $\epsilon\ll 1$. A closed form 
expression away from $\psi_{\rm crit}$, which describes well the full range (see 
Fig.~\ref{fig:capture}) reads 
\begin{equation}
t_{\rm roundtrip}\sim 8M \log(\cot{\left(\psi/2\right)})\,. \label{troundtrip}
\end{equation}
When averaging over $\psi$, the coordinate roundtrip time is then $32 M\,{\rm Cat}/\pi 
\approx 9.33 M$, for any $\epsilon\ll 1$, where ``Cat'' is Catalan's constant.
Remarkably, this result is independent of $\epsilon$ in the $\epsilon\to0$ limit.

In other words, part of the light coming from \emph{behind} a UCO is ``trapped'' by the 
photon sphere.
If the central object is a good absorber and illuminated with a source far away from it, an observer 
staring at the object sees a ``hole'' in the sky with radius $r_0=3\sqrt{3}M$, which corresponds to the critical impact 
parameter $K_{\rm esc}$.
On the other hand, radiation emitted near the surface of the object (as for example due to an accretion flow) can 
escape to infinity, with an escape angle that vanishes as $\Delta \Omega_{\rm esc}\sim 
\epsilon$ in the $\epsilon\to0$ limit. This simple discussion anticipates that the shadow of a non-accreting UCO can be very similar 
to that of a BH, and that the accretion flow from ECOs with $\epsilon\to0$ can also mimic 
that from an accreting BH~\cite{Vincent:2015xta}.

\subsection{The role of the spin}
While the overall picture drawn in the previous sections is valid also for rotating 
objects, angular momentum introduces qualitatively new features. Spin breaks 
spherical symmetry, introduces frame dragging, and breaks the degeneracy between co- and 
counter-rotating orbits. We focus here on two properties related to the spin which are important
for the phenomenology of ECOs, namely the existence of an ergoregion and the 
multipolar structure of compact spinning bodies.

\subsubsection{Ergoregion}
An infinite-redshift surface outside a horizon is called an \emph{ergosurface} and is the 
boundary of the so-called \emph{ergoregion}. In a stationary spacetime, this boundary is 
defined by the roots of $g_{tt}=0$. Since the Killing vector $\xi^{\mu}=(1,0,0,0)$ becomes 
spacelike in the ergoregion, $\xi^{\mu}\xi^{\mu}g_{\mu\nu}=g_{tt}>0$, the ergosurface is 
also the static limit: an observer within the ergoregion cannot stay still with respect to distant stars; the observer is forced to 
co-rotate with the spacetime due to strong frame-dragging effects.
Owing to this property, negative-energy (i.e., bound) states are possible within the 
ergoregion. This is the chief property that allows for energy and angular momentum 
extraction from a BH through various mechanisms, e.g. the Penrose's process, superradiant 
scattering, the Blandford-Znajek mechanism, etc.~\cite{Brito:2015oca}. 
An ergoregion necessarily exists in the spacetime of a stationary and 
axisymmetric BH and the ergosurface must lay outside the horizon or coincide with 
it~\cite{Brito:2015oca}. 
On the other hand, a spacetime with 
an ergoregion but without an event horizon is linearly unstable (see 
Sec.~\ref{sec:DynSpin}).

\subsubsection{Multipolar structure}\label{sec:multipoles}
As a by-product of the BH uniqueness and no-hair
theorems~\cite{Carter71,Hawking:1973uf} (see also~\cite{Heusler:1998ua,Chrusciel:2012jk,Robinson}),
the multipole moments of any stationary BH in isolation can be written
as~\cite{Hansen:1974zz},
\begin{equation}
 {\cal M}_\ell^{\rm BH}+i {\cal S}_\ell^{\rm BH}  
 ={M}^{\ell+1}\left(i\chi\right)^\ell\,, \label{nohair}
\end{equation}
where ${\cal M}_\ell$ (${\cal S}_\ell$) are the Geroch-Hansen mass (current) 
multipole moments~\cite{Geroch:1970cd,Hansen:1974zz}, the suffix ``BH'' refers to the 
Kerr metric, and
\begin{equation}
\chi\equiv\frac{{\cal S}_1}{{\cal M}_0^2}
\end{equation}
is the dimensionless spin. Equation~\eqref{nohair} implies that ${\cal M}_\ell^{\rm 
BH}$ (${\cal S}_\ell^{\rm BH}$) 
vanish when $\ell$ is odd (even), and that all moments with $\ell\geq2$ can be written 
only in terms of the mass ${\cal M}_0\equiv{M}$ and angular momentum 
${\cal S}_1\equiv{J}$ (or, equivalently, $\chi$) of the BH. Therefore, any 
independent measurement of three multipole moments (e.g. the mass, the spin and the mass 
quadrupole ${\cal M}_2$) provides a null-hypothesis test of the Kerr metric and, in turn, 
it might serve as a genuine strong-gravity confirmation 
of GR~\cite{Psaltis:2008bb,Gair:2012nm,Yunes:2013dva,Berti:2015itd,
Cardoso:2016ryw,Barack:2018yly,Sathyaprakash:2019yqt}.

The vacuum region outside a spinning object is not generically described by the Kerr 
geometry, due to the absence of an analog to Birkhoff's theorem in axisymmetry (for no-hair results around horizonless 
objects see Ref.~\cite{Raposo:2018xkf,Barcelo:2019aif,Quevedo:1991zz}). Thus, the multipole moments of an axisymmetric 
ECO will generically satisfy relations of the form
\begin{eqnarray}
{\cal M}_\ell^{\rm ECO} &=& {\cal M}_\ell^{\rm BH} +\delta {\cal M}_\ell \,, 
\label{mmECOM}\\
{\cal S}_\ell^{\rm ECO} &=& {\cal S}_\ell^{\rm BH} +\delta 
{\cal S}_\ell \,,\label{mmECOS}
\end{eqnarray}
where $\delta {\cal M}_\ell$ and $\delta {\cal S}_\ell$ are model-dependent corrections, whose precise value can be obtained by 
matching the metric describing the interior of the object to that of the exterior. 

For models of ECOs whose exterior is perturbatively close to Kerr, it has been 
conjectured that in the $\epsilon\to0$ limit, the deviations from the Kerr multipole moments (with 
$\ell\geq2$) vanish as~\cite{Raposo:2018xkf} 
\begin{eqnarray}
\frac{\delta {\cal M}_{\ell}}{M^{\ell+1}} &\to& 
a_\ell\frac{\chi^{\ell}}{\log\epsilon}+b_\ell\, \epsilon +...\,,\label{conjectureM}\\
\frac{\delta {\cal S}_{\ell}}{M^{\ell+1}} &\to& 
c_\ell\frac{\chi^{\ell}}{\log\epsilon}+d_\ell\, \epsilon +...\,, \label{conjectureS}
\end{eqnarray}
or \emph{faster}, where $a_\ell$, $b_\ell$, $c_\ell$, and $d_\ell$ are model-dependent numbers  which satisfy certain 
selection and $\mathbb{Z}_2$ rules~\cite{Raposo:2018xkf}. The coefficients $a_\ell$ and $c_\ell$ are related 
to the spin-induced contributions to the multipole moments and are typically of order unity or smaller, whereas the 
coefficients $b_\ell$ and $d_\ell$ are related to the nonspin-induced contributions. It is worth mentioning that, in 
all ECO models known so far, $b_\ell=d_\ell=0$. For example, for ultracompact gravastars $b_\ell=d_\ell=0$ 
for any $\ell$, $a_\ell=0$ ($c_\ell=0$) for odd (even) values of $\ell$, and the first nonvanishing terms are 
$a_2=-8/45$~\cite{Pani:2015tga} and $c_3=-92/315$~\cite{Glampedakis:2017cgd}.

In other words, the deviations of the multipole moments from their corresponding Kerr value must die 
sufficiently fast as the compactness of the object approaches that of a BH, or otherwise the curvature at the surface 
will grow and the perturbative regime breaks down~\cite{Raposo:2018xkf}. The precise way in which the multipoles die 
depends on whether they are induced by spin or by other moments.

Note that the scaling rules~\eqref{conjectureM}~and~\eqref{conjectureS} imply that in this case a quadrupole 
moment measurement will always be dominated by the spin-induced contribution, unless
\begin{equation}
 \chi\ll \sqrt{\epsilon\left|\frac{b_2}{a_2}\log\epsilon\right|}\,. \label{chicritquadrupole}
\end{equation}
For all models known so far, $b_\ell=0$ so obviously only the spin-induced contribution is important. Even 
more in general, assuming $b_2/a_2\sim{\cal O}(1)$, the above upper bound is unrealistically small when $\epsilon\to0$, 
e.g. $\chi\ll 10^{-19}$ when $\epsilon\approx10^{-40}$. This will always be the case, unless some fine-tuning of the 
model-dependent coefficients occurs.

\clearpage
\newpage


\newpage

\section{ECO taxonomy: from DM to quantum gravity\label{sec:taxonomy}}

A nonexhaustive summary of possible self-gravitating compact objects is shown in 
Table~\ref{tab:ECOs}. Different objects arise in different contexts. We refer the reader 
to specific works (e.g., Ref.~\cite{Carballo-Rubio:2018jzw}) for a more comprehensive 
review of the models.

%
\begin{table}[ht!]
\begin{scriptsize}
\begin{tabular}{@{}@{}l@{}|@{}c@{}@{}c@{}@{}c@{}@{}c@{}@{}c@{}@{}}
\hline \noalign{\smallskip}\hline \noalign{\smallskip}
%
Model           & Formation  & Stability           & EM signatures & GWs \\ 
\noalign{\smallskip}
\hline 
\\
Fluid stars     &\yes         &\yes                            &\yes & \yes\\
                &\cite{Shapiro:1983du}            & \cite{1985CQGra...2..219I,Kokkotas:1999bd,Cardoso:2014sna,Saida:2015zva,Stuchlik:2017qiz,Volkel:2017ofl,Volkel:2017kfj}   &   &\cite{Kokkotas:1999bd,Ferrari:2000sr,Cardoso:2014sna,Volkel:2017ofl} \\						
\\
Anisotropic stars & \no     &\yes                             &\yes & \yes\\
	                &         & \cite{Raposo:2018rjn,Dev:2003qd,Doneva:2012rd}   &\cite{Silva:2014fca,Yagi:2015upa,Yagi:2015cda}   &\cite{Yagi:2015upa,Yagi:2015cda,Raposo:2018rjn} \\						
\\
Boson stars \&   &\yes      &\yes                             & \yes &\yes \\
oscillatons      &\cite{Seidel:1991zh,Seidel:1993zk,Okawa:2013jba,Brito:2015yfh,Liebling:2012fv} & \cite{Gleiser:1988ih,Lee:1988av,Honda:2001xg,Cardoso:2007az,Brito:2015pxa,Macedo:2013jja}   &\cite{Vincent:2015xta,Cao:2016zbh,Shen:2016acv}  &\cite{Palenzuela:2007dm,Kesden:2004qx,Choptuik:2009ww,Macedo:2013qea,Cardoso:2016oxy,Cardoso:2017cfl,Sennett:2017etc,Maselli:2017cmm}\\
\\
Gravastars       &\no     & \yes                              &\yes & $\sim$\\
                 &        &\cite{Visser:2003ge,Cardoso:2007az}              & 
\cite{Sakai:2014pga,Uchikata:2015yma,Uchikata:2016qku} & 
\cite{Chirenti:2007mk,Pani:2009hk,Pani:2009ss,Pani:2010em,Chirenti:2016hzd,Cardoso:2016rao,Cardoso:2016oxy,
Uchikata:2016qku,Cardoso:2017cfl,Maselli:2017cmm,Volkel:2017ofl,Volkel:2017kfj}  \\
\\
AdS bubbles      &\no    & \yes                               &$\sim$ & \no\\
                 &       &\cite{Danielsson:2017riq}           & \cite{Danielsson:2017riq} &   \\
\\
Wormholes       &\no    & \yes                               & \yes & $\sim$\\
                &       &\cite{Gonzalez:2008wd,Gonzalez:2008xk,Bronnikov:2012ch,Cuyubamba:2018jdl}         &\cite{Nedkova:2013msa,Ohgami:2015nra,Abdujabbarov:2016efm,Zhou:2016koy}& \cite{Cardoso:2016rao,Cardoso:2017cfl,Maselli:2017cmm}  \\
\\
Fuzzballs        & \no 	&	\no                                & \no            &$\sim$\\
                 &                                & (but see \cite{Cardoso:2005gj,Chowdhury:2007jx,Eperon:2016cdd,Eperon:2017bwq})   &                                    &(but see 
								\cite{Cardoso:2016rao,Cardoso:2016oxy,Hertog:2017vod})\\
\\
Superspinars     &\no	 &\yes	                              & \no                         & $\sim$\\
                 &     & \cite{Cardoso:2008kj,Pani:2010jz}                  & (but see \cite{Patil:2015fua})&  \cite{Cardoso:2016rao,Cardoso:2016oxy} \\
\\
$2-2$ holes      & \no& \no	                               &\no                                & $\sim$\\
~                &                               & (but see \cite{Holdom:2016nek})                     & (but see \cite{Holdom:2016nek})   & \cite{Cardoso:2016rao,Cardoso:2016oxy}\\								
\\
Collapsed       & \no	 & \yes	                                 &\no                      & $\sim$\\
polymers        &   (but see \cite{Brustein:2016msz,Brustein:2017koc})   &\cite{Brustein:2018web}& ~\cite{Brustein:2017koc}\\								
\\
Quantum bounces /  & \no& \no	                                 &\no                                & $\sim$\\
Dark stars      & (but see~\cite{Bambi:2013caa,Barcelo:2007yk})                  &          & 	& ~\cite{Barcelo:2017lnx}\\								
\\
Compact quantum &\no&\no	                                               &\no                             &\yes\\
objects$^{*}$   &\cite{Dvali:2011aa,Dvali:2012rt,Giddings:2014ova}                                &                                                     &                                 &  \cite{Giddings:2019}\\
\\
Firewalls$^{*}$   &\no	     &\no	                                                &\no                                  &$\sim$ \\
~                 &                               &                                                    &                                     &\cite{Barausse:2014tra,Cardoso:2016oxy}  \\
\noalign{\smallskip}\hline \noalign{\smallskip} \hline
\end{tabular}
\end{scriptsize}
\vskip 8pt 
\centering \caption{Catalogue of some proposed horizonless 
compact objects. A $\yes$ tick means that the topic was addressed. With the exception of 
boson stars, however, most of the properties are not fully understood yet.
The symbol $\sim$ stands for incomplete treatment.
An asterisk $^*$ stands for the fact that 
these objects {\it are} BHs, but could have phenomenology similar to the other compact objects in the list.
%
}
\label{tab:ECOs}
\end{table}
%


\subsection{A compass to navigate the ECO atlas: Buchdahl's theorem}
\begin{figure}[th]
\begin{center}
\includegraphics[width=0.85\textwidth]{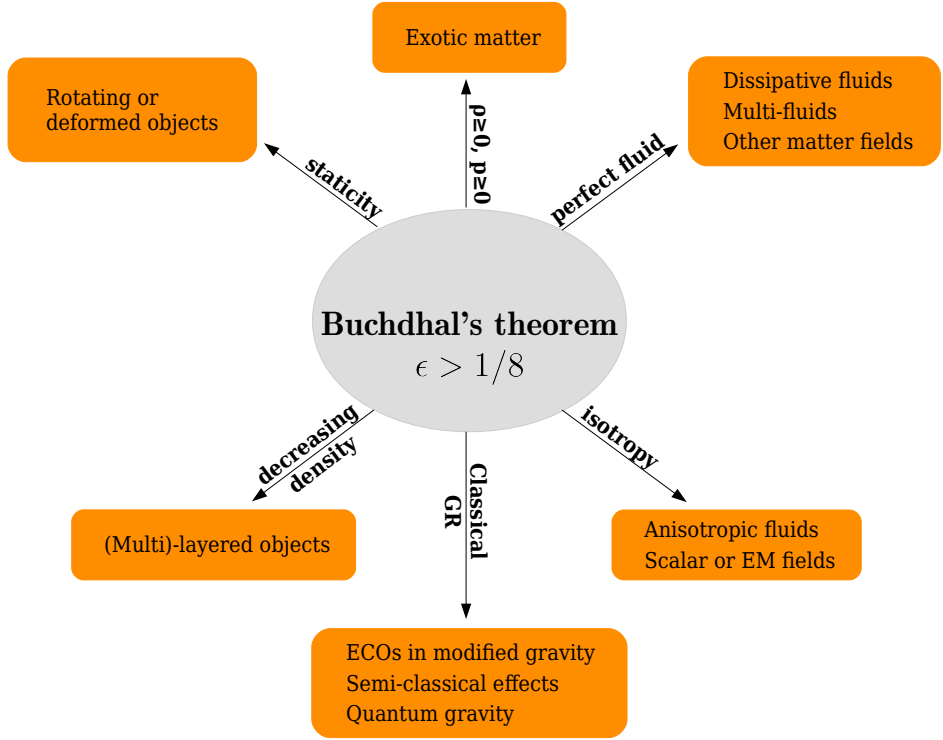}
\caption{
Buchdahl's theorem deconstructed.
\label{fig:Buchdahl}}
\end{center}
\end{figure}

Within GR, Buchdahl's theorem states that, under certain assumptions, the maximum 
compactness of a self-gravitating object is 
$M/r_0=4/9$ (i.e., $\epsilon\geq 1/8$)~\cite{Buchdahl:1959zz}. This result prevents the 
existence of ECOs with compactness arbitrarily close to that of a BH.
A theorem is only as good as its assumptions; one might ``turn it around'' and look at 
the assumptions of Buchdahl's theorem to find possible ways to evade it\footnote{A similar 
approach is pursued to classify possible extensions of GR~\cite{Berti:2015itd}.}.
More precisely, Buchdahl's theorem assumes that~\cite{Urbano:2018nrs}:
\begin{enumerate}
  \item GR is the correct theory of gravity;
  \item The solution is spherically symmetric;
  \item Matter is described by a single, perfect fluid;
  \item The fluid is either isotropic or mildly anisotropic, in the sense that the 
tangential pressure is smaller than the radial one, $P_r\gtrsim P_t$;
  \item The radial pressure and energy density are non-negative, $P_r\geq0$, $\rho\geq0$.
  \item The energy density decreases as one moves outwards, $\rho'(r)<0$.
\end{enumerate}
Giving up each of these assumptions (or combinations thereof) provides a way to 
circumvent the theorem and suggests a route to classify ECOs based on which of the 
underlying assumptions of Buchdahl's theorem they violate (see Fig.~\ref{fig:Buchdahl}).

%

\subsection{Self-gravitating fundamental fields}
One of the earliest and simplest known examples of a self-gravitating compact configuration is 
that of a (possibly complex) minimally-coupled massive scalar field $\Phi$, described by the action
\begin{eqnarray}
S=\int d^4x \sqrt{-g} \left( \frac{R}{16\pi} -g^{\mu\nu}\bar{\Phi}^{}_{,\mu}\Phi^{}_{,\nu} 
-\frac{\mu_S^2\bar{\Phi}\Phi}{2}
\right)\,.\label{eq:MFaction}
\end{eqnarray}
The mass $m_S$ of the scalar is related to the mass parameter as $m_{S}=\hbar\mu_{S}$, 
and the theory is controlled by the dimensionless coupling
\begin{equation}
\frac{G}{c\hbar} M\mu_{S} = 7.5\cdot 10^{9} \left(\frac{M}{M_{\odot}}\right) 
\left(\frac{m_{S}c^2}{\rm eV}\right)\,,\label{dimensionless_massparameter}
\end{equation}
where $M$ is the total mass of the bosonic configuration. 

Self-gravitating solutions for the theory above are broadly referred to as boson stars, 
and can be generalized through the inclusion
of nonlinear self-interactions~\cite{Kaup:1968zz,Ruffini:1969qy,Khlopov:1985jw,Seidel:1991zh,
Guth:2014hsa,Brito:2015pxa,Minamitsuji:2018kof}
(see Refs.~\cite{Jetzer:1991jr,Schunck:2003kk,Liebling:2012fv,Macedo:2013jja} for reviews). 
If the scalar is {\it complex}, there are {\it static}, spherically-symmetric  
geometries, while the field itself oscillates~\cite{Kaup:1968zz,Ruffini:1969qy} (for 
reviews, see Refs.~\cite{Jetzer:1991jr,Schunck:2003kk,Liebling:2012fv,Macedo:2013jja}). 
Analogous solutions for complex massive vector fields were also shown to 
exist~\cite{Brito:2015pxa}. Recently, multi-oscillating boson stars which are not exactly
static spacetimes were constructed, and these could represent intermediate states between static boson stars which underwent 
violent dynamical processes~\cite{Choptuik:2019zji}.
On the other hand, {\it real} scalars give rise to long-term 
stable oscillating geometries, but with a non-trivial time-dependent stress-energy 
tensor, called oscillatons~\cite{Seidel:1991zh}. Both solutions arise naturally as the end-state 
of gravitational collapse~\cite{Seidel:1991zh,Garfinkle:2003jf,Okawa:2013jba}, and both 
structures share similar features.

Static boson stars form a one-parameter family of solutions governed by the value of the bosonic field at the 
center of the star. The mass $M$ displays a maximum above which the configuration is 
unstable against radial perturbations, just like ordinary stars. The maximum mass 
and compactness of a boson star depend strongly on the boson self-interactions. As a rule 
of thumb, the stronger the self-interaction the higher the maximum compactness and mass 
of a stable boson stars~\cite{Schunck:2003kk,Liebling:2012fv} (see Table~\ref{tab:BSs}). 

\begin{table}[th]
\begin{center}
\scriptsize
\begin{tabular}{|c|c|c|}
\hline\hline
Model & \begin{tabular}{@{}c@{}}Potential \\ $V(|\Phi|^2)$\end{tabular}  & 
\begin{tabular}{@{}c@{}}Maximum mass \\ $M_{\max}/M_{\odot}$\end{tabular}\\
\hline
Minimal~\cite{Kaup:1968zz,Ruffini:1969qy} & $\mu^2|\Phi|^2$ & $8\left(\frac{10^{-11}{\rm 
eV}}{m_S}\right)$\\
Massive~\cite{Colpi:1986ye} & $\mu^2|\Phi|^2 + \frac{\alpha}{4}|\Phi|^4$ & 
$5\,\sqrt{\alpha\hbar}\left(\frac{0.1\,{\rm GeV}}{m_S}\right)^2$\\
Solitonic~\cite{Friedberg:1986tq} & 
$\mu^2|\Phi|^2\left[1-\frac{2|\Phi|^2}{\sigma_0^2}\right]^2$ & 
$5\left[\frac{10^{-12}}{\sigma_0}\right]^2\left(\frac{500\,{\rm GeV}}{m_S}\right)$\\
\hline\hline
\end{tabular}
\end{center}
\caption{Scalar potential and maximum mass for some scalar boson-star models. In 
our units, the scalar field $\Phi$ is dimensionless and the potential $V$ has dimensions 
of an inverse length squared. The bare mass of the scalar field is $m_S:=\mu\hbar$.
For minimal boson stars, the scaling of the maximum mass is exact.
For massive boson stars and solitonic boson stars, the scaling of the maximum mass is 
approximate and holds only when $\alpha\gg \mu^2$ and when $\sigma_0\ll1$, respectively. 
Adapted from Ref.~\cite{Cardoso:2017cfl}.}
\label{tab:BSs}
\end{table}

\begin{figure}[ht!]
\begin{center}
\begin{tabular}{cc}
\includegraphics[width=0.49\textwidth]{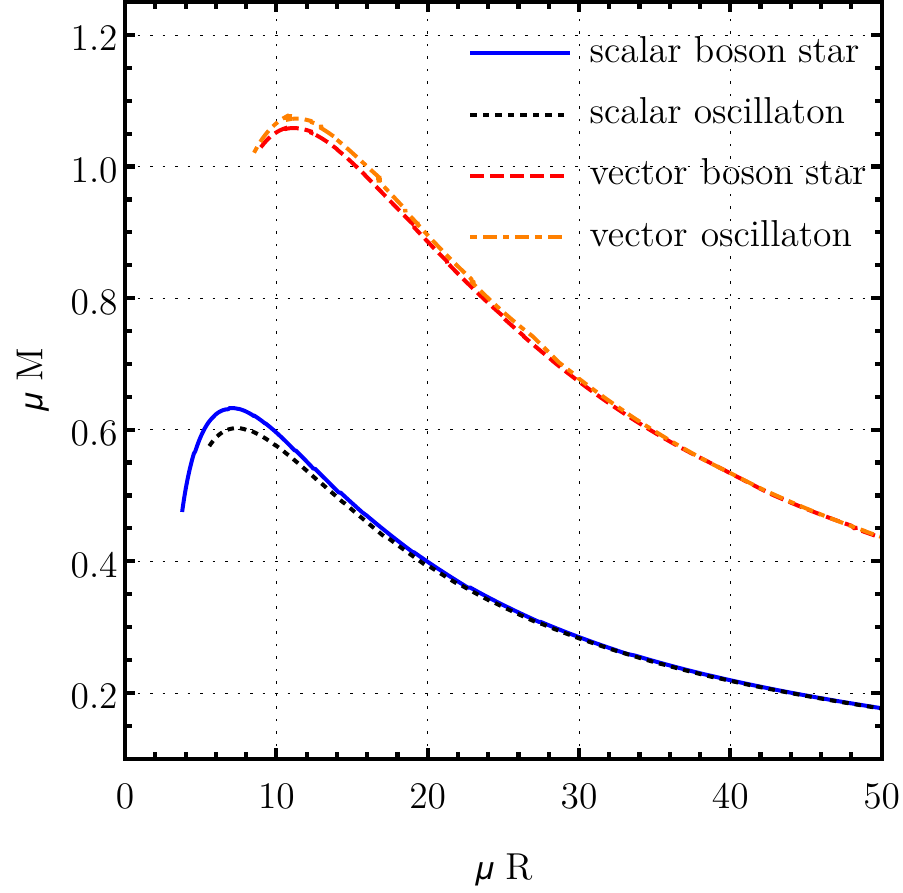}
\includegraphics[width=0.49\textwidth]{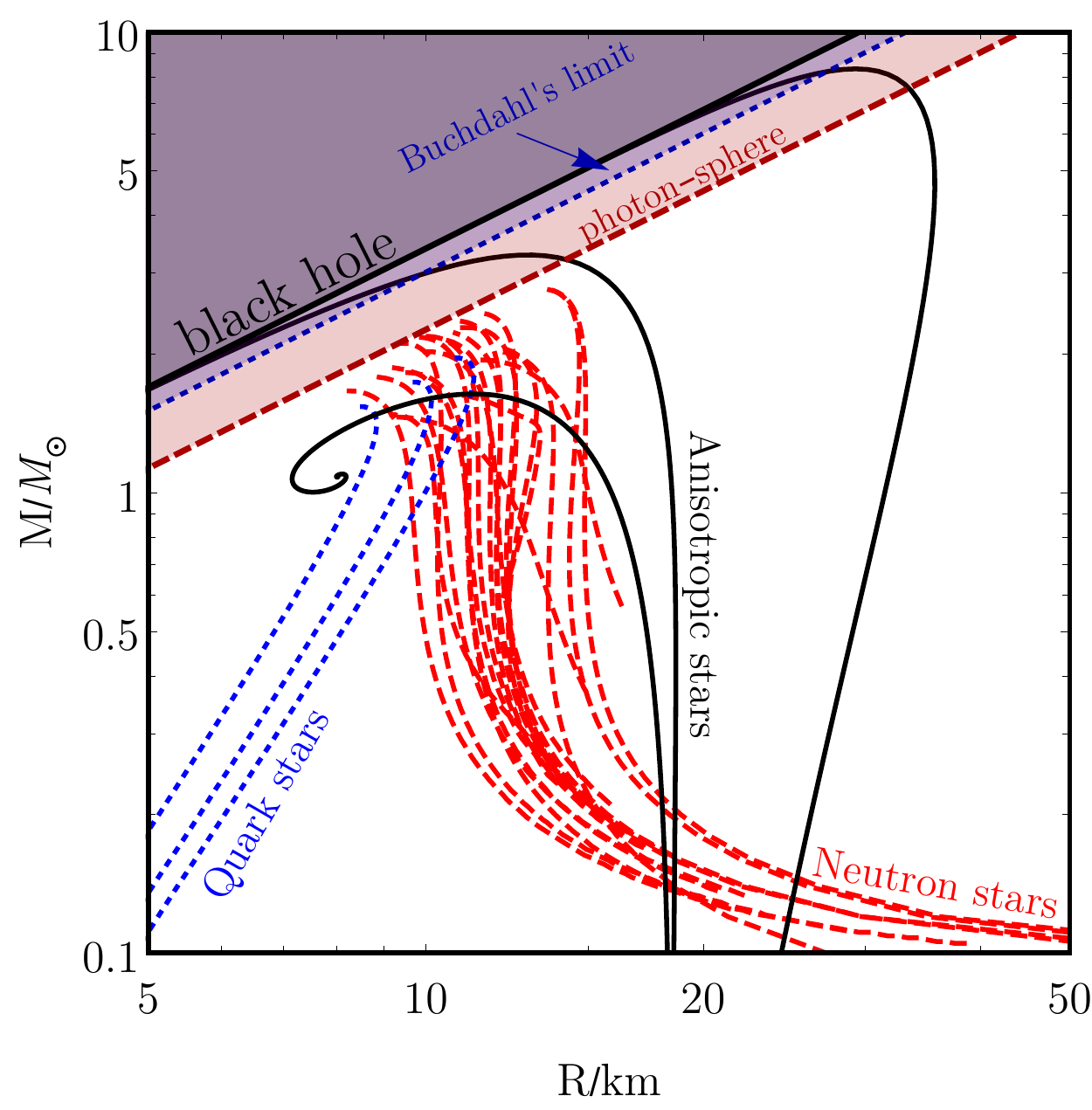}
\end{tabular}
\caption{{\bf Left:} Comparison between the total mass of a boson star ({\it complex} 
scalar or vector fields) and an oscillaton ({\it real} scalar or vector fields), as a 
function of their radius $R$. $R$ is defined as the radius containing 98\% of the total 
mass. The procedure to find the diagram is outlined in the main text. From 
Ref.~\cite{Brito:2015yfh}.
{\bf Right:} Mass-radius diagram for nonspinning fluid stars in GR. The red dashed (blue 
dotted) lines are ordinary NSs (quark stars) for several representative equations of 
state~\cite{Lattimer:2006xb,Ozel:2016oaf} (data taken from~\cite{OzelFreireWeb}); the 
black continuous lines are strongly-anisotropic stars~\cite{Raposo:2018rjn}. Note that 
only the latter have photon spheres in their exterior and violate 
Buchdahl's bound. \label{fig:MvsR}}
\end{center}
\end{figure}

The simplest boson stars are moderately
compact in the nonspinning case~\cite{Macedo:2013jja,Brito:2015pxa,Grandclement:2016eng}.
Their mass-radius relation is shown in Fig.~\ref{fig:MvsR}.
Once spin~\cite{Grandclement:2016eng} or nonlinear 
interactions~\cite{Colpi:1986ye,Macedo:2013jja,Friedberg:1986tq} are added, boson star 
spacetimes can have light rings and ergoregions. 
The stress-energy tensor of a self-interacting bosonic field 
contains anisotropies, which in principle allow to evade naturally Buchdahl's theorem.
However, there are no boson-star solutions which evade the Buchdahl's bound: in the 
static case, the most compact configuration has $r_0\approx 2.869M$ ($\epsilon\approx 
0.44$)~\cite{Kesden:2004qx}.

There seem to be no studies on the classification of 
such configurations (there are solutions known to display photon spheres, but it is 
unknown whether they can be as compact as ClePhOs)~\cite{Kesden:2004qx,Grandclement:2016eng}.

Because of their simplicity and fundamental character, boson stars are interesting on 
their own. A considerable interest in their properties arose with the understanding that 
light scalars are predicted to occur in different scenarios, and ultralight scalars can 
explain the DM puzzle~\cite{Hui:2016ltb}. Indeed, dilute bosonic configurations 
provide an alternative model for DM halos.

%

\subsection{Perfect fluids}
The construction of boson stars is largely facilitated by their statistics, which allow 
for a large number of bosons to occupy the same level. Due to Pauli's exclusion principle, 
a similar construction for fermions is therefore more challenging, and approximate 
strategies have been devised~\cite{Ruffini:1969qy,Shapiro:1983du}. 
In most applications, such fundamental description is substituted by an effective 
equation of state, usually of polytropic type, which renders the corresponding Einstein 
equations much easier to solve~\cite{Shapiro:1983du}.

When the stresses are assumed to be isotropic, static spheres in GR made of ordinary fluid 
satisfy the Buchdahl limit on their compactness, $2M/r_0<8/9$~\cite{Buchdahl:1959zz}; 
strictly speaking, they would not qualify as a ClePhO. However, GWs couple very weakly to 
ordinary matter and can travel unimpeded right down to the center of stars. Close to the 
Buchdahl limit, the travel time is extremely large, $\tau\sim \epsilon^{-1/2}M$, and in 
practice such objects would behave as ClePhOs~\cite{Pani:2018flj}. In addition, polytrope 
stars with a light ring (sometimes referred to as ultra-compact stars) {\it always} have 
superluminal sound speed~\cite{Saida:2015zva}. Neutron stars --~the only object in our 
list for which there is overwhelming evidence~-- are not expected to have light rings nor 
behave as ClePhOs for currently accepted equations of state~\cite{1985CQGra...2..219I}. The mass-radius relation for 
a standard neutron star is shown in Fig.~\ref{fig:MvsR}.

Some fermion stars, such as neutron stars, live in DM-rich environments. Thus, 
DM can be captured by the star due to gravitational deflection and a 
non-vanishing cross-section for collision with the star 
material~\cite{Press:1985ug,Gould:1989gw,Goldman:1989nd,Bertone:2007ae,Goldman:1989nd}. 
The DM material eventually thermalizes with the star, producing a composite 
compact object. Compact solutions made of both a perfect fluid and a massive 
complex~\cite{Henriques:1989ar,Henriques:1989ez,Lopes:1992np,Henriques:2003yr,
Sakamoto:1998aj,Pisano:1995yk} or real scalar or vector 
field~\cite{Brito:2015yga,Brito:2015yfh} were built, and model the effect of bosonic DM 
accretion by compact stars. Complementary to these studies, accretion of fermionic DM has 
also been considered, by modeling the DM core with a perfect fluid and constructing a 
physically motivated equation 
of state~\cite{Leung:2011zz,Leung:2013pra,Tolos:2015qra}.  
The compactness of such stars 
is similar to that of the host neutron stars, and does not seem to exceed the Buchdahl 
limit.

\subsection{Anisotropic stars}
The Buchdahl limit can be circumvented when the object is subjected to large anisotropic 
stresses~\cite{Andreasson:2007ck}. These might arise in a variety of contexts: at high 
densities~\cite{KippenhahnBook,Ruderman:1972aj,Canuto:1974gi}, when EM or fermionic fields 
play a role, or in pion condensed phase configurations in neutron 
stars~\cite{Sawyer:1973fv}, 
superfluidity~\cite{Carter:1998rn}, solid cores~\cite{KippenhahnBook}, etc. In fact, 
anisotropy is common and even a simple soap bubble support anisotropic 
stresses~\cite{Guven:1999wm}. Anisotropic stars were studied in GR, mostly at the level of 
static spherically symmetric 
solutions~\cite{1974ApJ...188..657B,Letelier:1980mxb,Bayin:1982vw,Dev:2000gt,Mak:2001eb,
Dev:2003qd,Herrera:2004xc,Andreasson:2007ck,1976A&A....53..283H,Doneva:2012rd,
Silva:2014fca,Yagi:2015hda,Yagi:2015upa,Yagi:2016ejg}. These studies are not covariant, 
which precludes a full stability analysis or nonlinear evolution of such spacetimes. 
Progress on this front has been achieved 
recently~\cite{Carloni:2017bck,Isayev:2018hqx,Raposo:2018rjn}.

The compactness of very anisotropic stars may be arbitrarily close to that of a BH;
compact configurations can exceed the Buchdahl limit, and some can be classified as 
ClePhOs. In some of these models, compact stars exist {\it across a wide range of 
masses}, evading one of the outstanding issues with BH mimickers, i.e. that most approach the BH compactness in a 
very limited range of masses, thus being unable to describe both stellar-mass 
and supermassive BH candidates across several orders of magnitude in mass~\cite{Raposo:2018rjn}. 
Such property of BHs in GR, visible in Fig.~\ref{fig:MvsR}, is a consequence of the scale-free character of the 
vacuum field equations. It is extremely challenging to reproduce once a scale is present, as expected for material 
bodies.
Fig.~\ref{fig:MvsR} summarizes the mass-radius relation for fluid stars.

\subsection{Quasiblack holes}
An interesting class of families of BH-mimickers, the quasiblack
holes, consist on extremal (charged and/or spinning) {\it
regular} spacetimes. These objects can be thought of as stars, on the
verge of becoming BHs~\cite{Lemos:2003gx,Lemos:2008cv}.

\subsection{Wormholes}
Boson and fermion stars discussed above arise from a simple theory, with relatively simple equations of motion,
and have clear dynamics. Their formation mechanism is embodied in the field equations and 
requires no special initial data.
On the other hand, the objects listed below are, for the most part, generic constructions 
with a well-defined theoretical motivation, but for which the formation mechanisms are 
not well understood.

Wormholes were originally introduced by Einstein and Rosen, as an attempt to describe particles~\cite{Einstein:1935tc}. They were (much)
later popularized as a useful tool to teach GR, its mathematical 
formalism and underlying geometric description of the 
universe~\cite{Morris:1988cz,Visser:1995cc,Lemos:2003jb}. Wormholes connect different 
regions of spacetime. Within GR they are not vacuum spacetimes and require matter. The realization 
that wormholes can be stabilized and constructed with possibly reasonable matter has 
attracted a considerable attention on these 
objects~\cite{Visser:1995cc,Lemos:2003jb,Maldacena:2018gjk}.

\begin{figure}[th]
\begin{center}
\epsfig{file=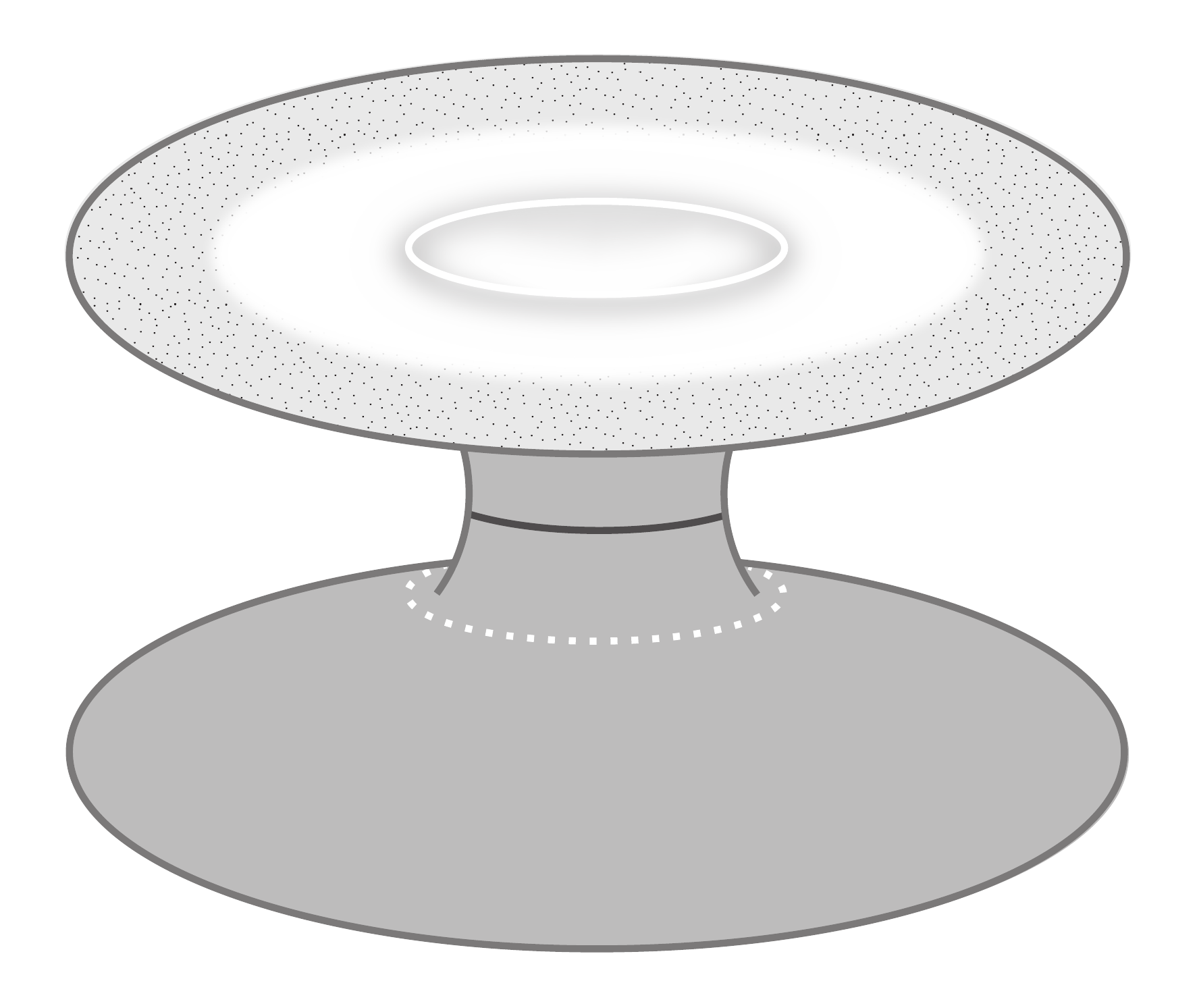,width=0.5\textwidth,angle=0,clip=true}
\caption{Embedding-like diagram of a wormhole connecting two different asymptotically-flat universes.
The black solid line denotes the wormhole's throat. There are two light rings in the spacetime, one for which universe.
\label{fig:embedding}}
\end{center}
\end{figure}
Different wormhole spacetimes can have very different properties. Since we are interested 
in understanding spacetimes that mimic BHs, consider the following two simple examples of 
a non-spinning geometries~\cite{Visser:1995cc,Damour:2007ap,Cardoso:2016rao}.
In the first example, we simply take the Schwarzschild geometry describing a mass $M$ down to a ``throat'' radius $r_0>2M$.
At $r_0$, we ``glue'' such spacetime to another copy of Schwarzschild. In Schwarzschild coordinates, the two metrics are identical and described by 
\begin{equation}
ds^2=-\left(1-\frac{2M}{r}\right) dt^2+\left(1-\frac{2M}{r}\right)^{-1} dr^2+r^2d\Omega^2\,.
\end{equation}
Because Schwarzschild's coordinates do not extend to $r<2M$, we use the tortoise 
coordinate $dr/dr_*=\pm \left(1-2M/r\right)$, 
to describe the full spacetime, where the upper and lower sign refer to the two different 
universes connected at the throat. Without loss of generality we assume $r_*(r_0)=0$, so 
that one domain is $r_*>0$ whereas the other domain is $r_*<0$. The surgery at the throat 
requires a thin shell of matter with surface density and surface 
pressure~\cite{VisserBook}
\begin{equation}
\sigma =-\frac{1}{2\pi r_0}\sqrt{1-2M/r_0}\,, \qquad p =\frac{1}{4\pi r_0} \frac{(1-M/r_0)}{\sqrt{1-2M/r_0}}\,,
\end{equation}
Although the spacetime is everywhere vacuum (except at the throat) the junction conditions force the pressure to be large 
when the throat is close to the Schwarzschild radius.

A similar example, this time of a non-vacuum spacetime, is the following geometry~\cite{Damour:2007ap}
\begin{equation}
ds^2=-\left(1-\frac{2M}{r}+\lambda^2\right)dt^2+\left(1-\frac{2M}{r}\right)^{-1}dr^2+r^2d\Omega^2\,.\label{solodukhin}
\end{equation}
The constant $\lambda$ is assumed to be extremely small, for example $\lambda\sim e^{-M^2/\ell_P^2}$ where $\ell_P$ is the Planck length. There is no event horizon at $r=2M$, such location is now the spacetime throat. Note that, even though such spacetime was constructed to be arbitrarily close to the Schwarzschild spacetime, the throat at $r=2M$ is a region of large (negative) curvature, for which the Ricci and Kretschmann invariant are, respectively,
\beq
R=-\frac{1}{8\lambda^2M^2}\,,\quad R_{abcd}R^{abcd}=\frac{1 + 24 \lambda^4}{64 \lambda^4 M^4}\,.
\eeq
Thus, such invariants diverge at the throat in the small $\lambda$-limit. A more general discussion on several 
wormholes models is presented in Ref.~\cite{Lemos:2008cv}.

The above constructions show that wormholes can be constructed to have any arbitrary mass 
and compactness. The procedure is oblivious to the formation mechanism, it is unclear if 
these objects can form without carefully tuned initial conditions, nor if they are stable.
Wormholes in more generic gravity theories have been constructed, some of which
can potentially be traversable~\cite{Shaikh:2016dpl,Chianese:2017qlx,Hohmann:2018shl,Shaikh:2018yku,Khaybullina:2018zcj}.
In such theories, energy conditions might be satisfied~\cite{Kanti:2011jz}. Generically however, wormholes are 
linearly unstable~\cite{Gonzalez:2008wd,Gonzalez:2008xk,Bronnikov:2012ch,Cuyubamba:2018jdl}.
\subsection{Dark stars}
Quantum field theory around BHs or around dynamic horizonless objects gives rise to phenomena such as particle creation.
Hawking evaporation of astrophysical BHs, and corresponding back-reaction on the geometry is negligible~\cite{Birrell:1982ix}.
Quantum effects on collapsing {\it horizonless} geometries (and the possibility of halting 
collapse to BHs altogether) are less 
clear~\cite{Visser:2009pw,Zeng:2016epp,Chen:2017pkl,Berthiere:2017tms,Buoninfante:2019swn,Terno:2019kwm,
Malafarina:2017csn}. There are arguments that semiclassical effects might suffice to halt collapse and to produce {\it 
dark stars}, even for macroscopic 
configurations~\cite{Barcelo:2009tpa,Kawai:2013mda,Barcelo:2015noa,Baccetti:2016lsb,Baccetti:2017oas,
Carballo-Rubio:2017tlh,Baccetti:2018qrp}, but see Ref.~\cite{Chen:2017pkl} for counter-arguments. For 
certain conformal fields, it was shown that a possible end-state are precisely wormholes 
of the form \eqref{solodukhin}. Alternative proposals, made to solve the information 
paradox, argue that dark stars could indeed arise, but as a ``massive remnant'' end state 
of BH evaporation~\cite{Giddings:1992hh,Unruh:2017uaw}.

\subsection{Gravastars}
Similar ideas that led to the proposal of ``dark stars'' were also in the genesis of a 
slightly different object, ``gravitational-vacuum stars'' or {\it 
gravastars}~\cite{Mazur:2001fv,Mazur:2004fk}. These are configurations supported by a 
negative pressure, which might arise as an hydrodynamical description of one-loop QFT 
effects in curved spacetime, so they do not necessarily require exotic new 
physics~\cite{Mottola:2006ew}. In these models, the Buchdahl limit is evaded both because the 
internal effective fluid is anisotropic~\cite{Cattoen:2005he} and because the pressure is negative (and thus violates 
some of the energy conditions~\cite{Mazur:2015kia}). Gravastars have been recently generalized to 
include anti-de Sitter cores, in what was termed {\it AdS bubbles}, and which may allow 
for holographic descriptions~\cite{Danielsson:2017riq,Danielsson:2017pvl}.
Gravastars are a very broad class of objects, and can have arbitrary compactness, 
depending on how one models the supporting pressure.
The original gravastar model was a five-layer construction, with an interior de Sitter 
core, a thin shell connecting it to a perfect-fluid region, and another thin-shell 
connecting it to the external Schwarzschild patch. 
A simpler construction that features all the main ingredients of the original gravastar 
proposal is the thin-shell gravastar~\cite{Visser:2003ge}, in which a de Sitter core is 
connected to a Schwarzschild exterior through a thin shell of perfect-fluid matter.
Gravastars can also be obtained as the BH-limit of constant-density stars, past the Buchdahl limit~\cite{Mazur:2015kia,Camilo:2018goy}.
It is interesting that such stars were found to be dynamically stable in this regime~\cite{Camilo:2018goy}.
It has been conjectured that gravastars are a natural outcome of the inflationary universe~\cite{Wang:2018cum},
or arising naturally within the gauge-gravity duality~\cite{Danielsson:2017riq,Danielsson:2017pvl}.

\subsection{Fuzzballs and collapsed polymers}
So far, quantum effects were dealt with at a semi-classical level only.
A proper theory of quantum gravity needs to be able to solve some of the inherent problems in BH physics, such as 
the lack of unitarity in BH evaporation or the origin and nature of the huge Bekenstein-Hawking entropy $S=k_{\rm B}c^3A/(4\hbar G)$ ($k_{\rm B}$ is Boltzmann's constant and $A$ is the BH area).
In other words, what is the statistical-mechanical account of BH entropy in terms of some microscopic degrees of freedom? String theory is able to provide a partial answer to some of these questions.
In particular, for certain (nearly) supersymmetric BHs, the Bekenstein-Hawking entropy,
as computed in the strongly-coupled supergravity description, can be reproduced in a weakly-coupled $D$-brane description
as the degeneracy of the relevant microstates~\cite{Strominger:1996sh,Peet:1997es,Das:2000su,David:2002wn,Bena:2007kg}.

Somewhat surprisingly, the geometric description of {\it individual microstates} seems to be regular and 
horizonless~\cite{Myers:1997qi,Mathur:2005zp,Bena:2007kg,Balasubramanian:2008da,Bena:2013dka}. This led to the ``fuzzball'' description of classical BH geometries, where a BH is dual to an ensemble of such microstates. In this picture, the BH geometry emerges 
in a coarse-grained description which ``averages'' over the large number of coherent superposition of microstates, producing an 
effective horizon at a radius where the individual microstates start to ``differ appreciably'' from one another~\cite{Lunin:2002qf,Lunin:2001jy}. In this description, quantum gravity effects are not confined close to the BH singularity, rather the entire interior of the BH is ``filled'' by fluctuating geometries -- hence this picture is often referred to as the ``fuzzball'' description of BHs.

Unfortunately, the construction of microstates corresponding to a fixed set of global charges has only been achieved in 
very special circumstances, either in higher-dimensional or in non asymptotically-flat spacetimes. Explicit regular, 
horizonless microstate geometries for asymptotically flat, four-dimensional spacetimes that could describe astrophysical 
bodies have not been constructed. Partly because of this, the properties of the geometries are generically unknown. 
These include the ``softness'' of the underlying microstates when interacting with GWs or light; the curvature radius or 
redshift of these geometries in their interior; the relevant lengthscale that indicates how far away from the 
Schwarzschild radius is the fuzziness relevant, etc.

A similar motivation led to the proposal of a very different BH interior in Refs.~\cite{Brustein:2016msz,Brustein:2017kcj};
the interior is described by an effective equation of state corresponding to a gas of highly excited strings close to 
the Hagedorn temperature. The behavior of such gas is similar to some polymers, and this was termed the 
``collapsed polymer'' model for BH interiors. In both proposals, large macroscopic BHs are described by objects with a 
regular interior, and the classical horizon is absent. In these models, our parameter $\epsilon$ is naturally of the 
order $\sim{\cal 
O}(\ell_P/M)\in(10^{-39},10^{-46})$ for masses in the range $M\in(10,10^8)M_\odot$.

\subsection{``Naked singularities'' and superspinars}
Classical GR seems to be protected by Cosmic Censorship, in that evolutions leading to spacetime singularities
also produce horizons cloaking them. Nevertheless, there is no generic proof that cosmic censorship
is valid, and it is conceivable that it is a fragile, once extensions of GR are allowed.
A particular impact of such violations was discussed in the context of the Kerr geometry describing spinning BHs.
In GR, the angular momentum $J$ of BHs is bounded from above by $J\leq GM^2/c$. In string theory however,
such ``Kerr bound'' does not seem to play any fundamental role and could conceivably receive large corrections. It is 
thus possible that there are astrophysical objects where it is violated. Such objects were termed {\it 
superspinars}~\cite{Gimon:2007ur}, but it is part of a larger class of objects which would arise if singularities (in 
the classical theory of GR) would be visible.
The full spacetime description of superspinars and other such similar objects is lacking: to avoid singularities and 
closed-timelike curves unknown quantum effects need to be invoked to create an effective surface somewhere in the 
spacetime.
There are indications that strong GW bursts are an imprint of such objects~\cite{Harada:2000jya}, but a complete 
theory is necessary to understand any possible signature.

\subsection{$2-2$ holes and other geons}
As we remarked already, the questioning of the BH paradigm in GR comes hand in hand with the search for
an improved theory of the gravitational interaction, and of possible quantum effects.
A natural correction to GR would take the form of higher-curvature terms in the Lagrangian ${\cal 
L}=R+c_1R^2+c_2R_{abcd}R^{abcd}+...$ with couplings $c_j$ suppressed by some 
scale~\cite{Stelle:1976gc,Voronov:1984kq,Holdom:2015kbf}. The study of (shell-like) matter configurations in such 
theories revealed the existence novel horizonless configurations, termed ``2-2-holes'', which closely matches the 
exterior Schwarzschild solution down to about a Planck proper length of the Schwarzschild radius of the 
object~\cite{Holdom:2016nek,Ren:2019afg}. In terms of the parameter $\epsilon$ introduced above, the theory predicts 
objects where $\epsilon\sim (\ell_P/M)^2\in(10^{-78},10^{-92})$~\cite{Holdom:2016nek,Ren:2019afg}. The existence and 
stability of proper star-like configurations was not studied. More generic theories result in a richer range of 
solutions, many of which are solitonic in nature and 
can be ultracompact (see e.g. Refs.~\cite{BeltranJimenez:2017doy,Afonso:2017aci,Afonso:2018bpv,Franzin:2018pwz,Sebastiani:2018ktb}.
Recently, a quantum mechanical framework to describe astrophysical, horizonless objects devoid of curvature 
singularities was put forward in the context of nonlocal gravity (arising from infinite derivative 
gravity)~\cite{Koshelev:2017bxd,Buoninfante:2018rlq,Buoninfante:2019swn}. The corresponding stars can be ultracompact, 
although never reaching the ClePhO category.

\subsection{Firewalls, compact quantum objects and dirty BHs}
Many of the existing proposals to solve or circumvent the breakdown of unitarity in BH evaporation 
involve changes in the BH structure, without doing away with the horizon.
Some of the changes could involve ``soft'' modifications of the near-horizon region, such 
that the object still looks like a regular GR BH~\cite{Giddings:2017mym,Giddings:2013kcj,Giddings:2019}. 
However, the changes could also be drastic and involve ``hard'' structures localized 
close to the horizon such as firewalls and other compact quantum objects~\cite{Almheiri:2012rt,Kaplan:2018dqx,Giddings:2019}. 
Alternatively, a classical BH with modified dispersion relations for the graviton could effectively appear as having a 
hard surface~\cite{Zhang:2017jze,Oshita:2018fqu}. A BH surrounded by some hard structure -- of quantum origin such as 
firewalls, or classical matter piled up close to the horizon -- behaves for many purposes as a compact horizonless 
object.

\begin{figure}[th]
\begin{center}
\includegraphics[width=1\textwidth]{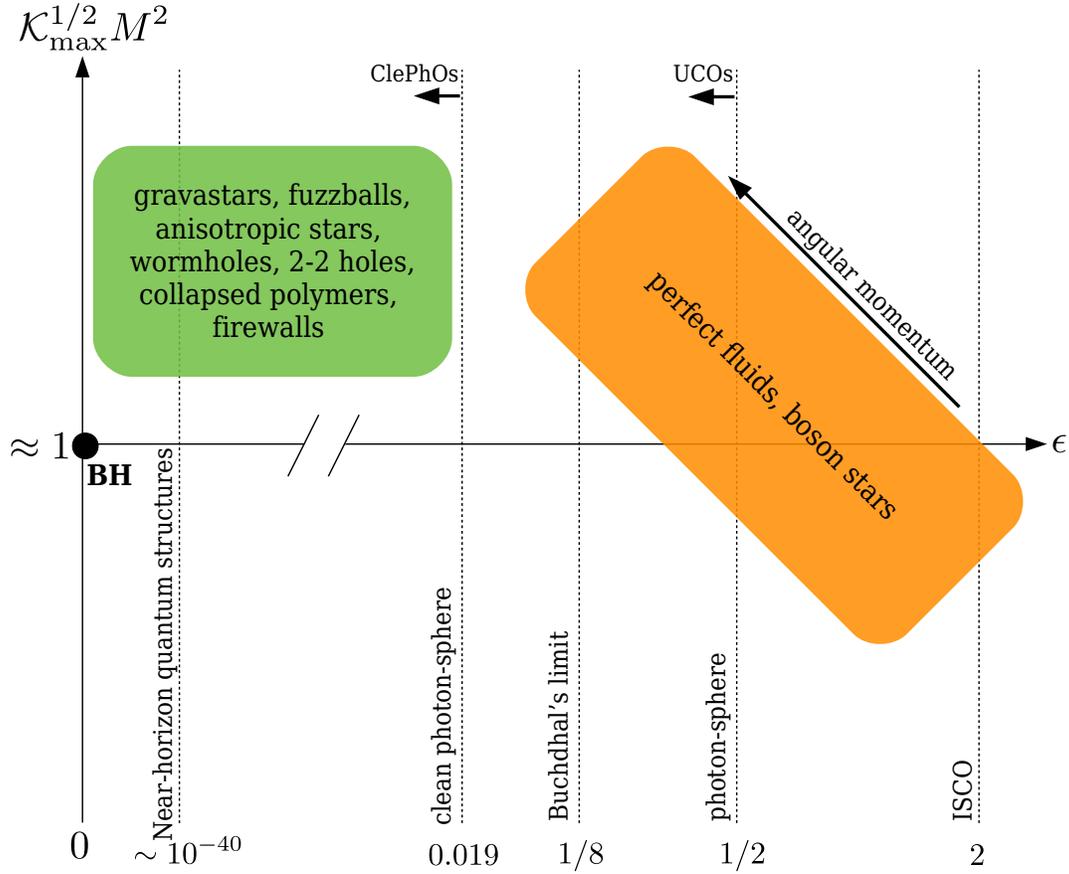}
\caption{Schematic representation of ECO models in a compactness-curvature diagram. The 
horizontal axis shows the compactness parameter $\epsilon$ associated to the 
object, which can be also mapped (in a model-dependent way) to a characteristic 
light-crossing timescale.
The vertical axis shows the maximum curvature (as measured by the Kretschmann scalar 
${\cal K}$) of the object normalized by the corresponding quantity for a BH with the same 
mass $M$. All known ECO models with $\epsilon\to0$ have large curvature in their 
interior, i.e. the leftmost bottom part of the diagram is conjectured to be empty. Angular 
momentum tends to decrease $\epsilon$ and to increase ${\cal K}_{\rm max}$.
\label{fig:zoo}}
\end{center}
\end{figure}

The zoo of compact objects is summarized in Fig.~\ref{fig:zoo}. In all 
these cases, both quantum-gravity or microscopic corrections at the horizon scale select 
ClePhOs as well-motivated alternatives to BHs. Despite a number of supporting arguments 
--~some of which urgent and well founded~-- it is important to highlight that there is no 
horizonless ClePhO for which we know sufficiently well the physics at the moment.

\clearpage
\newpage


\newpage

\section{Dynamics of compact objects}

\hskip 0.2\textwidth
\parbox{0.8\textwidth}{
\begin{flushright}
{\small 
\noindent {\it ``There is a crack in everything. That's how the light gets in.''}\\
Leonard Cohen, Anthem (1992) 
}
\end{flushright}
}

\vskip 1cm

EM observations of compact bodies are typically performed in a context where 
spacetime fluctuations are irrelevant, either due to the long timescales involved or 
because the environment has a negligible backreaction on the body itself. 
For example, EM observations of accretion disks around a compact object can be interpreted 
using a stationary background geometry. Such geometry is a solution to the field equations 
describing the compact body while neglecting the accretion disk, the dynamics of which 
is governed by the gravitational pull of the central object and by internal forces. This 
approximation is adequate since the total amount of energy density around 
compact objects is but a small fraction of the object itself, and the induced changes in 
the geometry can be neglected~\cite{Barausse:2014tra}. In addition, the wavelength of EM 
waves of interest for Earth-based detectors is always much smaller that any lengthscale 
related to coherent motion of compact objects: light can be treated as a null particle 
following geodesics on a stationary background. Thus, the results of the previous sections 
suffice to discuss EM observations of compact objects, as done in Sec.~\ref{sec:Tests} 
below.

For GW astronomy, however, it is the spacetime fluctuations themselves that are relevant. 
A stationary geometry approximation would miss GW emission entirely. In addition, GWs 
generated by the coherent motion of sources have a wavelength of the order of the size of 
the system. Therefore, the geodesic approximation becomes inadequate (although it can still be used as a guide).
Compact binaries are the preferred sources for GW detectors. Their GW signal is naturally divided
in three stages, corresponding to the different cycles in the evolution driven by GW 
emission~\cite{Buonanno:2006ui,Berti:2007fi,Sperhake:2011xk}:
the inspiral stage, corresponding to large separations and well approximated by 
post-Newtonian theory; the merger phase when the two objects coalesce and which can only 
be described accurately through numerical simulations; and finally, the ringdown phase 
when the merger end-product relaxes to a stationary, equilibrium solution of the field 
equations~\cite{Sperhake:2011xk,Berti:2009kk,Blanchet:2013haa}. All three stages provide independent, unique tests of 
gravity and of compact GW sources. 
Overall, GWs are almost by definition attached to highly dynamical spacetimes, such as the coalescence and merger 
of compact objects. We turn now to that problem.

\subsection{Quasinormal modes\label{sec:qnms}}
Consider first an isolated compact object described by a stationary spacetime. Again, we 
start with the spherically-symmetric case and for simplicity. Birkhoff's theorem then implies that the exterior 
geometry is Schwarzschild.
Focus on a small disturbance to such static spacetime, which could describe
a small moving mass (a planet, a star, etc), or the late-stage in the life of a coalescing binary
(in which case the disturbed ``isolated compact object'' is to be understood as the final state of the coalescence).
 
In the linearized regime, the geometry can be written as $g_{\mu\nu}=g_{\mu\nu}^{(0)}+h_{\mu\nu}$, where
$g_{\mu\nu}^{(0)}$ is the geometry corresponding to the stationary object, 
and $h_{\mu \nu}$ are the small deviations induced on it by whatever is causing the 
dynamics. The metric fluctuations can be combined in a single master function $\Psi$ which 
in vacuum is governed by a master partial differential equation of the 
form~\cite{Zerilli:1971wd,Berti:2009kk}
\begin{equation}
\frac{\partial^2 \Psi(t,z)}{\partial z^2}-\frac{\partial^2 \Psi(t,z)}{\partial 
t^2}-V(r)\Psi(t,z)=S(t,z)\,.\label{pde}
\end{equation}
where $z$ is a suitable coordinate. The source term $S(t,z)$ contains information about the cause of the disturbance 
$\Psi(t,z)$. The information about the angular dependence of the wave is encoded in the way the 
separation was achieved, and involves an expansion
in tensor harmonics. One can generalize this procedure and consider also scalar or vector 
(i.e., EM) waves. These can also be reduced to a master function of the type \eqref{pde}, 
and separation is achieved with spin-$s$ harmonics for different spins $s$ of field. These 
angular functions are labeled by an integer $l\geq|s|$. For a Schwarzschild spacetime, 
the effective potential is
\begin{equation}
V=f\left(\frac{l(l+1)}{r^2}+(1-s^2)\frac{2M}{r^3}\right)\,, \label{potential}
\end{equation}
with $s=0,\pm1,\pm2$ for scalar, vector or (axial) tensor modes. The $s=\pm2$ equation 
does not describe completely all of the gravitational degrees of freedom. There is an 
another (polar) gravitational mode (in GR, there are two polarizations for GWs), also 
described by Eq.~\eqref{pde} with a slightly more complicated 
potential~\cite{Chandrasekhar:1975zza,Kokkotas:1999bd,Berti:2009kk}. 
Note that such results apply only when there are no further degrees of 
freedom that couple to the GR modes~\cite{Blazquez-Salcedo:2016enn,Cardoso:2009pk,Tattersall:2017erk,Cardoso:2018ptl,Molina:2010fb}.

The solutions to Eq.~\eqref{pde} depend on the source term and initial conditions, just 
like for any other physical system. We can gain some insight on the general properties of 
the system by studying the source-free equation in Fourier space. 
This corresponds to studying the ``free'' compact object when the driving force died off. 
As such, it gives us information on the late-time behavior of any compact object. By 
defining the Fourier transform through $\Psi(t,r)=\frac{1}{\sqrt{2\pi}}\int e^{-i\omega 
t}\psi(\omega,r)d\omega$, one gets the following ODE
\begin{equation}
 \frac{d^2\psi}{dz^2}+\left(\omega^2-V\right)\psi=0\,. \label{ode}
\end{equation}
For a Schwarzschild spacetime, the ``tortoise'' coordinate $z$ is related to the original $r$ by $dr/dz=f$, i.e.
\begin{equation}
z=r+2M \log\left(\frac{r}{2M}-1\right)\,,
\end{equation}
such that $z(r)$ diverges logarithmically near the horizon.
In terms of $z$, Eq.~\eqref{ode} is equivalent to the time-independent Schr\"odinger 
equation in one dimension and it reduces to the wave equation governing a string when 
$M=l=0$. For a string of length $L$ with fixed ends,
one imposes Dirichlet boundary conditions and gets an eigenvalue problem for $\omega$. 
The boundary conditions can only be satisfied for a discrete set of \emph{normal} 
frequencies,  $\omega= n\pi/L$ ($n=1,2,...$). The corresponding wavefunctions are 
called normal modes and form a basis onto which one can expand any configuration of the 
system. The frequency is purely real because the associated problem is conservative.

\begin{figure*}[ht]
\begin{center}
\includegraphics[width=0.8\textwidth]{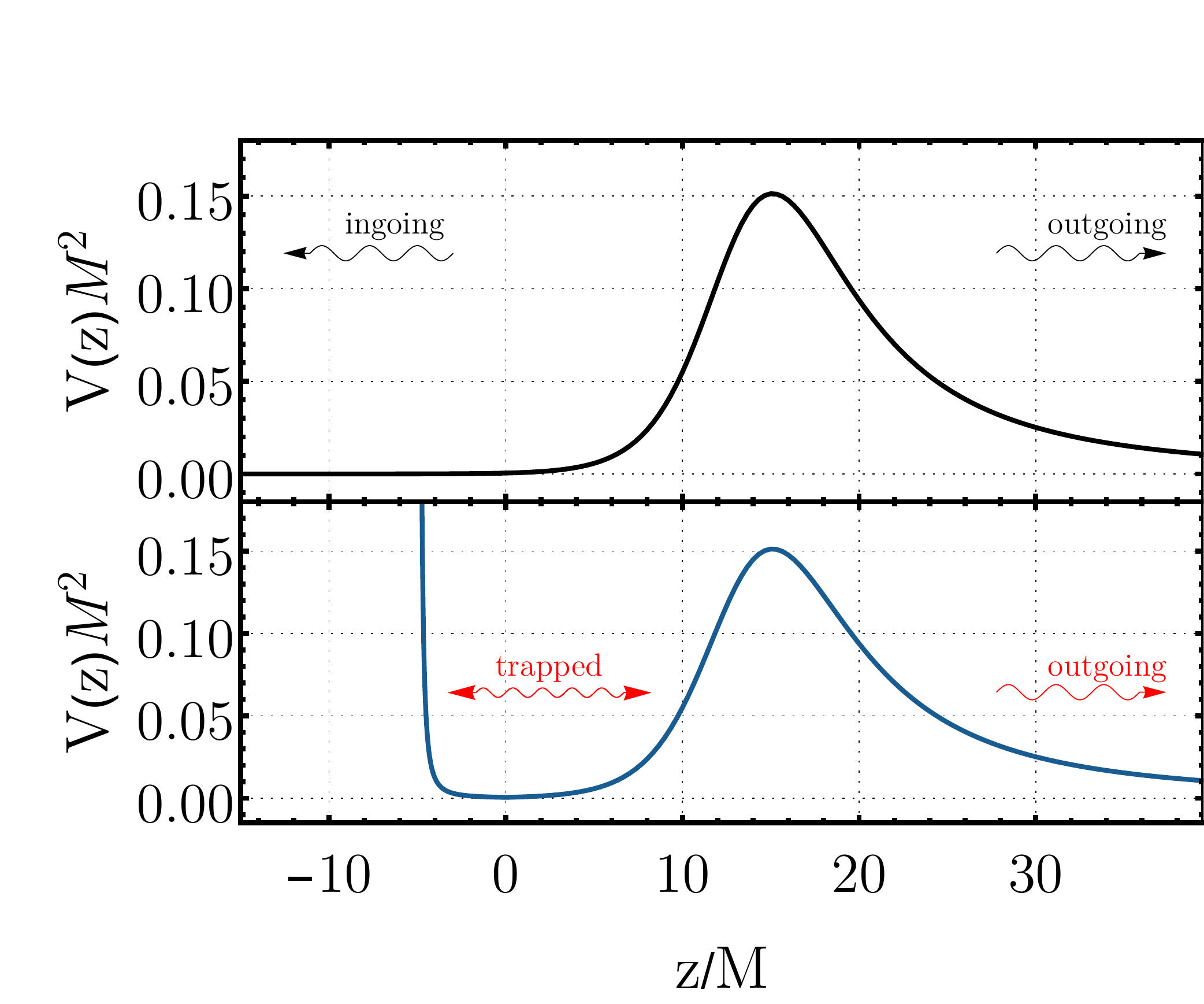}
\caption{Typical effective potential for perturbations of a Schwarzschild BH 
(top panel) and of an horizonless compact object (bottom panel). The effective potential is peaked
at approximately the photon sphere, $r\approx 3M$. For BHs, QNMs are waves 
which are outgoing at infinity ($z\to+\infty$) and ingoing at the horizon 
($z\to-\infty$), whereas the presence of a potential well (provided either by a partly reflective 
surface, a centrifugal barrier at the center, or by the geometry) supports quasi-trapped, 
long-lived modes.}
\label{fig:potential}
\end{center}
\end{figure*}

If one is dealing with a BH spacetime, the appropriate conditions (required by causality) 
correspond to having waves traveling outward to spatial infinity ($\Psi \sim e^{i\omega (
z-t)}$ as $z\to\infty$) and inwards to the horizon ($\Psi \sim e^{-i\omega (z+t)}$ as 
$z\to-\infty$) [see Fig.~\ref{fig:potential}]. 
The effective potential displays a maximum approximately at the photon sphere, $r\approx 3M$, the exact value depending on the type of perturbation and on the value of 
$l$ ($r_\to 3M$ in the $l\to\infty$ limit).
Due to backscattering off the effective potential~\eqref{potential}, the eigenvalues 
$\omega$ are not known in closed form, but they can be computed 
numerically~\cite{Chandrasekhar:1975zza,Kokkotas:1999bd,Berti:2009kk}. The fundamental 
$l=2$ mode (the lowest dynamical multipole in GR) of gravitational perturbations  
reads~\cite{rdweb}
\begin{equation}
M\omega_{\rm BH}\equiv M(\omega_R+i\omega_I)\approx 0.373672 -i 0.0889623\,.\label{bh_qnm}
\end{equation}
Remarkably, the entire spectrum is the same for both the axial or the polar gravitational 
sector; this property is ofter referred to as \emph{isospectrality}~\cite{Chandrasekhar:1975zza}. The frequencies 
are complex and are therefore called 
{\it quasinormal mode} (QNM) frequencies. Their imaginary component describes the decay in 
time of fluctuations on a timescale $\tau\equiv 1/|\omega_I|$, and hints at the stability 
of the geometry. Unlike the case of a string with fixed end, we are now dealing with an 
open system: waves can travel to infinity or down the horizon and therefore it is 
physically sensible that any fluctuation damps down.
The corresponding modes are QNMs, which in general do \emph{not} form a complete 
set~\cite{Kokkotas:1999bd}.

Boundary conditions play a crucial role in the structure of the QNM spectrum. If a reflective surface is placed at 
$r_0=2M(1+\epsilon)\gtrsim 2M$, where (say) Dirichlet or Neumann boundary conditions have to be imposed, the spectrum 
changes considerably. The QNMs in the $\epsilon\to0$, low-frequency limit 
read~\cite{Vilenkin:1978uc,Maggio:2017ivp,Maggio:2018ivz}
\begin{eqnarray}
M \omega_R&\simeq&-\frac{M\pi}{2|z_0|}\left(q+\frac{s(s+1)}{2}\right)\sim 
|\log\epsilon|^{-1}\,, \label{omegaRecho}\\
M\omega_I &\simeq&-\beta_{ls}\frac{M}{|z_0|}(2M\omega_R)^{2l+2}\sim -|\log\epsilon|^{-(2l+3)}\,, \label{omegaIecho}
\end{eqnarray}
where $z_0\equiv z(r_0)\sim 2M\log\epsilon$, $q$ is a positive odd (even) 
integer for polar (axial) modes (or equivalently for Dirichlet (Neumann) boundary 
conditions), and 
$\beta_{ls}=\left[\frac{(l-s)!(l+s)!}{(2l)!(2l+1)!!}\right]^2$~\cite{Starobinskij2,
Brito:2015oca,Maggio:2018ivz}. Note that the two gravitational sectors are no longer 
isospectral. More importantly, the perturbations have smaller frequency and are much 
longer lived, since a decay channel (the horizon) has disappeared. For example, for 
$\epsilon=10^{-6}$ we find numerically the fundamental scalar modes 
\begin{eqnarray}
 M\omega_{\rm polar}&\approx& 0.13377 -i\, 2.8385\times 10^{-7} \,,\\
 M\omega_{\rm axial}&\approx& 0.13109- i\, 2.3758\times 10^{-7}\,.
\end{eqnarray}
These QNMs were computed by solving the exact linearized equations numerically but agree well with 
Eqs.~\eqref{omegaRecho} and \eqref{omegaIecho}.

The above scaling with $\epsilon$ can be understood in terms of modes trapped between the peak of the 
potential~\eqref{potential} at $r\sim 3M$ and the ``hard surface'' at 
$r=r_0$~\cite{Cardoso:2016rao,Cardoso:2016oxy,Volkel:2017ofl,Mark:2017dnq,Maggio:2018ivz} 
[see Fig.~\ref{fig:potential}]. Low-frequency waves are almost trapped by the potential, 
so their wavelength scales as the size of the cavity (in tortoise coordinates), 
$\omega_R\sim1/z_0$, just like the normal modes of a string. The (small) imaginary part is 
given by waves which tunnel through the potential and reach infinity. The tunneling 
probability can be computed analytically in the small-frequency regime and scales as 
$|\mathcal{A}|^2\sim (M\omega_R)^{2l+2}\ll1$~\cite{Starobinskij2}. After a time $t$, a 
wave trapped inside a box of size $z_0$ is reflected $N=t/z_0$ times, and its amplitude 
reduces to $A(t)=A_0\left(1-|\mathcal{A}|^2\right)^N\sim 
A_0\left(1-t|\mathcal{A}|^2/z_0\right)$. 
Since, $A(t)\sim A_0 e^{-|\omega_I| t}\sim A_0(1-|\omega_I| t)$ in this limit,  we immediately obtain
\begin{equation}
 \omega_R\sim1/z_0\,,\qquad \omega_I\sim|\mathcal{A}|^2/z_0 \sim \omega_R^{2l+3}\,.
\end{equation}
This scaling agrees with exact numerical results and is valid for any $l$ and any type of 
perturbation. 

The reverse-engineering of the process, i.e., a reconstruction of the scattering potential $V$ from a mode measurement  was proposed in Ref.~\cite{Volkel:2017kfj,Volkel:2018hwb,Volkel:2018czg}. The impact of measurement error on such reconstruction is yet to be assessed.

Clearly, a perfectly reflecting surface is an idealization. In certain models, only 
low-frequency waves are reflected, whereas higher-frequency waves probe the internal 
structure of the specific object~\cite{Saravani:2012is,Mathur:2012jk}. In general, the 
location of the effective surface and its properties (e.g., its reflectivity) can depend 
on the energy scale of the process under consideration. Partial absorption is particularly important in the case of 
spinning objects, as discussed in Sec.~\ref{sec:DynSpin}. 

\subsection{Gravitational-wave echoes\label{sec:echoes}}
\subsubsection{Quasinormal modes, photon spheres, and echoes\label{sec:echoes_transient}}

The effective potential $V$ for wave propagation reduces to that for geodesic motion 
($V_{\rm geo}$) in the high-frequency, high-angular momentum (i.e., eikonal) regime. 
Thus, some properties of geodesic motion have a wave counterpart~\cite{Ferrari:1984zz,Cardoso:2008bp}.
The instability of light rays along the null circular geodesic translates into some properties of waves around objects 
compact enough to feature a photon sphere.
A wave description needs to satisfy ``quantization conditions'', which can be worked out in a WKB approximation. Since 
GWs are quadrupolar in nature, the lowest mode of vibration should satisfy 
\begin{equation}
M\omega_R^{\rm geo}=2\frac{\dot{\varphi}}{\dot{t}}=\frac{2}{3\sqrt{3}}\sim 0.3849\,.
\end{equation}
In addition the mode is damped, as we showed, on timescales $3\sqrt{3}M$. Overall then, the geodesic analysis predicts
\begin{equation}
M\omega^{\rm geo}\sim 0.3849-i\,0.19245\,.
\end{equation}
This crude estimate, valid in principle only for high-frequency waves, matches well even the fundamental mode of a 
Schwarzschild BH, Eq.~\eqref{bh_qnm}.

Nevertheless, QNM frequencies can be defined for any dissipative system, not only for 
compact objects or BHs. Thus, the association with photon spheres has limits, for instance it neglects possible 
coupling terms~\cite{Blazquez-Salcedo:2016enn}, nonminimal kinetic terms~\cite{Konoplya:2017wot}, etc.
Such an analogy is nonetheless enlightening
in the context of objects so compact that they have photon spheres and resemble Schwarzschild deep into the geometry, 
in a way that condition~\eqref{eps_crit} is satisfied~\cite{Cardoso:2016rao,Cardoso:2016oxy,Price:2017cjr,Ghersi:2019trn}.

\begin{figure*}[ht]
\begin{center}
\includegraphics[width=0.7\textwidth]{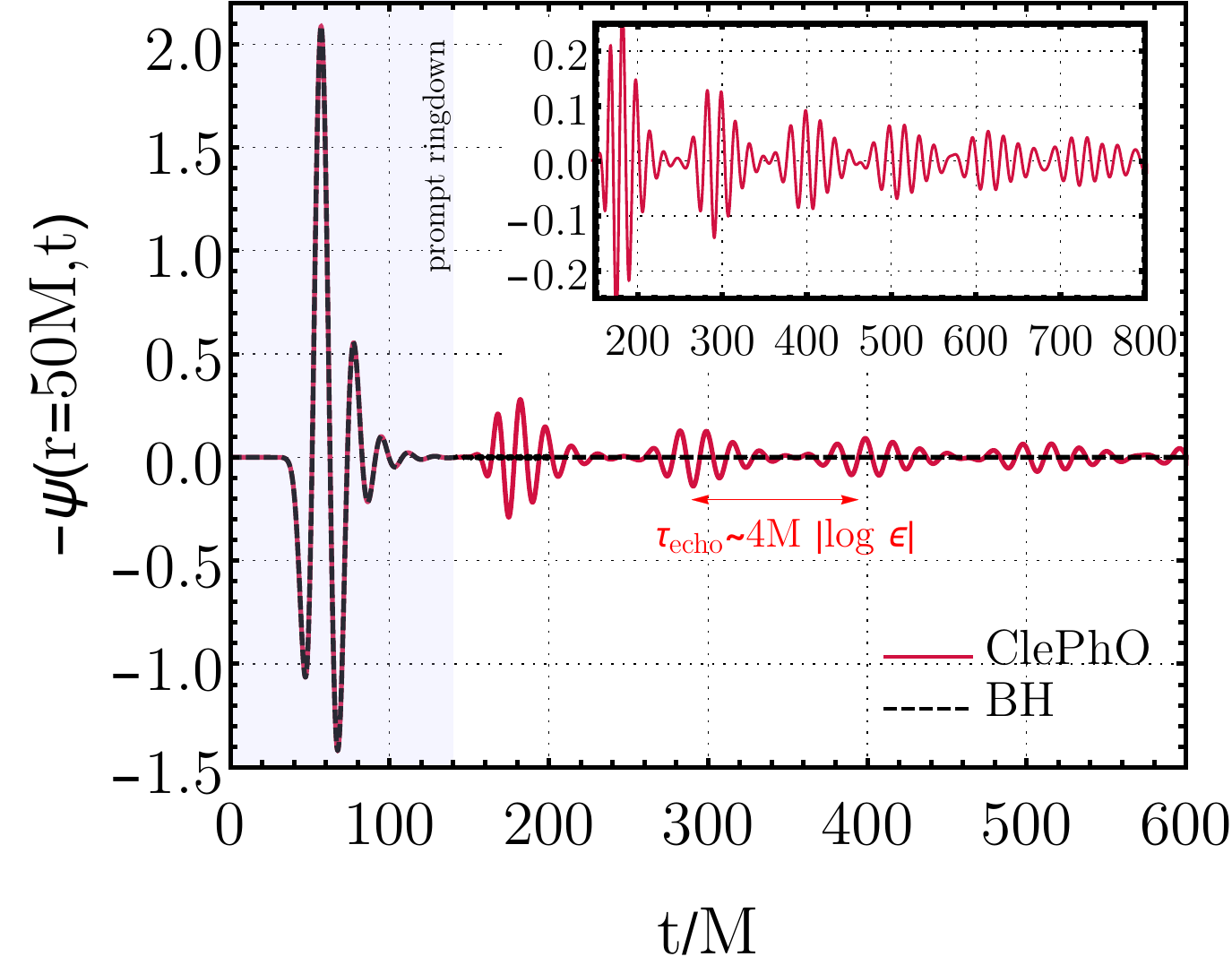}
\caption{Ringdown waveform for a BH (dashed black curve) compared to a ClePhO (solid red curve) with a reflective 
surface at $r_0=2M(1+\epsilon)$ with $\epsilon=10^{-11}$. We considered $l=2$ axial gravitational perturbations and a 
Gaussian wavepacket $\psi(r,0)=0$, $\dot\psi(r,0)=e^{-(z-z_m)^2/\sigma^2}$ (with $z_m=9M$ and $\sigma=6M$) as initial 
condition. 
Note that each subsequent echo has a smaller frequency content and that the damping of subsequent echoes is much larger 
than the late-time QNM prediction ($e^{-\omega_I t}$ with $\omega_I M\sim 4\times 10^{-10}$ for these parameters).
Data available online~\cite{rdweb}.
\label{fig:ringdown}}
\end{center}
\end{figure*}

For a BH, the excitation of the spacetime modes happens at the photon sphere~\cite{Davis:1971gg,Davis:1972ud,Ferrari:1984zz}. Such waves travel 
outwards to possible observers or down the event horizon. The structure of GW signals at late times is therefore 
expected to be relatively simple. This is shown in Fig.~\ref{fig:ringdown}, for the scattering of a 
Gaussian 
pulse of axial quadrupolar modes off a BH. The pulse crosses the photon sphere, and excites its modes. The ringdown 
signal, a fraction of which travels to outside observers, is to a very good level described by the lowest QNMs. The 
other fraction of the signal generated at the photon sphere travels downwards and into the horizon. It dies off and has 
no effect on observables at large distances.

\begin{figure*}[ht]
\begin{center}
\includegraphics[width=0.8\textwidth]{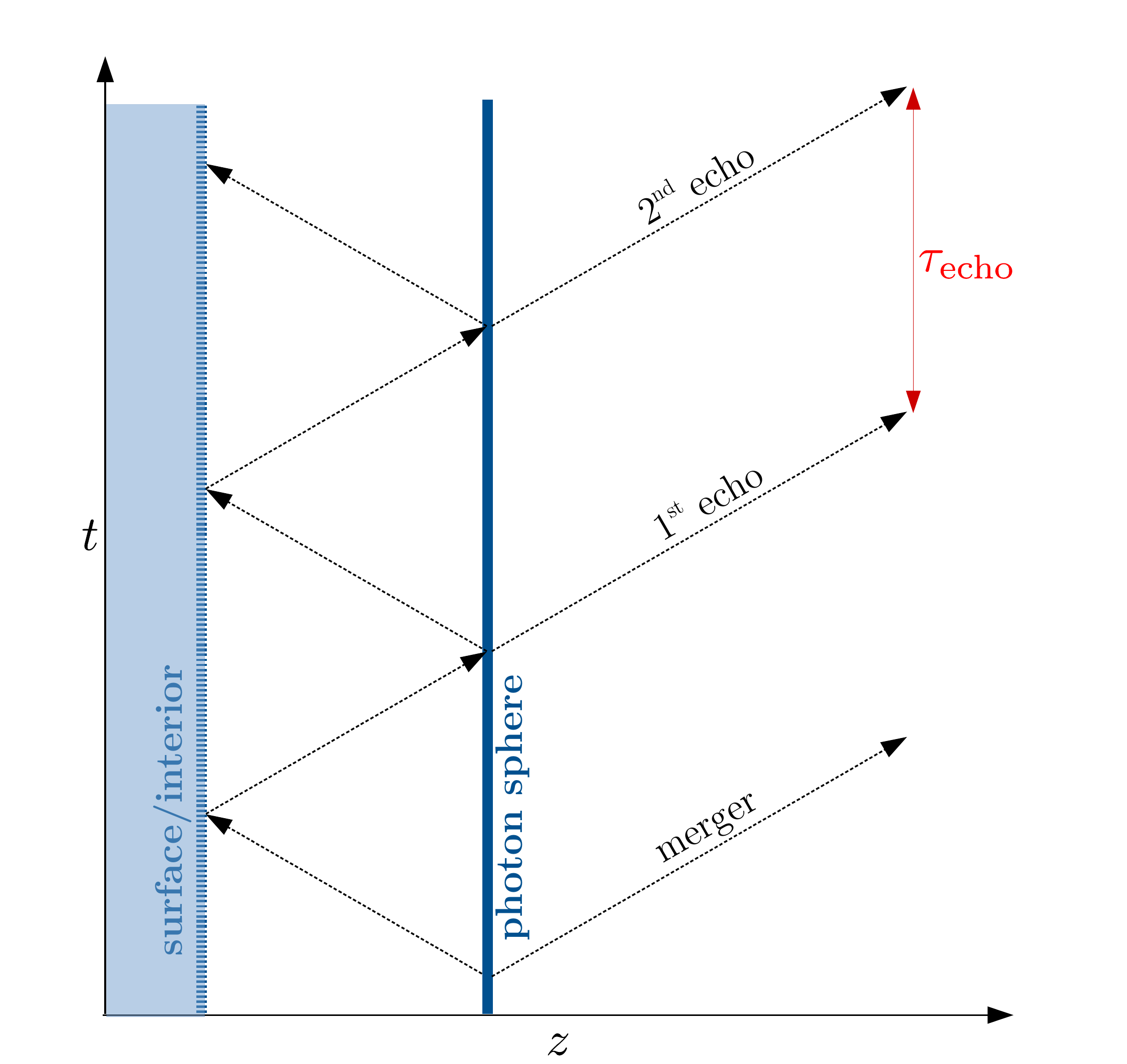}
\caption{Schematic Penrose diagram of GW echoes from an ECO. 
Adapted from Ref.~\cite{Abedi:2016hgu} (similar versions of this plot appeared in other contexts in 
Refs.~\cite{Vilenkin:1978uc,Mark:2017dnq}).
\label{fig:echo_Penrose}}
\end{center}
\end{figure*}

Contrast the previous description with the dynamical response of ultracompact objects for which 
condition~\eqref{eps_crit} is satisfied (i.e., a ClePhO) [cf.\ Fig.~\ref{fig:ringdown}]. The initial description of the 
photon sphere modes still holds, by causality. Thus, up to timescales of the order $|z_0|\sim-M\log\epsilon$ (the 
roundtrip time of radiation between the photon sphere and the surface) the signal is {\it identical} to that of 
BHs~\cite{Cardoso:2016rao,Cardoso:2016oxy}.
At later times, however, the pulse traveling inwards is bound to see the object and be reflected either at its surface 
or at its center. In fact, this pulse is semi-trapped between the object and the light ring. Upon each interaction with 
the light ring, a fraction exits to outside observers, giving rise to a series of {\it echoes} of ever-decreasing 
amplitude. From Eqs.~\eqref{omegaRecho}--\eqref{omegaIecho}, repeated reflections occur in a characteristic echo delay 
time~\cite{Cardoso:2016rao,Cardoso:2016oxy,Ghersi:2019trn} [see Fig.~\ref{fig:echo_Penrose}]
\begin{equation}
 \tau_{\rm echo}\sim 4M |\log\epsilon|\,. \label{tauecho}
\end{equation}
However, the main burst is typically generated at the photon sphere and has therefore a \emph{frequency content} of the 
same order as the BH QNMs~\eqref{bh_qnm}.
The initial signal is of high frequency and a substantial component is able to cross the 
potential barrier. Thus, asymptotic observers see a series of echoes whose amplitude is 
getting smaller and whose frequency content is also going down.
It is crucial to understand that echoes occur in a \emph{transient} regime; at {\it very late times}, the signal is 
dominated by the lowest-damped QNMs, described by Eqs.~\eqref{omegaRecho}--\eqref{omegaIecho}. 

We end this discussion by highlighting that GW echoes are a feature of very compact ECOs, but also arise in many other 
contexts: classical BHs surrounded by a ``hard-structure'' close to the 
horizon~\cite{Barausse:2014tra,Kaplan:2018dqx,Ramos:2018oku}, or far from it~\cite{Barausse:2014tra,Konoplya:2018yrp,Lin:2019qyx},
or embedded in a theory that effectively makes the graviton see a hard wall there~\cite{Zhang:2017jze,Oshita:2018fqu} 
will respond to incoming GWs producing echoes. Finally, as we described earlier, even classical but very compact neutron 
or strange quark stars may be prone to exciting 
echoes~\cite{Ferrari:2000sr,Raposo:2018rjn,Pani:2018flj,Mannarelli:2018pjb}. A simple picture of how echoes arise in 
a simple two-barrier system is provided in Ref.~\cite{Mirbabayi:2018mdm}.

\subsubsection{A black-hole representation and the transfer function}
The QNMs of a spacetime were defined as the eigenvalues of the homogeneous ordinary differential equation \eqref{ode}.
Their role in the full solution to the homogeneous problem becomes clear once we re-write 
Eq.~\eqref{pde} in Fourier space,
\begin{equation}
\frac{d^2\psi}{dz^2}+\left(\omega^2-V\right)\psi={\cal S}\,, \label{ode_source}
\end{equation}
where ${\cal S}$ is the Fourier transformed source term $S$.
Since the potential is zero at the boundaries, two independent homogeneous solutions are
\begin{equation}
\psi_-=\left\{\begin{array}{ll}
               e^{-i \omega  z}   \quad &z\to -\infty \\
               A_{\rm in}e^{-i\omega z}+A_{\rm out}e^{i\omega z}\quad &z\to +\infty
              \end{array} 
              \right.\,, \label{bc_left}
\end{equation}
and
\begin{equation}
\psi_+=\left\{\begin{array}{ll}
               B_{\rm in}e^{-i\omega z}+B_{\rm out}e^{i\omega z}  \quad &z\to -\infty \\
               e^{i \omega  z}					  \quad &z\to +\infty
              \end{array}
              \right.\,, \label{bc_right}
\end{equation}
Note that $\psi_{+}$ was chosen to satisfy outgoing conditions at large distances; this is the behavior we want to 
impose on a system which is assumed to be isolated. On the other hand, $\psi_-$ satisfies the correct near-horizon 
boundary condition \emph{in the case of a BH}. Define reflection and transmission coefficients,
\begin{equation}
 {\cal R}_{\rm BH} =\frac{B_{\rm in}}{B_{\rm out}}\,,\qquad   {\cal T}_{\rm BH} =\frac{1}{B_{\rm out}}\,. 
\label{refltrasmcoeff}
\end{equation}
Given the form of the ODE, the Wronskian $W\equiv\psi_-\psi'_+-\psi'_{-}\psi_+$ is a constant (here $'\equiv d/dz$), 
which can be evaluated at infinity to yield $W=2i\omega A_{\rm in}$. The general solution to our problem can be written 
as~\cite{Bender:1999}
\begin{equation}
\psi=\psi_{+}\int^z\frac{{\cal S}\psi_-}{W}dz+\psi_-\int^z\frac{{\cal 
S}\psi_{+}}{W}dz+A_1\psi_-+A_2\psi_{+}\,,\label{variation_constants_1}
\end{equation}
where $A_1,\,A_2$ constants. If we impose the boundary conditions appropriate for BHs, we find 
\begin{equation}
\psi_{\rm BH}=\psi_{+}\int_{-\infty}^z\frac{{\cal S}\psi_-}{W}dz+\psi_-\int_{z}^{\infty}\frac{{\cal 
S}\psi_{+}}{W}dz\,.\label{variation_constants_2}
\end{equation}
This is thus the response of a BH spacetime to some source. Notice that close to the horizon the first term drops and 
$\psi_{\rm BH}(r\sim r_+)\sim e^{-i\omega z}\int_{z}^{\infty}\frac{{\cal S}\psi_{+}}{W}dz$. For detectors located far 
away from the source, on the other hand $\psi_{\rm BH}(r\to\infty)\sim e^{i\omega z}\int_{-\infty}^{\infty}\frac{{\cal 
S}\psi_-}{W}dz$.
It is easy to see now that QNMs correspond to poles of the propagator~\cite{Berti:2009kk}, and hence they do indeed have 
a significant contribution to the signal, both at infinity and near the horizon.

Consider now that instead of a BH, there is an ultracompact object. Such object has a surface at $r_0$, corresponding 
to large negative tortoise $z_0$. Then, the boundary condition \eqref{bc_left} on the left needs to be changed to
\begin{equation}
\psi_{\rm ECO}\sim e^{-i \omega  z}+{\cal R}e^{i\omega z}\,,\quad z\to z_0\,,\label{bc_ECO}
\end{equation}
where we assume the compact object surface to have a (possible frequency-dependent) reflectivity ${\cal R}$.
We will now show that the spacetime response to an ultracompact object can be expressed in terms of the BH response 
and a transfer function~\cite{Mark:2017dnq} (for a previous attempt along these lines see Ref.~\cite{Nakano:2017fvh}).
First, notice that the ODE to be solved is exactly the same, with different conditions on one of the boundaries. We can 
thus still pick the two independent homogeneous solutions
\eqref{bc_left} and \eqref{bc_right}, but choose different integration constants in \eqref{variation_constants_1} so 
that the boundary condition is satisfied.
Adding a homogeneous solution ${\cal K}\psi_+ \int_{-\infty}^{\infty}\frac{{\cal S}\psi_{+}}{W}dz$ to 
\eqref{variation_constants_2} is allowed since it still satisfies outgoing conditions at large spatial distances.
We then find at large negative $z$
\begin{equation}
\psi_{\rm ECO}=\left(e^{-i\omega z}+{\cal K}(B_{\rm in}e^{-i\omega z}+B_{\rm out}e^{i\omega 
z})\right)\int_{z_0}^{\infty}\frac{{\cal S}\psi_{+}}{W}dz \,.
\end{equation}
where ${\cal K}$ is a constant. On the other hand, to obey the boundary condition \eqref{bc_ECO}, one must impose
$\psi_{\rm ECO}=k_0\left(e^{-i\omega z}+{\cal R}e^{i\omega z}\right)$ with an unknown constant $k_0$.
Matching outgoing and ingoing coefficients, we find
\begin{eqnarray}
{\cal K}=\frac{{\cal T}_{\rm BH}{\cal R}}{1-{\cal R}_{\rm BH}{\cal R}}\,. \label{transferfunction}
\end{eqnarray}

Thus, a detector at large distances sees now a signal 
\begin{equation}
\psi_{\rm ECO}(r\to\infty)=\psi_{\rm BH}(r\to\infty)+{\cal K}e^{2i\omega z}\psi_{\rm BH}(r\sim r_+)\,. 
\label{ECOresponse}
\end{equation}
In other words, the signal seen by detectors is the same as the one from a BH, modified by a piece that is controlled 
by the reflectivity of the compact object.

Following Ref.~\cite{Mark:2017dnq}, the extra term can be expanded as a geometric series
\begin{equation}
{\cal K}={\cal T}_{\rm BH}{\cal R}\sum_{n=1}^{\infty}\left({\cal R}_{\rm BH}{\cal 
R}\right)^{n-1} 
\,.\label{transfer_echoes}
\end{equation}
A natural interpretation emerges: a main burst of radiation is generated for example when an object crosses the 
light ring (where the peak of the effective potential is located). 
A fraction of this main burst is outward traveling and gives rise to the ``prompt'' response $\psi_{\rm 
BH}(r\to\infty)$, which is equivalent to the response of a BH. However, another fraction is 
traveling inwards. The first term is the result of the primary reflection of $\psi_{\rm BH}$ at $z_0$. Note the 
time delay factor $2(z-z_0)$ between the first pulse and
the main burst due the pulse's extra round trip journey between the boundary the peak of the scattering potential, 
close to the light ring at $z\sim 0$. When the pulse reaches the potential barrier, it is partially transmitted and emerges as a contribution to the signal. The successive terms are ``echoes'' of this first reflection which bounces an integer number of times between the potential barrier and the compact object surface. Thus, a mathematically elegant formulation gives formal support 
to what was a physically intuitive picture.

The derivation above assumes a static ECO spacetime, and a potential which vanishes at its surface. An extension of the 
procedure above to include both a more general potential and spin is worked out in Ref.~\cite{Conklin:2017lwb}.
Such a ``transfer-function'' representation of echoes was embedded into an effective-field-theory 
scheme~\cite{Burgess:2018pmm}, showing that linear ``Robin'' boundary conditions at $r=r_0$ dominate at low energies. 
In this method the (frequency dependent) reflection coefficient and the surface location can be obtained in terms of a 
single low-energy effective coupling. Recently, another model for the frequency-dependent reflectivity of quantum 
BHs has been proposed in Ref.~\cite{Oshita:2019sat}.

The previous description of echoes and of the full signal is reasonable and describes all the known numerical results.
At the technical level, more sophisticated tools are required to understand the signal: the intermediate-time response 
is dominated by the BH QNMs, which are {\it not} part of the QNM spectrum of an 
ECO~\cite{Cardoso:2016rao,Barausse:2014tra,Khanna:2016yow}. While this fact is easy to 
understand in the time domain due to causality (in terms of time needed 
for the perturbation to probe the boundaries~\cite{Cardoso:2016rao}), it is not at all obvious in the 
frequency domain. Indeed, the poles of the ECO Green's functions in the complex frequency plane are
different from the BH QNMs. The late-time signal is dominated by the dominant ECO poles, whereas the prompt 
ringdown is governed by the by the dominant QNMs of the corresponding BH spacetime.

\subsubsection{A Dyson-series representation}
The previous analysis showed two important aspects of the late-time behavior of very compact objects:
(i)~that it can be expressed in terms of the corresponding BH response if one uses a transfer function ${\cal K}$; 
(ii)~that the signal after the main burst and precursor are a sequence of echoes, trapped between the object and the 
(exterior) peak of the potential.

The response of any system with a non-trivial scattering potential and nontrivial boundary conditions includes 
echo-like  components.
To see that, let's use a very different approach to solve \eqref{ode_source}, namely the Lippman-Schwinger integral 
solution used in quantum mechanics~\cite{Correia:2018apm}.
In this approach, the setup is that of flat spacetime, and the scattering potential is treated as a {\it perturbation}. 
In particular, the field is written as
\begin{equation} \label{eq:integral}
\psi = \psi_0+ \int_{z_0}^{\infty} g(z,z') \, V(z') \psi(z') \, dz'\,,
\end{equation}
where 
\begin{equation} \label{eq:g}
g(z,z') = {e^{i \omega |z-z'|} + {\cal R} \, e^{i \omega (z+z')} \over 2 i \omega} \, ,
\end{equation}
is the Green's function of the free wave operator $d^2/d x^2 + \omega^2$ with boundary condition \eqref{bc_ECO}, and
$\psi_0 = \int_{z_0}^{\infty} \! g(z,z')  {\cal S}(z') dz'$
is the free-wave amplitude. The formal solution of Eq.~\eqref{eq:integral} is the Dyson series (sometimes also 
called Born or Picard series)
\begin{equation}
\psi= \sum_{k=1}^\infty \int_{z_0}^{\infty} \! g(z,z_1) \cdots g(z_{k-1},z_k) V(z_1) \cdots V(z_{k-1}) {\cal S}(z_k) 
dz_1 \cdots dz_k \, ,\label{eq:dyson}
\end{equation}
which effectively works as an expansion in powers of $V/\omega^2$, so we expect it to converge rapidly for high 
frequencies and to be a reasonable approximation also for fundamental modes. 
It is possible to reorganize \eqref{eq:dyson} and express it as a series in powers of ${\cal R}$.
We start by separating the Green's function \eqref{eq:g} into $g = g_o + {\cal R}g_r$, with
\be \label{eq:go}
g_o(z,z') = {e^{i \omega |z-z'|} \over 2 i \omega} \, ,
\end{equation}
the open system Green's function, and
\be 
g_r(z,z') = {e^{i \omega (z+z')} \over 2 i \omega} \, ,\label{eq:gr}
\end{equation}
the ``reflection'' Green's function.
We can then write \eqref{eq:integral} as
\begin{equation}
\psi= \int_{z_0}^{\infty} g_o(z,z') {\cal S}(z') dz' + {\cal R}\int_{z_0}^{\infty} g_r(z,z') {\cal 
S}(z') dz' + \int_{z_0}^{\infty} g(z,z') V(z') \psi(z') dz' \, .\label{eq:step0}
\end{equation}
Now, in the same way as a Dyson series is obtained, we replace the $\psi(\omega,x')$ in the third integral with the 
entirety of the rhs of Eq. \eqref{eq:step0} evaluated at $x'$. Collecting powers of ${\cal R}$ yields
\begin{eqnarray}
\psi&=&\int \! g_o {\cal S} + \int \!\!\int\! g_o V g_o {\cal S}+ {\cal R}\, \bigg[ \int \! g_r {\cal S} + 
\int \!\! \int\! (g_r V g_o + g_o V g_r)  {\cal S}\bigg]\nonumber\\
&+&  {\cal R}^2 \!\int\!\! \int\! g_r V g_r {\cal S} + \int\!\! \int \! g \, V g \, V   
\psi  \,.\label{eq:step1}
\end{eqnarray}

If we continue this process we end up with a geometric-like series in powers of $R$,
\begin{equation}
\psi= \psi_o + \sum_ {n=1}^\infty \psi_n \, ,\label{eq:tpsi}
\end{equation}
with each term a Dyson series itself:
\begin{equation}
\psi_o= \sum_{k=1}^\infty \int_{z_0}^{\infty} \! g_o(z,z_1) \cdots g_o(z_{k-1},z_k) V(z_1) \,\, \cdots V(z_{k-1}) {\cal 
S}(z_k) dz_1 \cdots dz_k \,.\label{eq:24a}
\end{equation}
The reflectivity terms can be re-arranged as:
\begin{eqnarray}
\psi_n &=& \sum_{k=n}^\infty {{\cal R}^n \over n! (k-n)!} \sum_{\sigma \in P_k} 
\int_{z_0}^{+ \infty} g_r(z_{\sigma(1)-1},z_{\sigma(1)}) \cdots 
 g_r(z_{\sigma(n)-1},z_{\sigma(n)})g_o(z_{\sigma(n+1)-1},z_{\sigma(n+1)}) \nonumber\\
&&\cdots  g_o(z_{\sigma(k)-1},z_{\sigma(k)}) \times V(x_1) \cdots V(z_{k-1})\, {\cal S}(z_k)\, dz_1 \cdots 
dz_k\,,\label{eq:Dyson}
\end{eqnarray}
where $x_0 \! :=\! x$, $P_k$ is the permutation group of degree $k$ and ${1 \over n! (k-n)!}  \! \sum_{\sigma \in P_k} 
$ 
represents the sum on all possible distinct ways of ordering $n$ $g_r$'s and $k-n$ $g_o$'s, resulting in a total of 
${|P_k| \over n! (k-n)!} = \binom{k}{n}$ terms~\cite{Correia:2018apm}.

Although complex-looking, Eq. \eqref{eq:Dyson} has a special significance, giving the amplitude of the 
(Fourier-transformed) $n$-th \emph{echo} of the initial burst~\cite{Correia:2018apm}.
When ${\cal R}=0$ then $\psi =\psi_o$, the open system waveform. There are no echoes as expected. When ${\cal R} \neq 0$ 
there are additional (infinite) Dyson-series terms. The series is expected to converge, (i.e., the contribution 
of $\psi_n$ becomes smaller for large $n$), because of two features of Eq.~\eqref{eq:Dyson}:
first, if $|{\cal R}| < 1$, ${\cal R}^n$ contributes to damp the contribution of large-$n$ terms.  
Moreover, the Dyson series starts at $k=n$. Since $g_o$ and $g_r$ are of the same order of magnitude, it is natural to 
expect that the series starting ahead (with less terms) has a smaller magnitude and contributes less to $\psi$ than the 
ones preceding them. This can be verified numerically.

Finally, an important outcome of this analysis is that echoes that arise later have a smaller frequency component than the 
first ones: the Dyson series is basically an expansion on 
powers of $V/\omega^2$; thus by starting at $k=n$, $\psi_n$ skips the high frequency contribution to the series until 
that term.
This is easily explained on physical grounds: high frequency components ``leak'' easily from the cavity (the cavity 
being formed by the ultracompact object and the potential barrier).
Lower frequency components are harder to tunnel out. Thus, at late times only low frequencies are present.

Recently, this approach was extended to ECOs modeled with a multiple-barrier filter near the surface, showing 
that the late-time ringdown exhibits mixing of echoes~\cite{Li:2019kwa}.

\subsubsection{Echo modeling}\label{sec:echotemplates}
The GW signal composed of echoes is a \emph{transient} signal, which captures the transition between
the photosphere ringdown. GW echoes are not well described by the QNMs of the ECO, which dominate 
the response only at very late times. Thus, a proper understanding of the signal in the ``echoing stage''
requires the full understanding of the theory and ensuing dynamics of the object. Unfortunately,
as we discussed, there is a plethora of proposed candidate theories and objects, with unknown properties.
Thus, the GW signal is known accurately for only a handful of special setups, and under very specific assumptions on the matter content~\cite{Cardoso:2016oxy,Price:2017cjr}. For this reason the echo signal is very rich, and different approaches have 
been recently developed to model it.

\paragraph{{\bf Templates for matched-filters.}}
The first phenomenological time-domain echo template was proposed in Ref.~\cite{Abedi:2016hgu}. It is based on a standard GR inspiral-merger-ringdown template ${\cal M} (t)$ and five extra free parameters,
\begin{equation}
 h(t) \equiv A\displaystyle\sum_{n=0}^{\infty}(-1)^{n+1}\eta^{n} {\cal M}(t+t_{\rm 
merger}-t_{\rm echo}-n\Delta t_{\rm echo},t_{0})\,, \label{templateAbedi} 
\end{equation}
with ${\cal M} (t, t_{0}) \equiv\Theta(t, t_{0}) {\cal M} (t)$ and where
\begin{equation}
\Theta(t, t_{0})\equiv\frac{1}{2}\left\{1+ \tanh\left[\frac{1}{2} \omega(t)(t-t_{\rm merger}-t_{0})\right]  \right\}\,,
\end{equation}
is a smooth cut-off function.
The parameters are the following: $\Delta t_{\rm echo}=2\tau_{\rm echo}$ is the 
time-interval in between successive echoes, see Eq.~\eqref{tauecho} for nonspinning objects and Eq.~\eqref{tauechospin} 
below when rotation is included; $t_{\rm echo}$ is the time of arrival of the first echo, which
can be affected by nonlinear dynamics near merger and does not necessarily coincide with $\Delta t_{\rm echo}$; $t_0$ 
is a cutoff time which dictates the part of the GR merger template
used to produce the subsequent echoes; $\eta\in [0,1]$ is the (frequency-dependent) damping factor of successive 
echoes; $A$ is the overall amplitude of the echo template with respect to the main burst at the merger (at $t=t_{\rm 
merger}$). Finally, $\omega(t)$ is a phenomenological time-dependent mode frequency that is used in standard 
inspiral-merger-ringdown phenomenological models~\cite{TheLIGOScientific:2016src}.
For a given model, the above parameters are not necessarily independent, as discussed below.
The $(-1)^{n+1}$ term in Eq.~\eqref{templateAbedi} is due to the phase inversion of the truncated model in each 
reflection. This implies that Dirichlet boundary conditions are assumed on the surface (or, more generally, that the 
reflection coefficient is real and negative, see discussion in Ref.~\cite{Testa:2018bzd}). The phase inversion does not 
hold for Neumann-like boundary conditions or for wormholes~\cite{Testa:2018bzd}.
This template was used in actual searches for echoes in the post-merger phases of LIGO/Virgo BH events, with 
conflicting claims discussed in Sec.~\ref{sec:testechoes}. Extensions of the original 
template~\cite{Abedi:2016hgu} have been developed and analyzed in Ref.~\cite{Wang:2018gin} and in 
Ref.~\cite{Uchikata:2019frs}.

A more phenomenological time-domain template, less anchored to the physics of echoes was proposed in 
Ref.~\cite{Maselli:2017tfq}, using a superposition of sine-Gaussians with several free parameters.
This template is very generic, but on the other hand suffers from a proliferation of parameters, which should not be in 
fact independent.

Note that the above two templates were directly modelled for spinning ECOs, since their underlying ingredients are very 
similar to the nonspinning case.

A frequency-domain template for nonspinning ECOs was built in Ref.~\cite{Testa:2018bzd} by approximating the BH 
potential with a P\"{o}schl-Teller potential~\cite{Poschl:1933zz,Ferrari:1984zz}, thus finding an analytical 
approximation to the transfer function defined in Eq.~\eqref{transfer_echoes}. 
The template construction assumes that the source is localized in space, which allows to solve for the 
Green's function analytically. The final form of the ECO response in the frequency domain reads
\begin{equation}
h(\omega) = h_{\rm BH}^{\rm ringdown}(\omega) \left[1+{\cal R}'\frac{\pi-e^{2 i 
\omega d}\Upsilon\cosh\left(\frac{\pi \omega_R}{\alpha}\right)}{\pi+e^{2 i 
\omega d}{\cal R}'\Upsilon\cosh\left(\frac{\pi \omega_R}{\alpha}\right)} \right] 
\,, \label{FINALTEMPLATEGEN}
 \end{equation}
where $d$ is the width of the cavity of the potential (i.e. the distance between the surface and the potential 
barrier), 
\begin{equation}
 {\cal R}'\equiv {\cal R}e^{2i\omega z_0} \,,\label{Rprime}
\end{equation}
is the ECO reflection coefficient defined as in Refs.~\cite{Mark:2017dnq,Testa:2018bzd} (notice the phase difference 
relative to that of Eq.~\eqref{bc_ECO}), $h_{\rm BH}^{\rm ringdown}$ is the standard BH ringdown template, 
$\omega_R$ is the real part of the QNMs of the corresponding BH, $\alpha$ is a parameter of the P\"{o}schl-Teller 
potential, defined by $\omega_R = \sqrt{V_0 - 
\alpha^2/4}$, $V_{\rm max}$ being the value of the exact potential at the maximum, and $
\Upsilon=\Gamma\left(\frac{1}{2}  - 
i\frac{\omega+\omega_R}{\alpha}\right)\Gamma\left(\frac{1}{2}  - 
i\frac{\omega-\omega_R}{\alpha}\right)\frac{\Gamma\left(1+\frac{i\omega}{\alpha}
\right)}{\Gamma\left(1-\frac{i\omega}{\alpha}\right)}$.
The above expression assumes that the source is localized near the surface, a more general 
expression is provided in Ref.~\cite{Testa:2018bzd}.
Notice that the quantity ${\cal R}'$ has a more direct physical meaning than ${\cal R}$. For example, Dirichlet 
and Neumann boundary conditions on $\psi$ correspond to ${\cal R}'=-1$ and ${\cal R}'=1$, respectively (see 
Eq.~\eqref{bc_ECO}.

The above template depends only on two physical inputs: the reflection coefficient ${\cal R}$ (or ${\cal R}'$) --~which 
can be in general a complex function of the frequency~-- and the width of the cavity $d$, which is directly related to 
the compactness of the object. For a given model of given compactness, ${\cal R}(\omega)$ and $d$ are fixed and the 
mode does not contain other free parameters. 
For example, the damping factor introduced in the previous template can be 
written in terms of ${\cal R}$ and the reflection coefficient of the BH potential, ${\cal R}_{\rm BH}$ (see 
Eq.~\eqref{refltrasmcoeff}) as $\eta=|{\cal R}{\cal R}_{\rm BH}|$~\cite{Testa:2018bzd}. Since ${\cal R}_{\rm BH}$ is 
frequency dependent so must be $\eta$, even in the case of perfect reflectivity ($|{\cal R}|=1$).
The time-domain waveform contains all the features previously 
discussed for the echo signal, in particular amplitude and frequency modulation and phase inversion of each echo 
relative to the previous one for certain boundary conditions~\cite{Testa:2018bzd}.

Note that practically all generic modeling of echoes which do not start from a first-principles calculation of the GW 
signal assume equal-spacing for the echoes. This seems certainly a good approximation for stationary geometries, but 
will fail for collapsing objects for example~\cite{Wang:2018mlp,Wang2019}. Furthermore, if the ECO reflective 
properties are modeled as a multiple-barrier filter --~as in certain scenarios motivated by BH area 
quantization~\cite{Bekenstein:1995ju,Cardoso:2019apo}~-- mixing of echoes occurs~\cite{Li:2019kwa}.

\paragraph{{\bf Wavelets for burst searches.}} 

Heuristic expressions for the echoing signal are useful, but the performance of template-based search techniques is
highly dependent on the (unknown, in general) ``faithfulness'' of such templates. Based on the excellent performance of wavelet analysis for glitch signals, Ref.~\cite{Tsang:2018uie} proposed a ``morphology-independent'' echo-search.
The analysis is based on generalized wavelets which are ``combs''
of sine-Gaussians, characterized by a time separation between
the individual sine-Gaussians as well as a fixed phase shift between
them, an amplitude damping factor, and a widening factor.
Even though actual echo signals are unlikely to resemble
any single generalized wavelet and may not even have well-defined
values for any of the aforementioned quantities, superpositions of generalized wavelets are expected to
capture a wide variety of physical echo waveforms. The comb is composed of a number $N_G$ of sine-Gaussians,
\be
h=\sum_{n=0}^{N_G}A\eta^n {\rm exp}\left(-(t-t_n)^2/(w^{2n}\tau^2)\right)\cos {\left(2\pi f_0 (t-t_n)+\phi_0+n\Delta\phi\right)}\,,
\ee
with $f_0$ a central frequency, $\tau$ is a damping time, $\Delta t$ the time between successive sine-Gaussians, 
$\Delta \phi$ is a phase difference between them, $\eta$ is a damping factor between
one sine-Gaussian and the next, and $w$ is a widening
factor. Here, $A$ is an overall amplitude, $t_0$ the central
time of the first echo and $\phi_0$ a reference phase.

\paragraph{{\bf Searches with Fourier windows.}} 
A similar but independent search technique was devised in Ref.~\cite{Conklin:2017lwb},
and uses the fact that echoes should pile up power at very specific frequencies
(those implied by the cavity delay time) which are nearly equally spaced (cf. Eq.~\eqref{omegaRecho}) (but see 
\cite{Wang:2018mlp,Wang2019}). 
Thus, the technique consists on producing a ``combing'' window in Fourier space, able to
match (maximizing over extrinsic parameters) the frequencies of the cavity. The specific shape of the tooth-comb
was found not to be determinant, as long as it is able to capture the power in the resonant mode. An extension of 
this strategy is discussed in Ref.~\cite{Conklin:2019fcs}.
\subsubsection{Echoes: a historical perspective}
There exist in the literature examples of works where the main gist of the idea behind echoes is present, albeit only for 
specific examples and without the full appreciation of the role of the light ring. Already in 1978, the study of the 
instability of  spinning horizonless compact objects (see Sec.~\ref{sec:ERinstability}) led to the understanding that 
the driving mechanism were the recurrent reflections of quasi-bound states within the ergoregion~\cite{Vilenkin:1978uc}. 
\emph{Mutatis mutandis}, these modes produce the echoes discussed above. Indeed, a Penrose diagram similar to that 
of Fig.~\ref{fig:echo_Penrose} was already shown in Ref.~\cite{Vilenkin:1978uc} (without a discussion of the GW 
emission slowly leaking from the potential barrier).

Probably the first example of echoes dates back to 1995, with the study of axial GWs emitted by perturbed (through 
Gaussian wavepackets) constant-density compact stars~\cite{Kokkotas:1999bd,Kokkotas:1995av}. This was later extended in 
the following years to include the scattering of point 
particles~\cite{Tominaga:1999iy,Andrade:1999mj,Tominaga:2000cs,Ferrari:2000sr,Andrade:2001hk}. In all these studies the 
GW signal shows a series of clear echoes after the main burst of radiation, which were identified as the excitation of 
quasi-trapped modes of ultracompact stars~\cite{Chandrasekhar449}. As we explained in 
Section~\ref{sec:echoes_transient}, the true trapped-mode behavior only sets in at much later times, and the correct 
description is that of echoes. The original references did not attempt to explain the pattern 
in the signal, but in hindsight these results fit perfectly 
in the description we provided above: axial modes do not couple to the fluid (nor polar modes, which couple only very weakly~\cite{Andersson:1995ez}) and travel free to the geometrical center of the star, which is therefore the effective surface in this particular case. The time delay of the echoes in Fig.~1 of Ref.~\cite{Ferrari:2000sr} is very well described by the GW's roundtrip time to the center, $\tau_{\rm echo}\sim \frac{27\pi}{8}\epsilon^{-1/2}M$, where $r_0=\frac{9}{4}M(1+\epsilon)$ is the radius of 
the star~\cite{Pani:2018flj} and $r_0=\frac{9}{4}M$ is the Buchdahl's limit~\cite{Buchdahl:1959zz}.
 
Shortly after, but in a very different context, the overall picture of echoes would emerge in the fuzzball program.
In Refs.~\cite{Lunin:2001jy,Giusto:2004ip}, the authors express the reflection coefficient of low-energy scalars
as a sum over the number of bounces at the ``throat'' of these geometries. The idea behind is similar to the 
expansion~\eqref{transfer_echoes}, and results in a series of ``echoes''~\cite{Giusto:2004ip}. A quantitative 
calculation of the response, as well as the role of the light ring, were left undone.

In the context of wormhole physics (particularly the geometry \eqref{solodukhin}), the main features of the 
response of ClePhOs were identified in Ref.~\cite{Damour:2007ap}. The postmerger train of echoes of the main burst was 
not addressed quantitatively.

Finally, in yet a different context, Ref.~\cite{Barausse:2014tra} discussed the late-time response of ``dirty'' BHs,
modeling environmental effects (such as stars, gas etc) and showed that there are  ``secondary pulses'' of radiation in 
the late-time response. These secondary pulses are just the echoes of a ``mirrored'' version of our original problem, 
where now it is the far region responsible for extra features in the effective potential, and hence the cavity is 
composed of the photosphere and the far region where matter is located.

\subsection{The role of the spin} \label{sec:DynSpin}

The previous sections dealt with static background spacetimes. Rotation introduces 
qualitatively new effects. For a Kerr BH, spinning with horizon angular velocity $\Omega$
along the azimuthal angle $\phi$, perturbations are well understood using the Newman-Penrose formalism and a 
decomposition in so-called spin-weighted spheroidal harmonics~\cite{Teukolsky:1972my,Teukolsky:1973ha,MTB}. It is still 
possible to reduce the problem to a PDE similar to Eq.~\eqref{pde}, but the effective potential is frequency-dependent;
breaks explicitly the azimuthal symmetry, i.e. it depends also on the azimuthal number $m$ 
(fluctuations depend on the azimuthal angle as $\sim e^{im\phi}$); is generically complex, although there exist transformations of the perturbation variables that make it real~\cite{1977RSPSA.352..381D,Maggio:2018ivz};
In particular, the explicit dependence on $m$ gives rise to a Zeeman splitting of the QNMs as functions of the spin, 
whereas the frequency dependence gives rise to the interesting phenomenon of \emph{superradiance} whereby modes with 
frequency $\omega$ are amplified when $\omega(\omega-m\Omega)<0$.
In particular, the potential is such that $V(r\to\infty)=\omega^2$, whereas $V(r\to r_+)=k^2$, with $k=\omega-m\Omega$.
The relation between null geodesics and BH QNMs in the eikonal limit is more involved but conceptually similar to the 
static case~\cite{Yang:2012he}.

Further features arise if the object under consideration is not a Kerr BH. In general, the vacuum region outside a spinning object is not described by the Kerr geometry. However, when $\epsilon\to0$ any deviation from the multipolar 
structure of a Kerr BH must die off sufficiently fast~\cite{Raposo:2018xkf,Glampedakis:2017cgd} (see Sec.~\ref{sec:multipoles}).
Explicit examples are given in Refs.~\cite{Pani:2015tga,Uchikata:2015yma,Uchikata:2016qku,Yagi:2015hda,Yagi:2015upa,Posada-Aguirre:2016qpz}. Therefore, if one is interested in the very small $\epsilon$ limit,
one can study a \emph{Kerr-like ECO}~\cite{Cardoso:2008kj,Abedi:2016hgu,Maggio:2017ivp,Nakano:2017fvh}, i.e. a geometry 
described by the Kerr metric
when $r>r_0=r_+(1+\epsilon)$ and with some membrane with model-dependent reflective properties at $r=r_0$.
Beyond the $\epsilon\to0$ limit, ECOs may have arbitrary multipole moments and even break equatorial 
symmetry~\cite{Raposo:2018xkf,Papadopoulos:2018nvd}. In such cases, it may not be possible to separate 
variables~\cite{Glampedakis:2017cgd,Glampedakis:2018blj,Allahyari:2018cmg,Pappas:2018opz} and the results below may not 
hold.

\subsubsection{QNMs of spinning Kerr-like ECOs}

Scalar, EM and gravitational perturbations in the exterior Kerr
geometry are described in terms of Teukolsky's master
equations~\cite{Teukolsky:1972my,Teukolsky:1973ha,Teukolsky:1974yv}
\begin{eqnarray}
\Delta^{-s} \frac{d}{dr}\left(\Delta^{s+1} \frac{d_{s}R_{lm}}{dr}\right)+ \left[\frac{K^{2}-2 i s (r-M) K}{\Delta}+4 i 
s \omega r -\lambda\right] \ _{s}R_{l m}&=&0\,,\label{wave_eq} \quad \\
\left[\left(1-x^2\right)~_{s}S_{l m,x}\right]_{,x}+ \bigg[(a\omega x)^2-2a\omega s x + 
s+~_{s}A_{lm}-\frac{(m+sx)^2}{1-x^2}\bigg]~_{s}S_{l m}&=&0\,, \label{angular}
\end{eqnarray}
where $a=\chi M$, $~_{s}S_{l m}(\theta)e^{im\phi}$ are spin-weighted spheroidal
harmonics, $x\equiv\cos\theta$, $K=(r^2+a^2)\omega-am$, and the separation
constants $\lambda$ and $~_{s}A_{l m}$ are related by $\lambda \equiv  ~_{s}A_{l
m}+a^2\omega^2-2am\omega$. When $\chi=0$, the angular eigenvalues are $\lambda=(l-s)(l+s+1)$, whereas for $\chi\neq0$ 
they can be computed numerically or with approximated analytical expansions~\cite{Berti:2005gp}. 

It is convenient to make a change of variables by introducing the function~\cite{1977RSPSA.352..381D}
\begin{equation}
 _{s}X_{lm} = \Delta^{s/2} \left(r^2+a^2\right)^{1/2} \left[\alpha \
_{s}R_{lm}+\beta \Delta^{s+1} \frac{d_{s}R_{lm}}{dr}\right]\,,\label{DetweilerX}
\end{equation}
where $\alpha$ and $\beta$ are certain radial functions. Introducing the
tortoise coordinate $r_*$, defined such that
$dr_*/dr=(r^2+a^2)/\Delta$, the master equation~\eqref{wave_eq} becomes
\begin{equation}
 \frac{d^2_{s}X_{lm}}{dr_*^2}- V(r,\omega) \,_{s}X_{lm}=0\,, \label{final}
\end{equation}
where the effective potential is
\begin{equation}
 V(r,\omega)=\frac{U\Delta }{(r^2+a^2)^2}+G^2+\frac{dG}{dr_*}\,,
\end{equation}
and
\begin{eqnarray}
G &=& \frac{s(r-M)}{r^2+a^2}+\frac{r \Delta}{(r^2+a^2)^2} \,, \\
U &=& \frac{2\alpha' + (\beta' \Delta^{s+1})'}{\beta \Delta^s} -\frac{1}{\Delta}\left[K^2-is\Delta'K+\Delta(2isK'-\lambda)\right] \,.
\end{eqnarray}
The prime denotes a derivative with respect to $r$ and the functions $\alpha$ and
$\beta$ can be chosen such that the resulting potential is \emph{purely real} (definitions of $\alpha$ and $\beta$
can be found in Ref.~\cite{1977RSPSA.352..381D} and ~\cite{Maggio:2018ivz} for EM and gravitational perturbations respectively).
It is natural to define the generalization of Eq.~\eqref{bc_ECO} as~\cite{Maggio:2018ivz}
\begin{equation}
X_{\rm ECO}\sim e^{-i k  z}+{\cal R}e^{i k z}\,,\quad z\to z_0\,,\label{bc_ECOspin}
\end{equation}
where now ${\cal R}$ generically depends on the frequency and on the spin and $z$ is the Kerr tortoise coordinate defined by $dz/dr=(r^2+a^2)/\Delta$.

\begin{figure*}[ht]
\begin{center}
\includegraphics[width=0.87\textwidth]{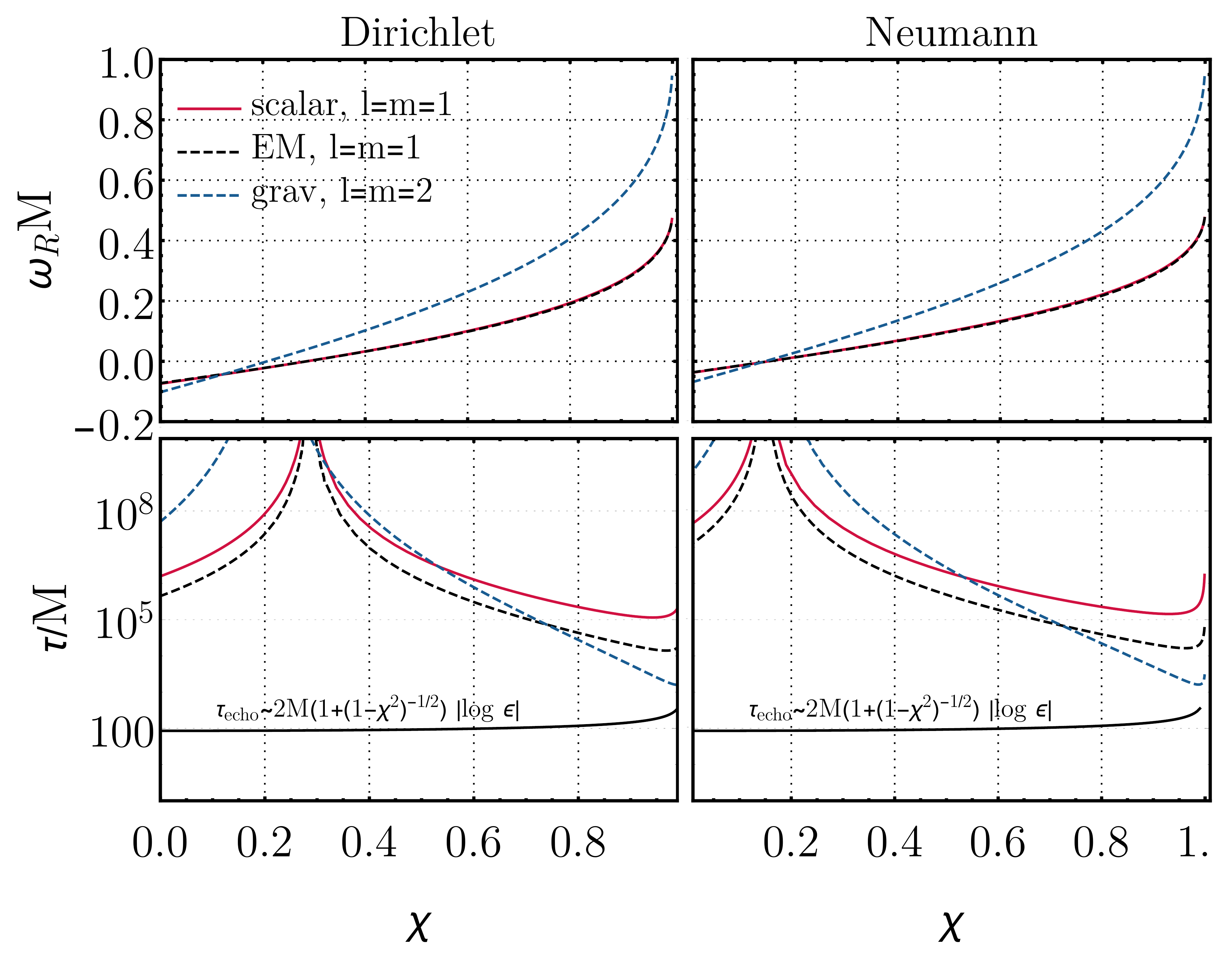}
\caption{Frequency (top panels) and damping/growing time (bottom panels) of scalar, EM, and gravitational QNMs of a 
Kerr-like ECO with a perfectly reflective surface for either Dirichlet (left panels) or Neumann (right panels) 
boundary conditions. We choose $\epsilon=10^{-10}$ [data adapted from Ref.~\cite{Maggio:2018ivz}]. 
Modes are stable (i.e., they decay in time) for $\omega_R<0$, whereas they turn unstable (i.e., they grow in time) when 
$\omega_R>0$. The damping/growing time diverges for marginally stable modes, when $\omega_R=0$.
In the bottom panels, the continuous black curve represents the characteristic echo delay time, much shorter than the instability time scale.).
\label{fig:KerrlikeQNMs}}
\end{center}
\end{figure*}
To search for the characteristic or QNMs of the system, Eq.~\eqref{final} is to be solved with boundary 
condition~\eqref{bc_ECOspin} at $z\sim z_0$ and (outgoing) $X\sim e^{i\omega z}$ at infinity~\cite{Maggio:2018ivz}.
A small frequency approximation yields~\cite{Starobinskij2,Maggio:2017ivp,Maggio:2018ivz} 
\begin{eqnarray}
M \omega_R&\simeq&m\Omega-\frac{M\pi}{2|z_0|}\left(q+\frac{s(s+1)}{2}\right)\,, \label{omegaRecho_spin}\\
M\omega_I &\simeq& 
-\frac{\beta_{ls}}{|z_0|}\left(\frac{2M^2r_+}{r_+-r_-}\right)\left[\omega_R(r_+-r_-)\right]^{2l+1}(\omega_R-m\Omega)\,, 
\label{omegaIechospin}
\end{eqnarray}
where now $z_0\sim M[1+(1-\chi^2)^{-1/2}]\log\epsilon$. 
This result shows how angular momentum can 
bring about substantial qualitative changes. The spacetime is unstable for $\omega_R(\omega_R-m\Omega)<0$ (i.e., in the 
superradiant regime~\cite{Brito:2015oca,Vicente:2018mxl}), on a timescale $\tau_{\rm inst}\equiv 1/\omega_I$. This phenomenon is called 
\emph{ergoregion instability}~\cite{Friedman:1978wla,Brito:2015oca,Moschidis:2016zjy,Vicente:2018mxl} (see Sec.~\ref{sec:ERinstability} 
below).
In the $\epsilon\to0$ limit and for sufficiently large spin, $\omega_R\sim m\Omega$ and $\omega_I\sim 
|\log\epsilon|^{-2}$. Note that, owing to the $\omega_R-m\Omega$ term in Eq.~\eqref{omegaIechospin}, polar 
and axial modes are not isospectral in the spinning case, even when $\epsilon\to0$: indeed, they have the same 
frequency but a slightly different time scale.
Numerical results, shown in Fig.~\ref{fig:KerrlikeQNMs}, are in excellent agreement with the 
above analytical approximations whenever $\omega M\ll1$, which also implies small rotation rates~\cite{Maggio:2018ivz}.
For very large spins there exists a more complex analytical approximation~\cite{Hod:2017cga}.
Note that in the superradiant regime the ``damping'' factor, $\omega_I/\omega_R>0$, so that, at very late times 
(when the pulse frequency content is indeed described by these formulas), the amplitude of the QNMs \emph{increases} due 
to the instability. This effect is small --~for example, $\omega_I/\omega_R\approx 4\times10^{-6}$ when 
$\epsilon=0.001$, $l=2$ and $\chi=0.7$~-- and, more importantly, it does not affect the first several echoes, since the 
latter appear on a timescale much shorter than the instability time scale (see Fig.~\ref{fig:KerrlikeQNMs}).

\subsubsection{Echoes from spinning ECOs}
GW echoes from spinning ECOs have been investigated actively~\cite{Abedi:2016hgu,Nakano:2017fvh,Bueno:2017hyj,Conklin:2017lwb,Vicente:2018mxl}. The overall picture is similar to the static case, with two notable differences. The echo delay 
time~\eqref{tauecho} now reads~\cite{Abedi:2016hgu}
\begin{equation}
\tau_{\rm echo}\sim 2M[1+(1-\chi^2)^{-1/2}]|\log\epsilon|\,, \label{tauechospin}
\end{equation}
in the $\epsilon\to0$ limit. This time scale corresponds to the period of the \emph{corotating} mode, $\tau_{\rm 
echo}\sim (\omega_R-m\Omega)^{-1}$. In addition, as we discussed above the spacetime is unstable over a time scale 
$\tau=1/|\omega_I|$. Such timescale is parametrically longer than $\tau_{\rm echo}$ 
(see~Fig.~\ref{fig:KerrlikeQNMs}) and does not affect the first $N\approx\tau/\tau_{\rm echo}\sim 
|\log\epsilon|$ echoes.
As we explained earlier, the signal can only be considered as a series of well-defined pulses at early 
stages, when the pulse still contains a substantial amount of high-frequency components. 
Thus, amplification occurs only at late times; the early-time evolution of the pulse generated at the photon sphere is 
more complex.

The transfer function of Eq.~\eqref{transferfunction} can be generalized to spinning 
``Kerr-like'' ECO subjected to boundary condition~\eqref{bc_ECOspin} near the surface. The final result 
reads formally the same, although ${\cal T}_{\rm BH}$ and ${\cal R}_{\rm BH}$ are defined in terms of the amplitudes of 
the waves scattered off the Kerr effective potential~\cite{Conklin:2017lwb,TestaInPrep}.
Echoes from Kerr-like wormholes (i.e., a spinning extension of the Damour-Solodukhin solution~\cite{Damour:2007ap}) 
have been studied in Ref.~\cite{Bueno:2017hyj}.
Phenomenological templates for echoes from Kerr-like objects were constructed 
in Refs.~\cite{Abedi:2016hgu,Nakano:2017fvh,Maselli:2017tfq,Wang:2018gin} and are discussed in 
Sec.~\ref{sec:echotemplates}.
%

\subsection{The stability problem}

\hskip 0.2\textwidth
\parbox{0.8\textwidth}{
\begin{flushright}
{\small 
\noindent {\it ``There is nothing stable in the world; uproar's your only music.''}\\
John Keats, Letter to George and Thomas Keats, Jan 13 (1818)
}
\end{flushright}
}

\vskip 1cm

Appealing solutions are only realistic if they form and remain as long-term stable solutions of the theory.
In other words, solutions have to be stable when slightly perturbed or they would not be observed (or they would not 
even form in the first place). There are strong 
indications that the exterior Kerr spacetime is stable, although a rigorous proof is still 
missing~\cite{Dafermos:2008en}. On the other hand, some --~and possibly most 
of~-- horizonless compact solutions are linearly or nonlinearly unstable. 

Some studies of linearized fluctuations of ultracompact objects are given in Table~\ref{tab:ECOs}. We will not discuss 
specific models, but we would like to highlight some general results.

\subsubsection{The ergoregion instability}\label{sec:ERinstability}
Several models of UCOs and ClePhOs are stable under \emph{radial} 
perturbations~\cite{1985CQGra...2..219I,Visser:2003ge} (see Table~\ref{tab:ECOs}). However,
UCOs (and especially ClePhOs) can develop negative-energy regions once spinning. In such a case, 
they develop a {\it linear} instability under \emph{non-radial} perturbations, which is dubbed as ergoregion 
instability. Such instability affects any horizonless 
geometry with an 
ergoregion~\cite{Friedman:1978wla,Kokkotas:2002sf,Moschidis:2016zjy,Cardoso:2007az,Oliveira:2014oja,Maggio:2017ivp,
Vicente:2018mxl} and is deeply connected to superradiance~\cite{Brito:2015oca}.
The underlying mechanism is simple: a negative-energy fluctuation in the ergoregion is forced to 
travel outwards; at large distances only positive-energy states exist, and energy conservation implies that the initial 
disturbance gives rise to a positive fluctuation at infinity plus a larger (negative-energy)
fluctuation in the ergoregion. Repetition of the process leads to a cascading instability. The only way to prevent such 
cascade from occurring is by absorbing the negative energy states, which BHs do efficiently (and hence Kerr BHs are 
stable against massless fields), but perfectly-reflecting horizonless objects must then be unstable.

This instability was discovered by Friedman for ultracompact slowly-rotating stars with an 
ergoregion~\cite{friedman1978}, and later extended in Refs.~\cite{CominsSchutz,1996MNRAS.282..580Y,Kokkotas:2002sf}.
Application to Kerr-like horizonless objects started in Ref.~\cite{Vilenkin:1978uc}, whereas an analysis for 
gravastars, boson stars, and other objects was done in 
Refs.~\cite{Cardoso:2005gj,Cardoso:2007az,Cardoso:2008kj,Chirenti:2008pf}.  More recently, 
Refs.~\cite{Maggio:2017ivp,Maggio:2018ivz} gave a detailed analysis of scalar, electromagnetic, and gravitational 
perturbations of a partially-reflective Kerr-like ECO in the $\epsilon\to0$ limit.

The overall summary of these studies is that the instability time scale depends strongly on the spin and on the 
compactness of the objects. The ergoregion-instability 
timescale can very long~\cite{Friedman:1978wla,Cardoso:2007az,Maggio:2017ivp}. For concreteness, for gravastars with 
$\epsilon\sim 0.1-1$ the ergoregion is absent even for moderately high spin~\cite{Chirenti:2008pf}. However, at least 
for perfectly-reflecting Kerr-like ECOs in the $\epsilon\to0$ limit, the critical spin above which the object is 
unstable is very low~\cite{Maggio:2018ivz} [see Eq.~\eqref{omegaIechospin}]
\begin{equation}
 \chi_{\rm crit} \sim \frac{\pi }{m|\log\epsilon|}\left(q+\frac{s(s+1)}{2}\right)\,. \label{chicrit}
\end{equation}
For example, a totally reflecting surface a Planck length outside the horizon of a 
$10M_{\odot}$ ECO ($\epsilon = l_P / r_+ \approx 5 \times 10^{-40}$) will generate an 
ergoregion instability if $\chi \gtrsim 0.07$ for $q=1$, $m=1$, and $s=-2$.
Note that the instability time scale can be very large near the instability threshold. 
From Eq.~\eqref{omegaIechospin}, we can estimate the timescale of the instability of a 
spinning ClePhO,
\begin{equation}
 \tau_{\rm 
inst}\equiv\frac{1}{\omega_I}\sim-|\log\epsilon|\frac{1+(1-\chi^2)^{-1/2}}{2\beta_{ls}}\left(\frac{r_+-r_-}{r_+ 
}\right) 
\frac{\left[\omega_R(r_+-r_-)\right]^{-(2l+1)}}{\omega_R-m\Omega}\,. \label{tauinstab}
\end{equation}
As previously discussed, a spinning ClePhO is (superradiantly) unstable only above a critical value of the spin. For 
example, for $l=m=s=2$ and $\chi=0.7$, the above formula yields
\begin{equation}
 \tau\in\left(5,1\right)\left(\frac{M}{10^6M_\odot}\right)\,{\rm yr}\quad {\rm when~~} \epsilon\in 
(10^{-45},10^{-22})\,.
\end{equation}
Generically, the ergoregion instability acts on timescales which are parametrically longer 
than the dynamical timescale, $\sim M$, of the object, but still short enough to be relevant in astrophysical 
scenarios. 
Although the evolution of this instability remains an open problem, it is likely that it will remove angular momentum 
from the object, spinning it down until the threshold condition, $\chi=\chi_{\rm crit}$, is reached~\cite{Barausse:2018vdb}. The 
phenomenological consequences of this phenomenon will be discussed in Sec.~\ref{sec:Tests}.

A possible way to quench the instability is by absorbing the negative-energy modes trapped within the ergoregion. Kerr 
BHs can absorb such modes efficiently and are indeed expected to be stable even if they have an ergoregion.
Given its long timescales, it is possible that the instability can be efficiently quenched by some dissipation 
mechanism of nongravitational nature, although this effect would be 
model-dependent~\cite{Maggio:2017ivp,Maggio:2018ivz}. 
Unfortunately, the effect of viscosity in ECOs is practically unknown~\cite{Cardoso:2014sna,Guo:2017jmi},
and so are the timescales involved in putative dissipation mechanisms that might quench this instability. 
It is also possible that, when spinning, a partially-absorbing object can support quasi-trapped superradiant modes 
with $\omega_R<m\Omega$, which might lead to an instability similar to that of massive bosonic fields around Kerr 
BHs~\cite{Brito:2015oca}.

Finally, there are indications that instabilities of UCOs are merely the equivalent of Hawking radiation for these 
geometries, and that therefore there might be a smooth transition in the emission properties when approaching the BH 
limit~\cite{Chowdhury:2007jx,Damour:2007ap}.

\subsubsection{Nonlinear instabilities I: long-lived modes and their backreaction}

Linearized gravitational fluctuations of any 
nonspinning UCO are extremely long-lived and decay no faster than 
logarithmically~\cite{Keir:2014oka,Cardoso:2014sna,Eperon:2016cdd,Eperon:2017bwq}. Indeed, such perturbations can be 
again understood in terms of modes quasi-trapped within the potential barrier shown in Fig.~\ref{fig:potential}: they 
require a photon sphere but are absent in the BH case (hence the photon sphere is sometimes referred to as ``loosely trapped'' or ``transversely trapping'' surface~\cite{Shiromizu:2017ego,Yoshino:2017gqv}). For a ClePhO, these modes are very well 
approximated by Eqs.~\eqref{omegaRecho}--\eqref{omegaIecho} in the static case and by their aforementioned extension 
in the spinning case.
The long damping time of these modes has led to the conjecture that any UCO is \emph{nonlinearly} unstable and may 
evolve through a Dyson-Chandrasekhar-Fermi type of mechanism~\cite{Keir:2014oka,Cardoso:2014sna}.
The endstate is unknown, and most likely depends on the equation of the state of the particular UCO: some objects may 
fragment and evolve past the UCO region into less compact configurations, via mass ejection, whereas other UCOs may be 
forced into gravitational collapse to BHs. 

The above mechanism is supposed to be active for any spherically symmetric UCO, and also on spinning solutions. 
However, it is nonlinear in nature and not well understood so far. For example, there are indications that a putative 
nonlinear instability would occur on very long timescales only; a model problem predicts an exponential dependence 
on the size of the initial perturbation~\cite{FritzJohn}.

\subsubsection{Nonlinear instabilities II: causality, hoop conjecture, and BH formation\label{sec:hoop}}
The teleological nature of horizons leads to possible spacelike behavior in the
way they evolve. In turn, this has led to constraints on the possible compactness of 
horizonless objects. Ref.~\cite{Carballo-Rubio:2018vin} finds the conceptual bound
\begin{equation}
\epsilon\leq 4\dot{M}\,,
\end{equation}
based on a special accreting geometry (so-called Vaidya spacetime) and on the requirement that the surface of the 
accreting ECO grows in a timelike or null way. The assumptions behind such result are relatively strong: the accreting 
matter is a very particular null dust, eternally accreting at a constant rate and without pressure. In addition, 
superluminal motion for the ECO {\it surface} is not forbidden, and may well be a rule for such compact geometries.

A different, but related, argument makes use of the hoop conjecture~\cite{Chen:2019hfg} (see also 
Ref.~\cite{Addazi:2019bjz} for similar work). In broad terms, the hoop conjecture states 
that if a body is within its Schwarzschild radius, then it must be a BH~\cite{Thorne:1972,Choptuik:2009ww}). Take two 
ECOs of mass $m_2\ll m_1$, inspiralling to produce a single ECO. The burst of energy emitted in ringdown modes is of 
order~\cite{Berti:2007fi}
\begin{equation}
\frac{E_{\rm ringdown}}{m_1+m_2}\sim 0.44\left(\frac{m_1m_2}{(m_1+m_2)^2}\right)^2 \,,\label{rd_energy}
\end{equation}
This estimate holds for BHs, and it seems plausible that 
it would approximately holds also for ECOs.
A similar amount of energy goes {\it inwards}. Then, when the small body crosses the photon sphere of the large ECO, an 
amount of mass \eqref{rd_energy} is emitted inwards and is swallowed by the large ECO increasing its mass to 
$m_1+E_{\rm ringdown}$. The hoop conjecture implies that $2(m_1+E_{\rm ringdown})\leq 2m_1(1+\epsilon)$, or
\begin{equation}
\epsilon \gtrsim 0.44 \frac{m_2^2}{m_1^2}\,,
\end{equation}
to avoid BH formation. Thus, $\epsilon$ of Planckian order are not allowed. There are issues with this type of 
arguments: The GWs are not spherical and not localized (their wavelength is of the order or larger than the ECO itself), 
thus localizing it on a sphere of radius $2m_1(1+\epsilon)$ is impossible. Furthermore, the argument assumes that all 
energy reaching the surface is accreted, whereas it might be efficiently absorbed by other channels.

The above argument can be made more powerful, making full use of the hoop conjecture: take two ECOs and boost them to 
large enough energies. Since all energy gravitates and is part of the hoop, the final object must be inside its 
Schwarzschild radius, hence it must be a BH (indeed, at large enough center of mass energies, the structure of 
the colliding objects is irrelevant~\cite{Eardley:2002re,Choptuik:2009ww,Sperhake:2012me}). It is very challenging to bypass this 
argument \emph{at the classical level}. Nevertheless, it is important to highlight a few points: (i)~Most of 
the arguments for ECO formation (and existence) rely directly or indirectly on unknown {\it quantum} effects 
associated with horizon or singularity 
formation~\cite{Giddings:1992hh,Mazur:2004fk,Mathur:2005zp,Mathur:2008nj,Giddings:2009ae,Mathur:2009hf,
Giddings:2012bm,Barcelo:2015noa,Giddings:2017mym,Barcelo:2017lnx}. Thus, it is 
very likely that horizons may form classically but that 
such picture is blurred by quantum effects (on unknown timescales and due to unknown 
dynamics)\footnote{In this respect, a parallel can drawn with neutron stars, which can be well described within GR by 
a simple self-gravitating perfect fluid, but whose formation process is significantly more complex than the 
gravitational collapse of a perfect fluid. Incidentally, such processes involve complex microphysics and quantum 
effects such as those occurring in a supernova collapse. In other words, the fact that an equilibrium solution can 
be well described by simple matter fields does not necessarily mean that its formation is equally simple nor does it 
exclude more complex formation processes.}; (ii)~even classically, the argument does not forbid the existence of ECOs, 
it merely forces their interaction at high energy to result in BH formation (indeed, the same argument can be applied 
if the two objects are neutron stars).

\subsection{Binary systems} \label{Sec:binaries}
Consider a compact binary of masses $m_i$ ($i=1,2$), total mass $m=m_1+m_2$, mass ratio $q=m_1/m_2\geq1$, and 
dimensionless spins $\chi_i$.
In a post-Newtonian~(PN) approximation (i.e. a weak-field/slow-velocity expansion of Einstein's equations), dynamics is 
driven by energy and angular momentum loss, and particles are endowed with a series of multipole moments and with 
finite-size tidal corrections~\cite{Blanchet:2006zz}. 
Up to $1.5$PN order, the GW phase depends only on $m_i$ and $\chi_i$ and is oblivious to the compactness of the binary 
components.
Starting from $2$PN order, the nature of the inspiralling objects is encoded in:
\begin{itemize}
 \item[(i)] the way they respond to their own gravitational field --~i.e., on their own multipolar 
structure~\cite{Krishnendu:2017shb,Krishnendu:2018nqa,Kastha:2018bcr};
 \item[(ii)] the way they respond when acted upon by the external gravitational 
field of their companion~-- through their tidal Love numbers (TLNs)~\cite{PoissonWill};
 \item[(iii)] on the amount of radiation that they possibly absorb, i.e. on tidal 
heating~\cite{Hartle:1973zz,PhysRevD.64.064004}.
\end{itemize}
These effects are all included in the waveform produced during the inspiral, and can be incorporated in the 
Fourier-transformed GW signal as 
\begin{equation}
\tilde{h}(f)={\cal A}(f)e^{i(\psi_\tn{PP}+\psi_\tn{TH}+\psi_\tn{TD})}\ , \label{waveform}
\end{equation}
where $f$ and ${\cal A}(f)$ are the GW frequency and amplitude, $\psi_\tn{PP}(f)$ is the ``pointlike'' 
phase~\cite{Blanchet:2006zz}, whereas $\psi_\tn{TH}(f),\,\psi_\tn{TD}(f)$ are the contributions of the tidal heating 
and the tidal deformability, respectively.

\subsubsection{Multipolar structure\label{Sec:multipolar}}

Spin-orbit and spin-spin interactions are included in $\psi_\tn{PP}$, the latter also depending on all higher-order 
multipole moments. The dominant effect is that of the spin-induced quadrupole moment, $M_2$, which yields a $2$PN 
contribution to the phase~\cite{Krishnendu:2017shb}
\begin{equation}
 \psi_{\rm quadrupole} = \frac{75}{64} \frac{\left(m_2 M_2^{(1)}+m_1 M_2^{(2)}\right)}{(m_1 
m_2)^2}\frac{1}{v}\,,\label{phaseM2}
\end{equation}
where the expansion parameter $v=(\pi m f)^{1/3}$ is the orbital velocity. By introducing the dimensionless 
spin-induced, quadrupole moment,
$\bar M_2^{(i)}=M_2^{(1)}/(\chi_i^2 m_i^3)$, it is clear that the above correction is quadratic in the spin.
For a Kerr BH, $\bar M_2^{(i)}=-1$, whereas for an ECO there will be 
generic corrections that anyway are bound to vanish as $\epsilon\to0$~\cite{Raposo:2018xkf} (Sec.~\ref{sec:multipoles}).

\subsubsection{Tidal heating\label{Sec:tides}}
A spinning BH absorbs radiation of frequency $\omega>m\Omega$, 
but amplifies radiation of smaller frequency~\cite{Brito:2015oca}.
In this respect, BHs are dissipative systems which behave just like a Newtonian viscous 
fluid~\cite{Damour_viscous,Poisson:2009di,Cardoso:2012zn}.
Dissipation gives rise to various interesting effects in a binary system --~such as tidal 
heating~\cite{Hartle:1973zz,PhysRevD.64.064004}, tidal acceleration, and tidal locking, as in the Earth-Moon system, 
where dissipation is provided by the friction of the oceans with the crust.

For low-frequency circular binaries, the energy flux associated to tidal heating at the 
horizon, $\dot E_H$, corresponds to the rate of change of the BH 
mass~\cite{Alvi:2001mx,Poisson:2004cw},
\begin{eqnarray}
 \dot M = \dot E_H \propto \frac{\Omega_K^5}{M^2}(\Omega_K-\Omega)\,, \label{Edot}
\end{eqnarray}
where $\Omega_K\ll 1/M$ is the orbital angular velocity and the (positive) prefactor 
depends on the masses and spins of the two bodies. Thus, tidal heating is stronger for 
highly spinning bodies relative to the nonspinning by a factor $\sim \Omega/\Omega_K\gg1$.

The energy flux~\eqref{Edot} leads to a potentially observable phase shift of GWs emitted 
during the inspiral. The GW phase $\psi$ is governed by $d^2\psi/df^2 =2\pi(dE/df)/\dot E$, where 
$E\sim v^2$ is the binding energy of the binary. To the leading order, this yields (for circular orbits and spins 
aligned with the orbital angular momentum)~\cite{Maselli:2017cmm}
\begin{equation}\label{Hphase}
\psi_\tn{TH}^{\rm BH} = \psi_{\tn{N}}\left(F(\chi_i,q) v^5\log v+G(q) v^8 [1-3\log v]\right)\,, 
\end{equation}
where $\psi_{\tn{N}}\sim v^{-5}$ is the leading-order contribution to the point-particle phase (corresponding to the 
flux ${\dot  E}_{\rm GW}$), and
\begin{eqnarray}
 F(\chi_i,q)&=&-\frac{10 \left(q^3 \left(3 \chi_1^3+\chi_1\right)+3 \chi_2^3+\chi_2\right)}{3 (q+1)^3} \,,\\
G(\chi_i,q)&=& \frac{10}{27 (q+1)^5} \left[q^5A_1+A_2 +q^4B_1+qB_2+q^3C_1+q^2C_2\right]\,,
\end{eqnarray}
with
\begin{eqnarray}
A_i&=&2 \left(3 \chi_i^2+1\right) \left(3-10 \chi_i^2+3 \Delta_i\right)\,,\\
B_i&=&3 \left(3 \chi_i^2+1\right) \left(2-5 \chi_i^2+2 
\Delta_j-5 \chi_i \chi_j\right)\,,\\
C_i&=&-20 \left(3 \chi_i^2+1\right) \chi_i\chi_j\,,
\end{eqnarray}
$\Delta_i\equiv \sqrt{1-\chi_i^2}$ and $j\neq i$. Therefore, absorption at the horizon introduces a $2.5$PN 
($4$PN)$\times\log 
v$ correction to the GW phase of spinning  (nonspinning) binaries, relative to the leading term.

Thus, it might be argued that an ECO binary can be distinguished 
from a BH binary, because $\dot E_H=0$ for the former. However, the 
trapping of radiation in ClePhOs can efficiently mimic the effect of a 
horizon~\cite{Maselli:2017cmm}. In order for absorption to affect the orbital motion, it 
is necessary that the time radiation takes to reach the companion, $T_{\rm rad}$, be much 
longer than the radiation-reaction time scale due to heating, $T_{\rm RR}\simeq E/\dot 
E_H$, where $E\simeq -\frac{1}{2} M(M\Omega_K)^{2/3}$ is the binding energy of the binary 
(assuming equal masses).
For BHs, $T_{\rm rad}\to\infty$ because of time dilation, so that the condition $T_{\rm 
rad}\gg T_{\rm RR}$ is always satisfied. 
For ClePhOs, $T_{\rm rad}$ is of the order of the GW echo delay time, 
Eq.~\eqref{tauechospin}, 
and therefore increases logarithmically as $\epsilon\to0$. Thus, an effective tidal 
heating might occur even in the absence of a horizon if the object is sufficient 
compact. The critical value of $\epsilon$ increases strongly as a function of the spin. 
For orbital radii larger than the ISCO, the condition $T_{\rm rad}\gg T_{\rm RR}$ 
requires 
$\epsilon\ll10^{-88}$ for $\chi\lesssim0.8$, and therefore even Planck corrections at the 
horizon scale are not sufficient to mimic tidal heating. This is not necessarily true for 
highly spinning objects, for example $T_{\rm rad}\gg T_{\rm RR}$ at the ISCO requires 
$\epsilon\ll10^{-16}$ for $\chi\approx0.9$.

\subsubsection{Tidal deformability and Love numbers}\label{sec:TLNs}

Finally, the nature of the inspiralling objects is also encoded in the way they respond 
when acted upon by the external gravitational field of their companion~-- through their 
tidal Love numbers~(TLNs)~\cite{PoissonWill}. An intriguing result in classical GR is that the TLNs of BHs are zero. This result holds: (i)~in the nonspinning case for 
weak tidal fields~\cite{Damour_tidal,Binnington:2009bb,Damour:2009vw} and also for tidal fields of arbitrary 
amplitude~\cite{Gurlebeck:2015xpa}; (ii)~in the spinning case~\cite{Poisson:2014gka,Pani:2015hfa,Landry:2015zfa} for 
weak tidal fields, at least in the axisymmetric case to second order in the spin~\cite{Pani:2015hfa} and generically to 
first order in the spin~\cite{Landry:2015zfa}.
On the other hand, the TLNs of ECOs 
are small but finite~\cite{Pani:2015tga,Uchikata:2016qku,Porto:2016zng,Cardoso:2017cfl,Wade:2013hoa,Giddings:2019}. 

In spherical symmetry, the TLNs can be defined as the proportionality factor between the induced mass 
quadrupole moment, $M_2$, and the (quadrupolar) external tidal field, $E_2$. Let us consider the mutual tides induced 
on the two bodies of a binary system at orbital distance $r$ due to the presence of a companion. In this case
\begin{equation}
 M_2^{(1)} = \lambda_1 E_2^{(2)} \qquad M_2^{(2)} = \lambda_2 E_2^{(1)}\,, \label{M2tidal}
\end{equation}
where $\lambda_i$ is the tidal deformability parameter of the $i$-th body. At Newtonian order, the external tidal 
field produced by the $i$-th object on its companion is simply
\begin{equation}
 E_2^{(i)} \sim\frac{m_i}{r^3}\propto v^6\,. \label{E2}
\end{equation}
%
The above results can be used to compute the contributions of the tidal deformability to the binding energy of the 
binary, $E(f)$, and to the energy flux dissipated in GWs, $\dot E$. The leading-order corrections 
read~\cite{Vines:2011ud}
\begin{eqnarray}
      E(f) &=&  -\frac{m q}{2(1+q)^2}v^2\left(1- \frac{6q(k_1 q^3+k_2)}{(1+q)^5} v^{10}\right)\,, \\
 \dot E(f) &=&  -\frac{32}{5}\frac{q^2}{(1+q)^4}v^{10}\left(1+ \frac{4 \left(q^4 (3+q) k_1+(1+3 q)
   k_2\right)}{(1+q)^5} v^{10} \right)\,, \label{Edot2}
\end{eqnarray}
where $k_i$ is the (dimensionless) TLN of the $i-$th object, defined as $\lambda_i=\frac{2}{3}k_i m_i^5$.
By plugging the above equations in $\frac{d^2\psi(f)}{df^2}=\frac{2\pi}{\dot E}\frac{dE}{df}$, we can solve for the 
tidal phase to leading order,
\begin{equation}
\psi_\tn{TD}(f)=- \psi_{\tn{N}}\frac{624\Lambda}{m^{5}} v^{10}\,, \label{tidalphase}
\end{equation}
where $39\Lambda=\left(1+{12}/{q}\right)m_1^5 k_1+(1+12 q)m_2^5 k_2$ is the weighted tidal deformability. 
Thus, the tidal deformability of the binary components introduces a $5$PN correction (absent in the BH case) to the GW 
phase relative to the leading-order GW term. 
This can be understood by noticing that the $v^6$ term in Eqs.~\eqref{M2tidal} and \eqref{E2} multiplies the $1/v$ term 
in Eq.~\eqref{phaseM2}, giving an overall factor $v^5$ which is a $5$PN correction relative to $\psi_{\tn{N}}\sim 
v^{-5}$.
This derivation is valid for nonspinning objects, the effect of spin is suppressed by 
a further 1.5PN order and introduces new classes of \emph{rotational TLNs}
\cite{Poisson:2014gka,Pani:2015nua,Landry:2015zfa,Abdelsalhin:2018reg,Jimenez-Forteza:2018buh}.

The TLNs of a nonspinning ultracompact object of mass $M$ and radius $r_0=2M(1+\epsilon)$ (with $\epsilon\ll1$) in 
Schwarzschild coordinates vanish logarithmically in the BH limit~\cite{Cardoso:2017cfl}, $k\sim 1/|\log\epsilon|$, 
opening the way to probe horizon scales. This scaling holds for 
any ECO whose exterior is governed (approximately) by vacuum-GR equations, and with generic Robin-type boundary 
conditions on the Zerilli function $\Psi$ at the surface, $a\Psi+b\frac{d\Psi}{dz}=c$~\cite{Maselli:2018fay}. 
In this case, in the $\epsilon\to0$ limit one gets
\begin{equation}
 k \sim \frac{2 (4 a-3 c)}{15 a \log\epsilon} \,. \label{Robin}
\end{equation}
Particular cases of the above scaling are given in Table~I of Ref.~\cite{Cardoso:2017cfl}.
Thus, the only exception to the logarithmic behavior concerns the zero-measure case $a=\frac{3}{4} c$, for which $k\sim 
\epsilon/\log\epsilon$. No ECO models described by these boundary conditions are known.

Such generic logarithmic behavior acts as a \emph{magnifying glass} to probe near-horizon quantum structures~\cite{Cardoso:2017cfl,Maselli:2017cmm}. Since $k\sim{\cal 
O}(10^{-3}-10^{-2})$ when $\epsilon\sim\ell_P/M$. As a comparison, for a typical neutron star $k_{\rm NS}\approx 200$, 
and probing quantum structures near the horizon will require a precision about $4$ orders of magnitude better 
than current LIGO constraints~\cite{Abbott:2018exr}. Prospects to detect this effect are discussed in 
Sec.~\ref{sec:test_TLNs}.
The logarithmic mapping between $k$ and $\epsilon$ makes it challenging constraint $\epsilon$ from measurements of the 
TLNs, because measurements errors propagate exponentially~\cite{Maselli:2017cmm,Addazi:2018uhd}. 
Nevertheless, this does not prevent to distinguish ECOs from BHs using TLNs, nor to perform model selection between 
different ECO models all with new microphysics at the Planck scale~\cite{Maselli:2018fay}.
\begin{figure}[th]
	\centering		
\includegraphics[width=.5\textwidth]{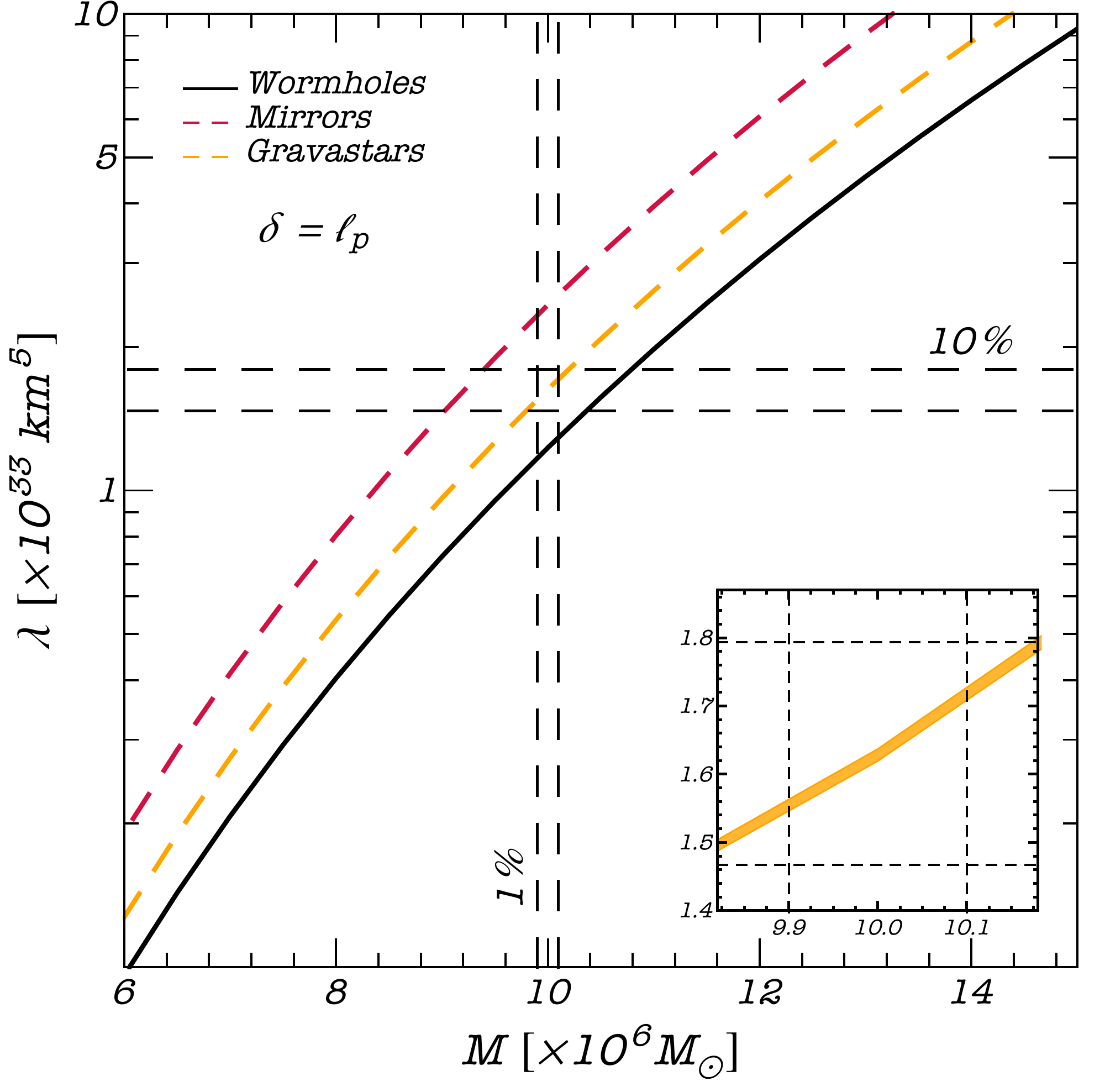}
\caption{Tidal deformability $\lambda$ as a function of the mass for three toy models of ECOs. For 
all models the surface is at Planckian distance from the Schwarzschild radius, $r_0-2M=\ell_P$. The dashed lines refer 
to a putative measurement of the TLN at the level of $10\%$ for an object with 
$M=10^7M_\odot$, which would allow to distinguish among different models at more than $90$\% 
confidence level. The zoomed inset resolves the thickness of each curve, with a width given by 
the intrinsic error due to the quantum uncertainty principle~\cite{Addazi:2018uhd}. Adapted from 
Ref.~\cite{Maselli:2018fay}.}
\label{fig:lambdaM}
\end{figure}
This is shown in Fig.~\ref{fig:lambdaM} --~inspired by standard analysis to discriminate among 
neutron-star equations of state~\cite{Hinderer:2010ih,Maselli:2013hl}. The figure shows 
the tidal deformability $\lambda=\frac{2}{3}M^5|k|$ as a function of the 
object mass for three different toy models (gravastars, wormholes, and perfectly-reflecting Schwarzschild-like ECOs) 
characterized by the same Planckian scale of the correction, $\delta\equiv r_0-2M={\ell}_P\approx1.6\times 
10^{-33}\,{\rm cm}$.

To summarize, finite-size effects in the inspiral waveform provide three different null-hypothesis tests of BHs.
BHs have vanishing TLNs but introduce a nonzero tidal heating ($\psi_\tn{TD}=0$, $\psi_\tn{TH}\neq0$), while ECOs have 
(logarithmically small) TLNs but zero tidal heating ($\psi_\tn{TD}\neq0$, $\psi_\tn{TH}=0$). In addition BHs have a 
very well defined set of multipole moments which depend on only two parameters (mass and angular momentum), whereas 
ECOs have in principle limitless possibilities.
In addition, it is possible that the inspiral excites the characteristic modes of each of the objects, i.e., their QNMs.
The extent to which this happens, and its impact on the inspiral stage are still to be understood~\cite{VegaInPrep}.

The TLNs were computed for boson stars~\cite{Mendes:2016vdr,Cardoso:2017cfl,Sennett:2017etc}, very compact anisotropic fluid stars~\cite{Raposo:2018rjn}, and gravastars~\cite{Uchikata:2016qku,Cardoso:2017cfl}. The TLNs of simple-minded ultracompact Schwarzschild-exterior spacetimes with a stiff equation of state at the surface were computed in Ref.~\cite{Cardoso:2017cfl}.
The TLNs of spacetimes mimicking ``compact quantum objects'' were recently investigated \cite{Giddings:2019}.

\subsubsection{Accretion and drag in inspirals around and inside DM objects}\label{sec:BinaryDM1}
When an object moves through any medium, it will be subject to (at least) two types of drag.
One is direct and caused by accretion: the accreting object grows in mass and slows down.
In addition, the moving body exerts a gravitational pull on all the medium, the backreaction of which 
produces dynamical friction (known also as ``gravitational drag''), slowing the object down. To quantify these 
effects, it is important to know how the medium behaves. Collisionless media cause, generically, a gravitational drag 
different from that of normal fluids~\cite{Macedo:2013qea,Macedo:2013jja}. The gravitational drag caused by media which 
is coherent on large scales may be suppressed~\cite{Hui:2016ltb}, but further work is necessary to understand this 
quantitatively.

Consider now a binary of two compact objects, in which one is made of DM. At large separations inspiral will be driven 
mostly by GW emission. However, at small distances, the dynamics will generically be dominated by accretion and 
gravitational drag. The phase evolution of a binary, taking gravitational radiation, accretion and drag was studied 
when a small BH or neutron star inspirals around {\it and inside} a massive boson 
star~\cite{Macedo:2013qea,Macedo:2013jja}.
These results can also be directly translated to inspirals within a DM 
environment~\cite{Barausse:2014tra,Macedo:2013qea,Macedo:2013jja,Eda:2014kra,Yue:2017iwc,Hannuksela:2018izj}. Full 
nonlinear simulations of the inspiral and merger of boson stars, oscillatons and axion stars include GW emission, drag 
and accretion and tidal deformations. Although considerably more difficult to systematize and perform, such studies have 
been undertaken 
recently~\cite{Bezares:2017mzk,Palenzuela:2017kcg,Bezares:2018qwa,Helfer:2018vtq,Dietrich:2018bvi,Clough:2018exo}.

\subsubsection{GW emission from ECOs orbiting or within neutron stars}\label{sec:BinaryDM2}
It is conceivable that ECOs play also a role in GW (as well as EM)
emission when orbiting close to neutron stars or white dwarfs. This might arise via two different possible ways.
ECOs can form via gravitational collapse of DM or unknown quantum effects, and cluster around compact stars through 
tidal dissipation mechanisms. Alternatively, compact stars evolving in DM-rich environments may accrete a significant 
amount of DM in their interior:
DM is captured by the star due to gravitational deflection and a non-vanishing
cross-section for collision with the star material~\cite{Press:1985ug,Gould:1989gw,Goldman:1989nd,Bertone:2007ae,Brito:2015yga}. 
The DM material eventually thermalizes with the star, and accumulates inside a finite-size 
core~\cite{Brito:2015yfh,Brito:2015yga,Gould:1989gw,Goldman:1989nd}. 

Interaction of the core with the surrounding star may lead to characteristic EM signatures~\cite{Brito:2015yfh,Brito:2015yga}.
Alternatively, a more generic imprint of such ECOs is GW emission, either via standard inspiralling 
processes~\cite{Maselli:2017vfi,Horowitz:2019aim} or by small oscillations of such ECOs {\it inside} neutron stars or 
white dwarfs~\cite{Horowitz:2019aim,Ellis:2017jgp}.

\subsection{Formation and evolution}\label{sec:formation}

In the context of DM physics, the formation and existence of ECOs is very reasonable~\cite{Giudice:2016zpa}. We know 
that DM exists, that it interacts gravitationally and that its coupling to Standard Model fields is very weak. 
Therefore, gravitationally bound structures made of DM particles are dark (by definition) and can potentially be 
compact. Examples which are well understood include boson stars, made of scalars or vectors, which constitute one 
notable exception to our ignorance on the formation of ECOs. These configurations can arise out 
of the gravitational collapse of massive scalars (or vectors). Their interaction and mergers can be studied by evolving 
the Einstein-Klein-Gordon (-Maxwell) system, and there is evidence that accretion of less massive boson stars makes 
them grow and cluster around the configuration of maximum mass. In fact, boson stars have efficient {\it gravitational 
cooling} mechanisms that allow them to avoid collapse to BHs and remain very compact after 
interactions~\cite{Seidel:1991zh,Seidel:1993zk,Brito:2015yfh,DiGiovanni:2018bvo}. Similar studies and similar 
conclusions hold for axion stars, where the coupling to the Maxwell field is taken into 
account~\cite{Widdicombe:2018oeo}. The cosmological formation of such dark compact solitons, their gravitational 
clustering and strong interactions such as scattering and mergers was recently investigated~\cite{Amin:2019ums}. If DM 
is built out of dark fermions, then formation should parallel that of standard neutron stars, and is also a well 
understood process. Collisions and merger of compact boson stars~\cite{Liebling:2012fv,Bezares:2017mzk}, boson-fermion 
stars~\cite{Bezares:2018qwa,Bezares:2019jcb}, and axion stars~\cite{Helfer:2016ljl,Clough:2018exo} have been studied in 
detail.

On the other hand, although supported by sound arguments, the vast majority of the alternatives to BHs are, at best, 
incompletely described. Precise calculations (and often even a rigorous framework) incorporating the necessary physics 
are missing.
Most models listed in Table~\ref{tab:ECOs} were built in a phenomenological way or they arise as solutions 
of Einstein 
equations coupled to exotic matter fields. For example, models of quantum-corrected objects do not include all the 
(supposedly large) local or nonlocal quantum effects that could prevent collapse from occurring. In the absence of a 
complete knowledge of the missing physics, it is unlikely that a ClePhO forms out of the merger of two ClePhOs. These 
objects are so compact that at merger they will be probably engulfed by a common apparent horizon. The end product is, 
most likely a BH as argued in Section~\ref{sec:hoop}.
On the other hand, if large quantum effects do occur, they would probably act on short timescales
to prevent apparent horizon formation possibly in all situations. Thus, for example quantum backreaction has been 
argued to lead to wormhole solutions rather than BHs~\cite{Berthiere:2017tms}. In some models, Planck-scale dynamics 
naturally leads to abrupt changes close to the would-be horizon, without fine tuning~\cite{Holdom:2016nek}. Likewise, in 
the presence of (exotic) matter or if GR is classically modified at the horizon scale, Birkhoff's theorem no longer 
holds, and a star-like object might be a more natural outcome than a BH. However, some studies suggest that compact 
horizonless bodies may form naturally as the result of gravitational collapse~\cite{Beltracchi:2018ait}. The generality 
of such result is unknown.

An important property of the vacuum field equations is their scale-invariance, inherited by BH solutions. Thus, the scaling properties
of BHs are simple: their size scales with their mass, and if a non-spinning BH of mass $M_1$ is stable, then a BH of 
mass $M_2$ is stable as well, the timescales being proportional to the mass.
Such characteristic is summarized in 
Fig.~\ref{fig:MvsR}. Once matter is added, this unique property is lost. Thus, it is challenging to find theories able 
to explain, with horizonless objects, all the observations of dark compact objects with masses ranging over more than 
seven orders of magnitude, although some ECO models can account for that~\cite{Raposo:2018rjn}. Such ``short blanket'' 
problem is only an issue if one tries to explain away {\it all} the dark compact objects with horizonless alternatives. 
If particle physics is a guidance, it is well possible that nature offers us a much more diverse universe content.

\clearpage
\newpage


\newpage

\section{Observational evidence for horizons} \label{sec:Tests}

\hskip 0.2\textwidth
\parbox{0.8\textwidth}{
{\small 
\noindent {\it ``It is well known that the Kerr solution provides the unique solution for 
stationary BHs in the universe.
But a confirmation of the metric of the Kerr spacetime (or some aspect of it) cannot even 
be contemplated in the foreseeable future.''}
\begin{flushright}
S. Chandrasekhar, The Karl Schwarzschild Lecture,\\ Astronomische Gesellschaft, Hamburg 
(September 18, 1986)
\end{flushright}
}
}

\vskip 1cm

Horizons act as perfect sinks for matter and radiation. The existence of a hard or smooth surface
will lead in general to clear imprints. Classically, EM waves are the traditional tool to investigate astrophysical 
objects. 
There are a handful of interesting constraints on the location of the surface of ECOs using 
light
\cite{Narayan:1997xv,Narayan:2002bn,McClintock:2004ji,Broderick:2005xa,Broderick:2007ek,Narayan:2008bv,Broderick:2009ph,
Lu:2017vdx}.
However, testing the nature of dark, compact objects with EM observations is
challenging. Some of these challenges, as we will discuss now, are tied
to the incoherent nature of the EM radiation in astrophysics, and the amount of modeling 
and uncertainties associated to such emission. Other problems are connected to the 
absorption by the interstellar medium. As discussed in the previous section, testing quantum or 
microscopic corrections at the horizon scale with EM probes is nearly impossible.
Even at the semiclassical level, Hawking radiation is extremely weak to detect
and not exclusive of BH spacetimes~\cite{Paranjape:2009ib,Barcelo:2010xk,Harada:2018zfg}.

The historical detection of GWs~\cite{Abbott:2016blz} opens up the exciting 
possibility of testing gravity in extreme regimes with unprecedented 
accuracy~\cite{TheLIGOScientific:2016src,Yunes:2013dva,Barausse:2014tra,Berti:2015itd,Giddings:2016tla,Yunes:2016jcc,Maselli:2017cmm}. GWs are generated by coherent motion of massive sources, 
and are therefore subjected to less modeling uncertainties (they depend on far fewer 
parameters) relative to EM probes. The most luminous GWs come from very dense sources, but 
they also interact very feebly with matter, thus providing the cleanest picture of the 
cosmos, complementary to that given by telescopes and particle detectors.

Henceforth we will continue using the parameter $\epsilon$ defined by Eq.~\eqref{eq_epsilon_def} to quantify the 
constraints that can be put on the presence/absence of a horizon.
The current and projected bounds discussed below are summarized in Table~\ref{tab:constraints} at the end of this 
section.

\subsection{Tidal disruption events and EM counterparts}

Main-sequence stars can be driven towards ECOs through different mechanisms, including two-body or resonant relaxation 
or other processes~\cite{Alexander:2005jz,binney2011galactic}. At sufficiently short orbital distances, stars 
are either tidally disrupted (if they are within the Roche limit of the central ECO), or swallowed whole.
In both cases, strong EM emission is expected for ECOs with a hard surface relative to the case of a BH~\cite{Abramowicz:2016lja,Malafarina:2016rdm,Zhang:2016kyq,Benavides-Gallego:2018htf}.
If the ECO mass is above $\sim 10^{7.5}M_{\odot}$, such emission 
should be seen in broad surveys and produce bright optical and UV transients. Such an emission has been ruled out by 
Pan-STARRS $3\pi$ survey~\cite{Chambers:2016jzn} at $99.7\%$
confidence level, if the central massive objects have a hard surface at radius larger than 
$2M(1+\epsilon)$ with~\cite{Lu:2017vdx}
\begin{equation}
\epsilon\approx 10^{-4.4}\,.
\end{equation}

The limit above was derived under the assumption of spherical symmetry, isotropic equation of state, and dropping some 
terms in the relevant equation. It assumes in addition that the infalling matter clusters at the surface (thereby 
excluding from the analysis those ECO models made of weakly interacting matter (e.g., boson stars) for which 
ordinary matter does not interact with the surface and accumulates in the interior).

\subsection{Equilibrium between ECOs and their environment: Sgr~A$^*$}
The previous results used a large number of objects and -- in addition to the caveats just pointed -- assume that all 
are horizonless. The compact radio source Sgr~A$^*$ at the center of galaxy is --~due to its proximity~-- a good 
candidate to improve on the above. Sgr A$^*$ has an estimated mass $M\sim 4\times 10^6M_{\odot}$, and is 
currently accreting at an extremely low level, with (accretion disk) luminosity $L_{\rm disk}\sim 10^{36} {\rm erg\, s}^{-1}$ (peaking 
at wavelength $\sim 0.1\,{\rm  mm}$), about 
$10^{-9}$ times the Eddington luminosity for the central mass~\cite{Johannsen:2015mdd,Eckart:2017bhq}. The
efficiency of the accretion disk at converting gravitational energy to radiation is less
than 100\%, which suggests a lower bound on the accretion rate $\dot{M} \geq L_{\rm disk} \sim 10^{15} \,{\rm g}\,{\rm 
s}^{-1}$
($10^{-24}$ in geometric units).

Assume now that the system is in steady state, and that there is a hard surface at $r_0=2M (1+\epsilon)$. In such a 
case, the
emission from the surface has a blackbody spectrum with temperature $T^4=\dot{M}/(4\pi\sigma r_0^2) \sim 3.5 \times 10^3\,{\rm K}$
and bright in the infrared (wavelength $\sim 1 \mu{\rm m}$)~\cite{Carballo-Rubio:2018vin}. However,
measured infrared fluxes at $1-10 \mu{\rm m}$ from Sgr A$^*$ are one to two orders of magnitude below this prediction.
Initial studies used this to place an extreme constraint, $\epsilon\lesssim 
10^{-35}$~\cite{Broderick:2005xa,Broderick:2007e}. 
However, the argument has several flaws~\cite{Cardoso:2017njb,Cardoso:2017cqb}:

\begin{itemize}
\item[i.] It assumes that a thermodynamic and dynamic equilibrium must be established between the accretion disk and 
the central object, on relatively short timescales.  However, strong lensing prevents this from happening; consider 
accretion disk matter, releasing isotropically (for simplicity) scattered radiation on the surface of the object.
As discussed in Section~\ref{sec:shadow}, only a fraction $\sim \epsilon$ is able to escape during the first 
interaction with the star, cf.\ Eq.~\eqref{solidangle}. The majority of the radiation will fall back onto the surface 
after a time $t_{\rm roundtrip} \sim 9.3 M$ given by the average of Eq.~\eqref{troundtrip}~\footnote{One might wonder if 
the trapped radiation bouncing back and forth the surface of the object might not interact with the accretion disk. As 
we showed in Section~\ref{sec:shadow}, this does not happen, as the motion of trapped photons is confined to within the 
photosphere.}. Suppose one injects, instantaneously,
an energy $\delta M$ onto the object. Then, after a time $T_a$, the energy emitted to infinity during $N=T_a/t_{\rm 
roundtrip}$ interactions reads
\begin{equation}
 \Delta E \sim  \left[1 - (1 - \epsilon)^N\right]{\delta M} \approx \epsilon \left(\frac{T_a}{t_{\rm 
roundtrip}}\right) \delta M\,.\label{delta E}
\end{equation}
where the last step is valid for $\epsilon N\ll1$.

We can assume $T_a=\tau_{\rm Salpeter}\approx 4.5\times 10^7\,{\rm yr}$ and $\dot M= f_{\rm 
Edd} \dot M_{\rm Edd}$, where $\dot M_{\rm Edd}\approx 1.3 \times10^{39}(M/M_\odot)\,{\rm erg/s}$ is the Eddington mass 
accretion rate onto a BH. Then, from Eq.~\eqref{delta E} we get
\begin{equation}
\dot{E}\sim 10^{-25}\left(\frac{\epsilon}{10^{-15}}\right)\left(\frac{f_{\rm Edd}}{10^{-9}}\right)\,. 
\label{EdotSgrA}
\end{equation}
where we have normalized the fraction of the Eddington mass accretion rate, $f_{\rm Edd}$, to its typical value for 
Sgr~A*. Requiring this flux to be compatible with the lack of observed flux from the central spot ($\dot E\lesssim 
10^{-25}$), one finds $\epsilon\lesssim 10^{-15}$.

Assuming $L\sim \dot E$ and using the Stefan-Boltzmann law, Eq.~\eqref{EdotSgrA} yields an estimate for 
the effective surface temperature of Sgr~A* if the latter had a hard surface,
\begin{equation}
 T\sim 7.8\times 
10^3\,\left(\frac{4\times 10^6 M_\odot}{M}\right)^{1/2}\left(\frac{\epsilon}{10^{-15}}\right)^{1/4}\left(\frac{\delta 
M}{10^{-7}M}\right)^{1/4}\,{\rm K}\,.\label{T_eff}
\end{equation}


\item[ii.] It assumes that the central object is returning in EM radiation most of the energy that it is taking in from 
the disk. However, even if the object were returning all of the incoming radiation on a sufficiently short timescale, a 
sizable fraction of this energy could be in channels other than EM.
For freely-falling matter on a radial trajectory, its four-velocity $v_{(1)}^{\mu}=(E/f,-\sqrt{E^2-f},0,0)$.
Particles at the surface of the object have $v_{(2)}^{\mu}=(\/\sqrt{f},0,0,0)$. When these two collide,
their CM energy reads~\cite{Banados:2009pr},
\begin{equation}
E_{\rm CM}=m_0\sqrt{2}\sqrt{1-g_{\mu\nu}v_{(1)}^\mu v_{(2)}^\mu}\sim \frac{m_0\sqrt{2E}}{\epsilon^{1/4}}\,, \label{Ecm}
\end{equation}
Thus, even for only moderately small $\epsilon$, the particles are already relativistic. 
At these CM energies, all known particles (photons, neutrinos, gravitons, etc) should be emitted ``democratically,''
and in the context of DM physics, new degrees freedom can also be excited. Even without advocating new physics beyond 
the $10\,{\rm TeV}$ scale, extrapolation of known hadronic interactions to large energies suggests that about $20\%$ of 
the collision energy goes into neutrinos, whose total energy is a sizable fraction of that of the photons emitted in 
the process~\cite{Kelner:2006tc}. To account for these effects, we take 
\begin{equation}
\epsilon\lesssim 10^{-14}\,,\label{SgrA_accretion}
\end{equation}
as a reasonable conservative bound coming from this equilibrium argument. 

If only a fraction of the falling material interacts with the object (for example, if it is made of DM with a small interaction cross-section), then the above constraint would deteriorate even further.

\item[iii.] The estimate \eqref{SgrA_accretion} was reached without a proper handling of the 
interaction between
the putative outgoing radiation and the disk itself, and assumes spherical symmetry. Thus, there might be large 
systematic uncertainties associated (and which occur for any astrophysical process where incoherent motion of the 
radiating charges play a key role).

\end{itemize}
%

\subsection{Bounds with shadows: Sgr~A$^*$ and M87}
Recent progress in very long baseline interferometry allows 
for direct imaging of the region close to the horizon, with the potential
to provide also constraints on putative surfaces. These images are also referred to as ``shadows'' since they
map sky luminosity to the source (typically an accretion disk), see Sec.~\ref{sec:shadow}.
Two supermassive BHs have been studied, namely the Sgr~A$^*$ source and the BH at the center of 
M87, whose imaging requires the lowest angular resolution
\cite{Doeleman:2008qh,Doeleman:2012zc,Loeb:2013lfa,Goddi:2016jrs,Abuter:2018drb,Amorim:2019xrp,
Johannsen:2015hib,Akiyama:2019cqa}.

In particular, the Event Horizon Telescope Collaboration has very recently obtained a radio image of the 
supermassive BH candidate in M87~\cite{Akiyama:2019cqa} and similar results for Sgr~A$*$ are expected soon.
The Event Horizon Telescope images of Sgr A$^*$ and M87$^*$ in the millimeter wavelength so far are consistent with a 
point source of radius $r_0=(2-4)\,M$~\cite{Doeleman:2008qh,Doeleman:2012zc,Johannsen:2015hib,Akiyama:2019cqa}, or
\begin{equation}
\epsilon \sim 1\,. \label{constraintshadow}
\end{equation}
This corresponds to the size of the photon sphere, which as we described in Section~\ref{sec:stage} will be the 
dominant relevant strong-field region for these observations. The absence of an horizon will influence the observed 
shadows, since some photons are now able to directly cross the object, or be reflected by it. There are substantial 
differences between the shadows of BHs and some horizonless objects (most notably boson 
stars~\cite{Cunha:2015yba,Cunha:2017wao,Cunha:2018acu}). Nevertheless, because of 
large astrophysical uncertainties and the focusing effect for photons when $\epsilon\to0$ [Eq.~\eqref{solidangle}], all 
studies done so far indicate that it is extremely challenging to use such an effect to place a constraint 
much stronger than Eq.~\eqref{constraintshadow}~\cite{Vincent:2015xta,Cunha:2018gql,Cardenas-Avendano:2019pec}.

In principle, the accretion flow can be very different in the absence of a horizon, when accreted matter can accumulate in the interior, possibly producing a bright spot within the object's shadow~\cite{Olivares:2018abq}. However, in practice this bright source may be too small to be resolved. Assuming matter is accreted at a fraction $f_{\rm Edd}$ of the Eddington rate, the relative 
angular size of the matter accumulated at the center relative to the size of central object is
\begin{equation}
 \frac{\Delta m_{\rm accr}}{M}\sim f_{\rm Edd}\frac{T_{\rm age}}{\tau_{\rm Salpeter}}\approx 3\times 
10^{-2}\left(\frac{f_{\rm Edd}}{10^{-4}}\right)\,,
\end{equation}
where in the last step we conservatively assumed that the central object is accreting at constant rate for $T_{\rm 
age}=T_{\rm Hubble}\approx 300 \tau_{\rm Salpeter}$ and have normalized $f_{\rm Edd}$ to the current value predicted 
for Sgr A$^*$~\cite{Quataert_1999}. Similar mass accretion rates are predicted for M87$^*$~\cite{DiMatteo:2002hif}.
Therefore, a resolution at least $\approx 100$ times better than current one is needed to possibly resolve the effect 
of matter accumulated in the interior of these sources. This is beyond what VLBI on Earth can achieve.

In a similar spirit, tests based on strong-lensing events~\cite{Nandi:2018mzm,Shaikh:2019itn} (in fact, a 
variant of shadows) or quantum versions of it~\cite{Sabin:2017dvx} have been proposed. Adding to the list of possible 
discriminators, Ref.~\cite{Gracia-Linares:2016gvy} studied the impact of
supersonic winds blowing through BHs and boson stars. The conclusion is that, while qualitatively the stationary regime 
of downstream wind distribution is similar, the density may defer by almost an order of magnitude depending on the boson 
star configuration. At an observational level, these differences would show up presumably as friction
on the compact object. However, quantitative tests based on observations are challenging to devise.

Finally, ``hotspots'' orbiting around supermassive objects can also provide information 
about near-horizon signatures~\cite{2005MNRAS.363..353B,Broderick:2005jj}. Recently, the first detection of these 
orbiting features at the ISCO of Sgr~A* was reported~\cite{2018A&A...618L..10G}, implying a bound of the same order as
Eq.~\eqref{constraintshadow}. 

\subsection{Tests with accretion disks}
Tests on the spacetime geometry can also be performed by monitoring how {\it matter} moves and radiates
as it approaches the compact object. Matter close to compact objects can form an accretion 
disk~\cite{LyndenBell:1969yx,Novikov:1973kta,page1974disk}, in which each element approximately moves in circular, Keplerian orbits. The 
disk is typically ``truncated'' at the ISCO (cf. Fig.~\ref{fig:diagram2}), which represents a transition point in the 
physics of the accretion disk. 
It is in principle possible to extract the ISCO location and angular velocity --~and hence infer properties of the 
central object such as the mass, spin, and quadrupole moment~--
from the EM signal emitted (mostly in the X-ray band) by the accreting matter, either for a stellar-mass BH or for a 
supermassive BH~\cite{Bambi:2015kza}.  In practice, the physics of accretion disks is very complex and 
extracting such properties with a good accuracy is challenging.

A promising approach is the analysis of 
the \emph{iron $K\alpha$ line}~\cite{fabian1989x}, one of the brightest components of the X-ray emission from accreting 
BH 
candidates. This line is broadened and skewed due to Doppler and (special and 
general) relativistic effects, which determine its characteristic shape. An analysis of this shape (assuming that the 
spacetime is described by the Kerr metric) provides a measurement of the BH spin and the inclination of the accretion 
disk~\cite{Reynolds:2013qqa}. Although limited by systematic effects~\cite{Bambi:2015kza}, this technique has been used 
also to test the 
spacetime metric~\cite{Johannsen:2010xs,Johannsen:2012ng,Bambi:2012at,Jiang:2014loa,Johannsen:2015rca,Moore:2015bxa,
Hoormann:2016dhy} and to distinguish boson stars from BHs~\cite{Cao:2016zbh,Shen:2016acv}.
Another approach is the study of the thermal component of the spectrum from stellar-mass BHs using the so-called 
\emph{continuum-fitting method} ~\cite{Li:2004aq,McClintock:2013vwa,Reynolds:2013qqa}, which can provide information 
about the ISCO location and hence the BH spin~\cite{McClintock:2013vwa}.
The method can be also used to test the spacetime 
geometry~\cite{Johannsen:2010xs,Bambi:2011jq,Bambi:2012tg,Kong:2014wha,Bambi:2014sfa,Johannsen:2015rca,Moore:2015bxa,
Hoormann:2016dhy} but is limited by the fact that deviations from the Kerr geometry are typically degenerate with 
the ISCO properties, e.g. with the spin of the object~\cite{Bambi:2015kza,Johannsen:2016uoh}. 
Finally, an independent approach is the study of the \emph{quasi-periodic oscillations} observed in the X-ray flux 
emitted by accreting compact objects~\cite{Stella:1998mq,Stella:1999sj,Abramowicz:2001bi}.
The underlying mechanism is not well understood yet, but these frequencies are believed to originate in the 
innermost region of the accretion 
flow~\cite{vanderKlis:2000ca}, and they might carry information about the spacetime near compact objects. Some of the 
proposed models try to explain such phenomena with combinations of the orbital and epicyclic frequencies of geodesics around the 
object. Based on these models, constraints on boson stars have been discussed in Ref.~\cite{Franchini:2016yvq}.

These approaches are helpful in providing indirect tests for the nature of the accreting central object, but are by 
construction unable to probe directly the existence of a surface. A possible 
alternative is the study of the time lag (``\emph{reverberation}'') between 
variability in the light curves in energy bands, corresponding to directly observed continuum emission from the corona 
around the BH and to X-rays reflected from the accretion disc~\cite{Wilkins:2012ty}. Such technique was explored 
assuming the central object to be a BH; the impact of a different central object or of a putative hard surface is 
unknown.

\subsection{Signatures in the mass-spin distribution of dark compact objects\label{subsec:regge}}
%
\begin{figure}[th]
	\centering		
\includegraphics[width=.7\textwidth]{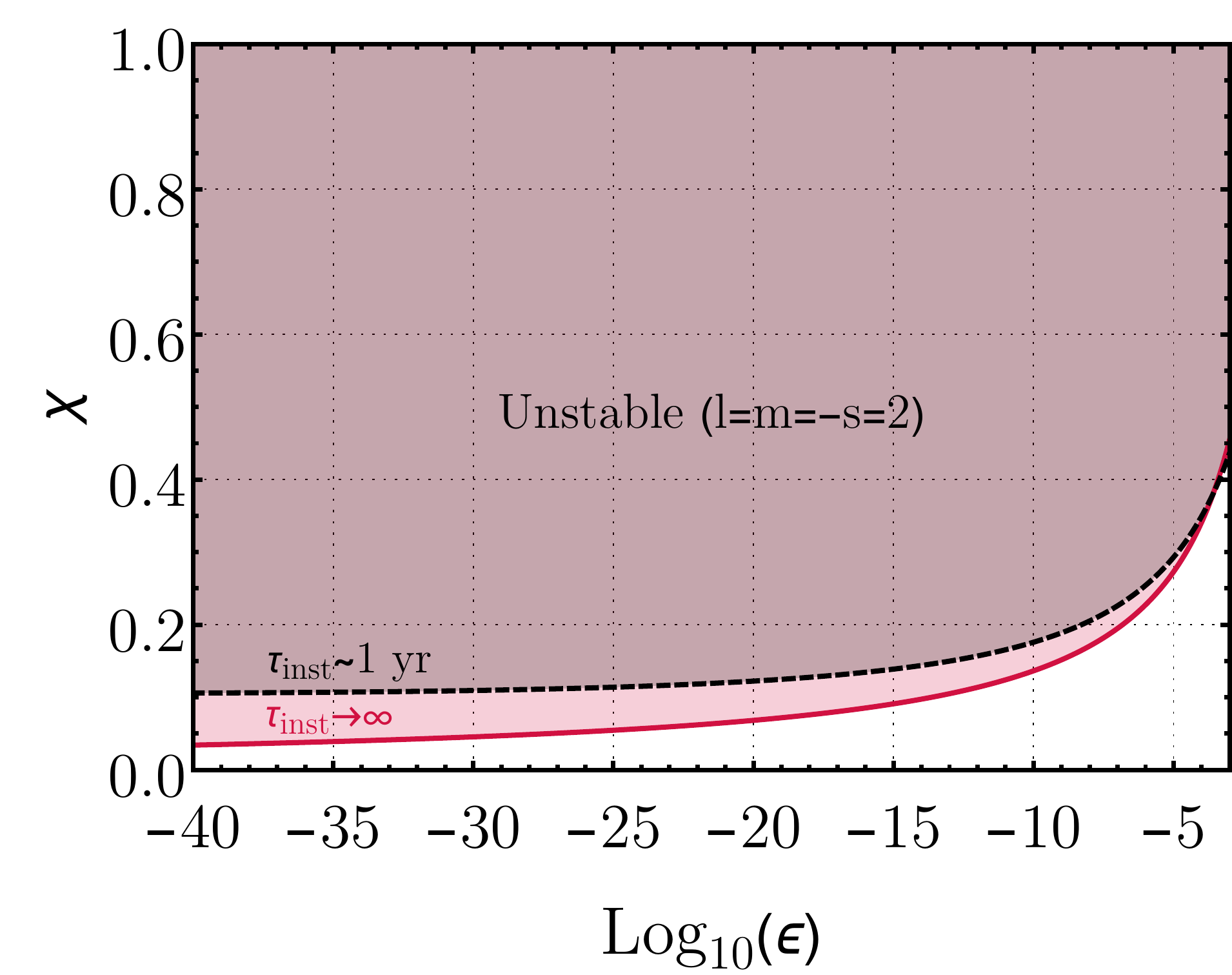}
\caption{Exclusion plot in the $\chi-\epsilon$ plane due to the ergoregion instability of ECOs, assumed to
be described by the Kerr geometry in their exterior. Shaded areas represent regions where a perfectly-reflecting ECO is 
unstable against gravitational perturbations with $l=m=2$, as described by Eqs.~\eqref{tauinstab} and \eqref{chicrit}.}
\label{fig:exclusion}
\end{figure}
The previous tests were based exclusively on EM measurements. 
There are tests which can be done either via EM or GW signals.
An exciting example concerns the measurement of the spin of compact objects, which can be performed either via the 
aforementioned EM tests or from GW detections of binary inspirals and mergers~\cite{LIGOScientific:2018mvr}. 
This requires a large population of massive objects to have been detected and their spins estimated to some accuracy.
EM or GW observations indicating statistical prevalence of slowly-spinning compact objects, 
across the entire mass range, indicate either a special formation channel for BHs, or could signal
that such objects are in fact horizonless: the development of the ergoregion instability is expected to deplete angular momentum from 
spinning ClePhOs, independently of their mass, as we discussed in Section~\ref{sec:ERinstability}. Thus, the spin-mass 
distribution of horizonless compact objects skews towards low spin.  
Although the effectiveness of such process 
is not fully understood, it would lead to slowly-spinning objects as a final state, see Fig.~\ref{fig:exclusion}.
On the other hand, observations of highly-spinning BH candidates can be used to constrain ECO models.

Spin measurements in X-ray binaries suggest that some BH candidates are highly spinning~\cite{Middleton:2015osa}. 
However, such measurements are likely affected by unknown systematics; in several cases different techniques 
yield different results, c.f. Table~1 in~\cite{Middleton:2015osa}. Furthermore, the very existence of the 
ergoregion instability in ECOs surrounded by gas has never been investigated in detail, and the backreaction of the 
disk mass and angular momentum on the geometry, as well as the viscosity of the gas, may change the character and 
timescale of the instability. Finally, as discussed at the end of Sec.~\ref{sec:ERinstability}, dissipation within the 
object might also quench the instability completely~\cite{Maggio:2017ivp,Maggio:2018ivz}.

\subsection{Multipole moments and tests of the no-hair theorem}\label{Sec:EMRIs}

\subsubsection{Constraints with comparable-mass binaries}
\begin{figure}[ht]
\begin{center}
\includegraphics[width=0.7\textwidth]{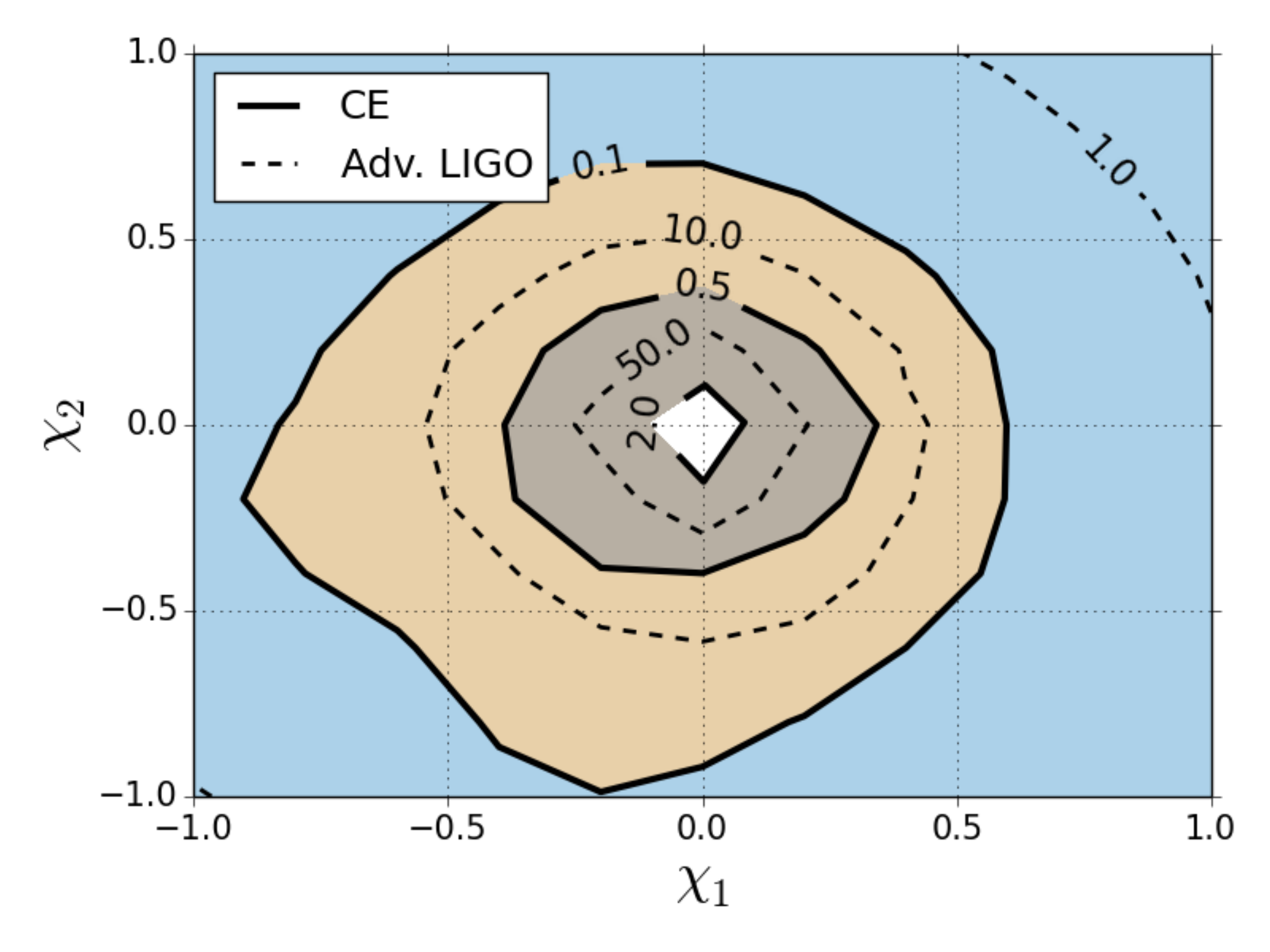}
\caption{Errors on the spin induced quadrupole 
moment ($\kappa_s=(\kappa_1+\kappa_2)/2$) of a binary system with a total mass of $(10+9)M_\odot$ in the dimensionless 
spin parameters plane 
($\chi_1-\chi_2$), assuming the binary components are BHs, i.e. $\kappa_s=1$. Here $\kappa_i$ 
are the spin induced quadrupole moment parameters of the binary constituents, i.e. $M_2^{(i)}=-\kappa_i \chi^2_i 
m_i^3$. The binary is assumed to be optimally oriented at a luminosity distance of $100\,{\rm Mpc}$; (extended 
from~\cite{Krishnendu:2017shb,Krishnendu:2018nqa}).}
\label{fig:quadrupole}
\end{center}
\end{figure}
An estimate of the bounds on the spin-induced quadrupole moment from GW detection of compact-binary inspirals was 
performed in Refs.~\cite{Krishnendu:2017shb,Krishnendu:2018nqa} (see Fig.~\ref{fig:quadrupole}). As we discussed, this correction enters at $2$PN order in the GW inspiral phase and is quadratic in the spin. Therefore, it requires relatively low-mass binaries (which perform many cycles in band before merger) and high spins.
The quadrupole moment of the binary was parametrized as ${\cal M}_2^{(i)}=-\kappa_i 
\chi^2_i m_i^3$ ($i=1,2$), where 
$\kappa_i=1$ for a Kerr BH. 

For moderately large values of the spin $(\chi_i\approx 0.5$) and a binary at 
$500\,{\rm Mpc}$, the projected bounds with Advanced LIGO are roughly $\kappa_s\equiv (\kappa_1+\kappa_2)/2\approx 
50$. This constraint will become approximately $50$ times more stringent with third-generation~(3G) GW detectors (such 
as the Einstein Telescope~\cite{Punturo:2010zz} and Cosmic Explorer~\cite{PhysRevD.91.082001}). Similar constraints 
could be placed by the space detector LISA~\cite{Audley:2017drz} for spinning supermassive binaries at luminosity 
distance of $3\,{\rm Gpc}$~\cite{Krishnendu:2017shb}.
Assuming an ECO model, a bound on $\kappa_i$ can be mapped into a constraint on $\epsilon$. The correction to the 
spin-induced quadrupole relative to the Kerr value for a generic class of ECO models (whose exterior is perturbatively 
close to Kerr) is given by the first term in Eq.~\eqref{conjectureM}. This yields $\kappa=1+a_2/\log\epsilon$, where 
$a_2\sim {\cal O}(1)$ is a model-dependent parameter. Therefore, based on a constraint on $\Delta\kappa\equiv 
|\kappa-1|$, we can derive the upper bound 
\begin{equation}
\epsilon\lesssim e^{-\frac{|a_2|}{\Delta\kappa}}\,,
\end{equation}
which (assuming $a_2\sim{\cal O}(1)$) gives $\epsilon\lesssim 1$ with Advanced LIGO and a factor 3 more stringent with 
3G and LISA. For a gravastar, $a_2=-8/45$~\cite{Pani:2015tga} and we obtain approximately $\epsilon\lesssim1$ with all 
detectors.
These constraints require highly-spinning binaries and the 
analysis of Ref.~\cite{Krishnendu:2017shb,Krishnendu:2018nqa,Kastha:2018bcr} assumes that the quadrupole moment is purely quadratic in 
the spin. This property is true for Kerr BHs, but not generically; for example, the 
quadrupole moment of highly-spinning boson stars contains $O(\chi^4)$ and higher corrections~\cite{Ryan:1996nk}, 
which are relevant for highly-spinning binaries.

In addition to projected bounds, observational bounds on parametrized 
corrections to the $2$PN coefficient of the inspiral waveform from binary BH coalescences can be 
directly translated --~using Eq.~\eqref{phaseM2}~-- into a bound on (a symmetric combination of) the spin-induced 
quadrupole moments of the binary components~\footnote{Unfortunately, for the majority of binary BH events detected so 
far~\cite{LIGOScientific:2018mvr}, either the spin of the binary component is compatible to zero, or the event had a 
low signal-to-noise ratio~(SNR) in the early inspiral, where the PN approximation is valid.
The most promising candidate for this test would be GW170729, for which the measured effective binary spin parameter is 
$\chi_{\rm eff}\approx 0.36^{+0.21}_{-0.25}$~\cite{LIGOScientific:2018mvr,Chatziioannou:2019dsz}. However, for such 
event no parametrized-inspiral test has been performed 
so far~\cite{LIGOScientific:2019fpa}. If confirmed, the recent claimed detection~\cite{Zackay:2019tzo} of a
highly-spinning BH binary would be ideal to perform tests of the spin-induced quadrupole moment.}.
This parametrized PN analysis has been recently done for various BH merger events, the combined constraint on the 
deviation of the $2$PN coefficient reads $\delta \varphi_2\lesssim 0.3$ at $90\%$ confidence 
level~\cite{LIGOScientific:2019fpa}. However, the component spins of these sources
are compatible with zero so these constraints cannot be translated into an upper bound on the spin-induced quadrupole 
moment in Eq.~\eqref{conjectureM}. They might be translated into an upper bound on the non-spin induced quadrupole 
moment, which is however zero in all ECO models proposed so far.

\subsubsection{Projected constraints with EMRIs}

Extreme-mass ratio inspirals~(EMRIs) detectable by the future space mission LISA will probe the spacetime around the 
central supermassive object with exquisite precision~\cite{Gair:2012nm,Berti:2015itd,Barack:2018yly}. These binaries 
perform $\sim m_1/m_2$ orbits before the plunge, the majority of which are very close to the ISCO. 
The emitted signal can be used to constrain the multipole moments of the central object. In particular, preliminary 
analysis (using kludge waveforms and a simplified 
parameter estimation) have placed the projected 
constrain $\delta {\cal M}_2/M^3<10^{-4}$~\cite{Babak:2017tow,Barack:2006pq}. 
In order to translate this into a bound on $\epsilon$, we need to assume a model for ECOs. Assuming the 
exterior to be described by vacuum GR and that $\delta {\cal M}_2$ is spin induced, from 
Eq.~\eqref{conjectureM} we can derive the following bound on $\epsilon$
\begin{eqnarray}
\epsilon &\lesssim& \exp\left(-\frac{10^{4}}{\zeta}\right)\,,\label{boundepsEMRIs1}
\end{eqnarray}
where we defined $\zeta\equiv \frac{\delta {\cal M}_2/M^3}{10^{-4}}$. Note that this is the best-case scenario, 
since we assumed saturation of Eq.~\eqref{conjectureM} 
(with an order-unity coefficient). Other models can exist in which $\delta {\cal M}_2\sim \epsilon^n$, which would lead 
to much less impressive constraints. On the other hand, Eq.~\eqref{boundepsEMRIs1} applies to certain models, e.g. 
gravastars. Notice how stringent the above bound is for those models~\cite{Raposo:2018xkf}.
For this reason, it is important to extend current analysis with more accurate waveforms (kludge waveforms 
are based on a PN expansion of the field equations~\cite{Barack:2006pq} but the PN series converges very slowly in the 
extreme mass ratio limit~\cite{Fujita:2011zk}, so results based on these waveforms are only indicative when 
$m_1/m_2>10^3$).

Model dependent studies on the ability of EMRIs to constrain quadrupolar deviations from Kerr have been presented in 
Refs.~\cite{Ryan:1996nk,Vigeland:2009pr,Moore:2017lxy}.

\subsection{Tidal heating}
Horizons absorb incoming high frequency 
radiation, and serve as sinks or amplifiers for low-frequency radiation able to tunnel in, see Sec.~\ref{Sec:tides}.
UCOs and ClePhOs, on the other hand, are not expected to absorb any significant amount of 
GWs. Thus, a ``null-hypothesis'' test consists on using the phase of GWs to measure 
absorption or amplification at the surface of the objects~\cite{Maselli:2017cmm}.

Because horizon absorption is related to superradiance and the BH area theorem~\cite{Brito:2015oca}, testing this 
effect is an indirect proof of the second law of BH thermodynamics. While this effect is too 
small to be detectable from a single event with second-generation detectors, a large number ($\approx 10^4$) of 
LIGO-Virgo detections might support Hawking's area theorem at $90\%$ confidence level~\cite{Lai:2018pmi}.

\begin{figure*}[th]
\centering
\includegraphics[width=0.49\textwidth]{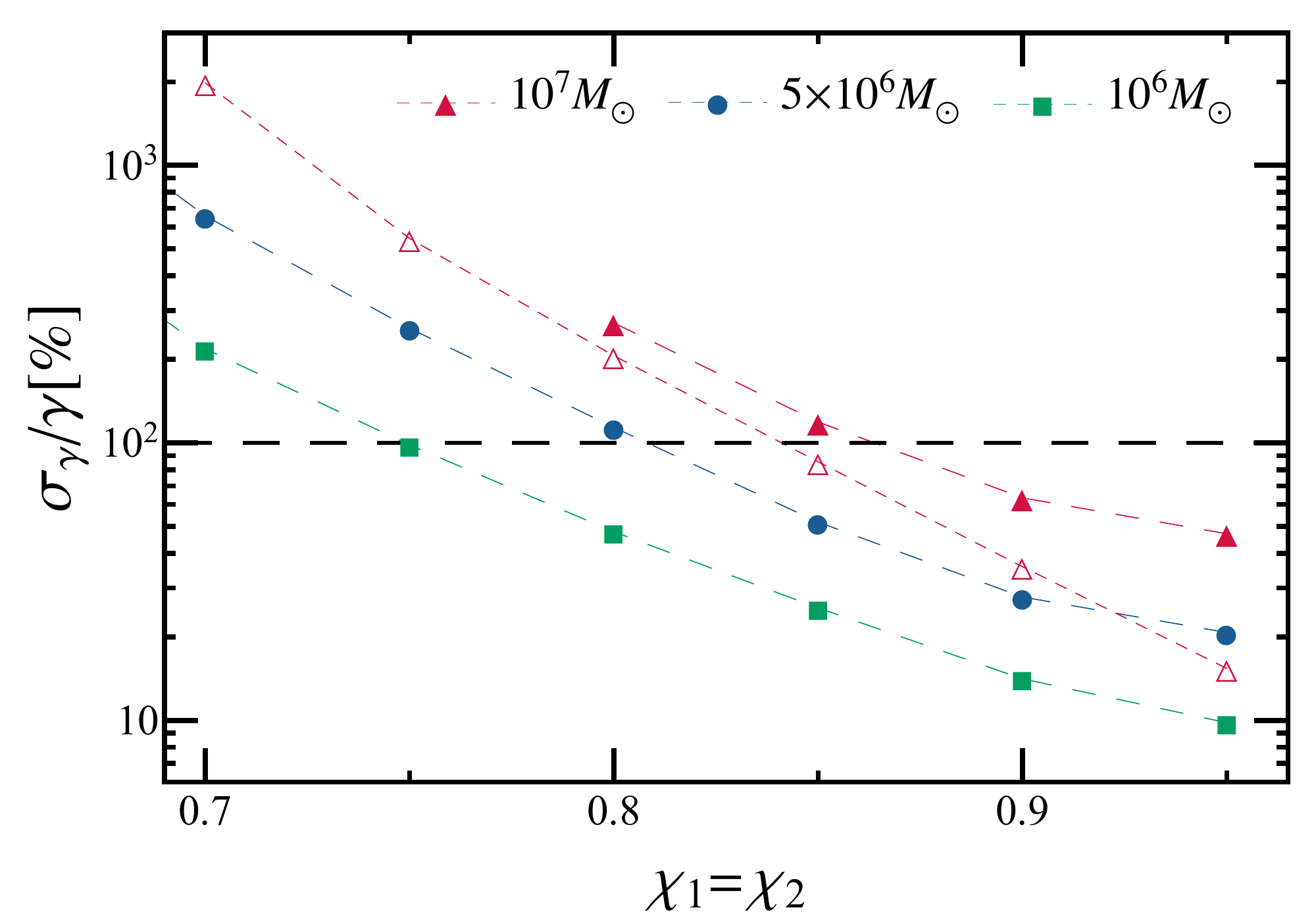}
\includegraphics[width=0.49\textwidth]{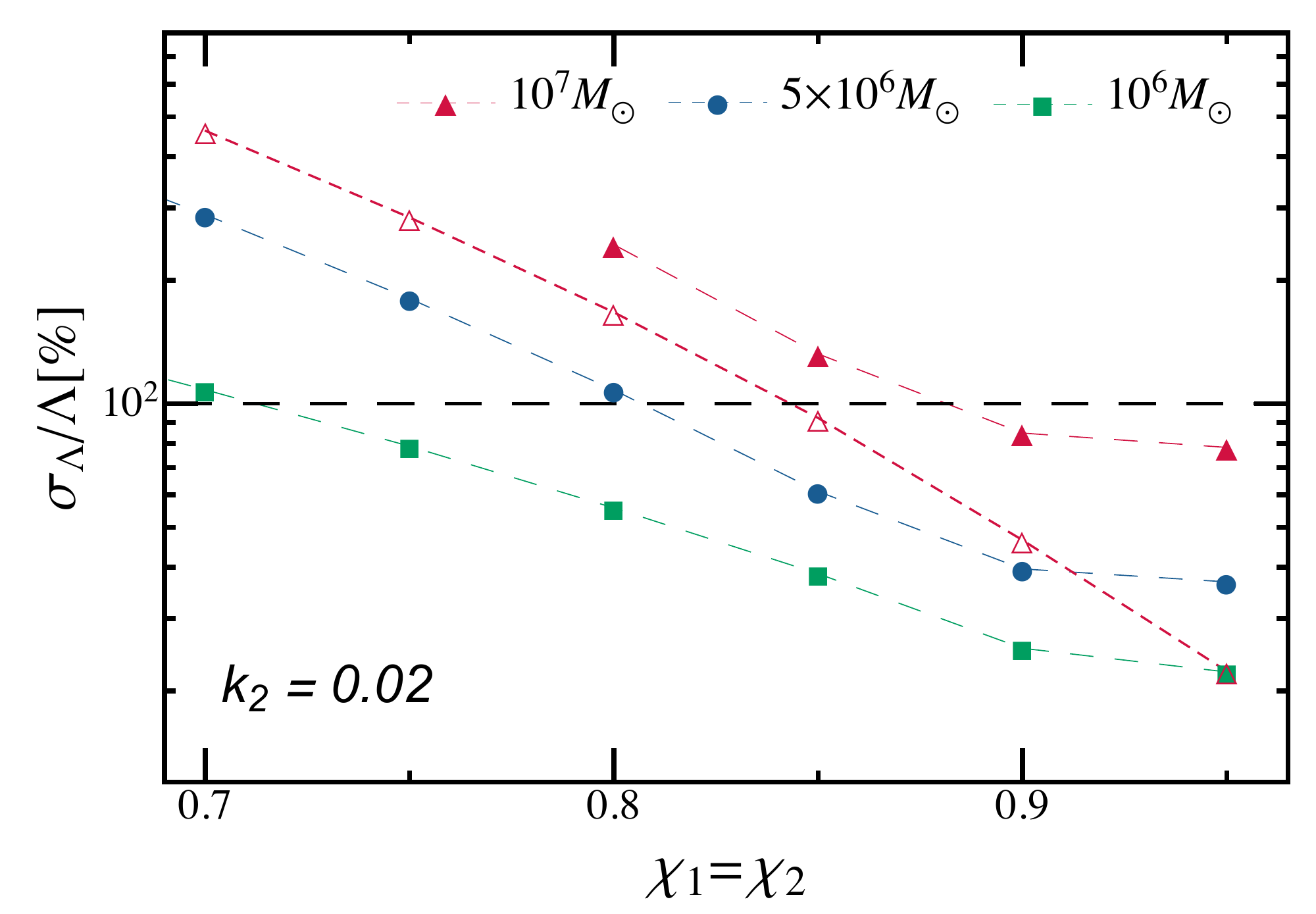}
\caption{Percentage relative projected errors on the tidal-heating parameter $\gamma$ (left panel) 
and on the average tidal deformability $\Lambda$ (right panel) as a function of the 
spin parameter $\chi_{1}=\chi_2$, 
for different values of the central mass $m_1=(10^6,5\times 10^6,10^7)M_\odot$ assuming a future detection with LISA.
In the left and right panel we considered negligible tidal deformability ($\Lambda=0$) and negligible tidal heating 
($\gamma=0$), respectively. 
Full (empty) markers refer to mass ratio $m_1/m_2=1.1$ ($m_1/m_2=2$).
Points below the horizontal line correspond to detections that can distinguish between a BH and an ECO at better than 
$1\sigma$ level.
We assume binaries at luminosity 
distance $2\,{\rm Gpc}$; $\sigma_a$ scales with the inverse luminosity distance, and $\sigma_\Lambda$ scales with 
$1/\Lambda$ when $k\ll1$. Taken from Ref.~\cite{Maselli:2017cmm}.
}
\label{fig:heating}
\end{figure*}

On the other hand, highly-spinning supermassive binaries detectable with a LISA-type GW interferometer will have a 
large SNR and will place stringent constraints on this 
effect, potentially reaching Planck scales near the horizon~\cite{Maselli:2017cmm}. This is shown in the 
left panel of Fig.~\ref{fig:heating}, which presents the bounds on parameter $\gamma$ defined by 
adding the tidal-heating term in the PN phase as $\gamma \psi_\tn{TH}^{\rm BH}$ (see Eq.~\eqref{Hphase}). For a BH 
$\gamma=1$, whereas $\gamma=0$ for a perfectly reflecting ECO. Notice that the effect is linear in the spin and it 
would be suppressed by two further PN orders in the nonspinning case. 

\begin{figure}[ht]
\begin{center}
\includegraphics[width=0.48\textwidth]{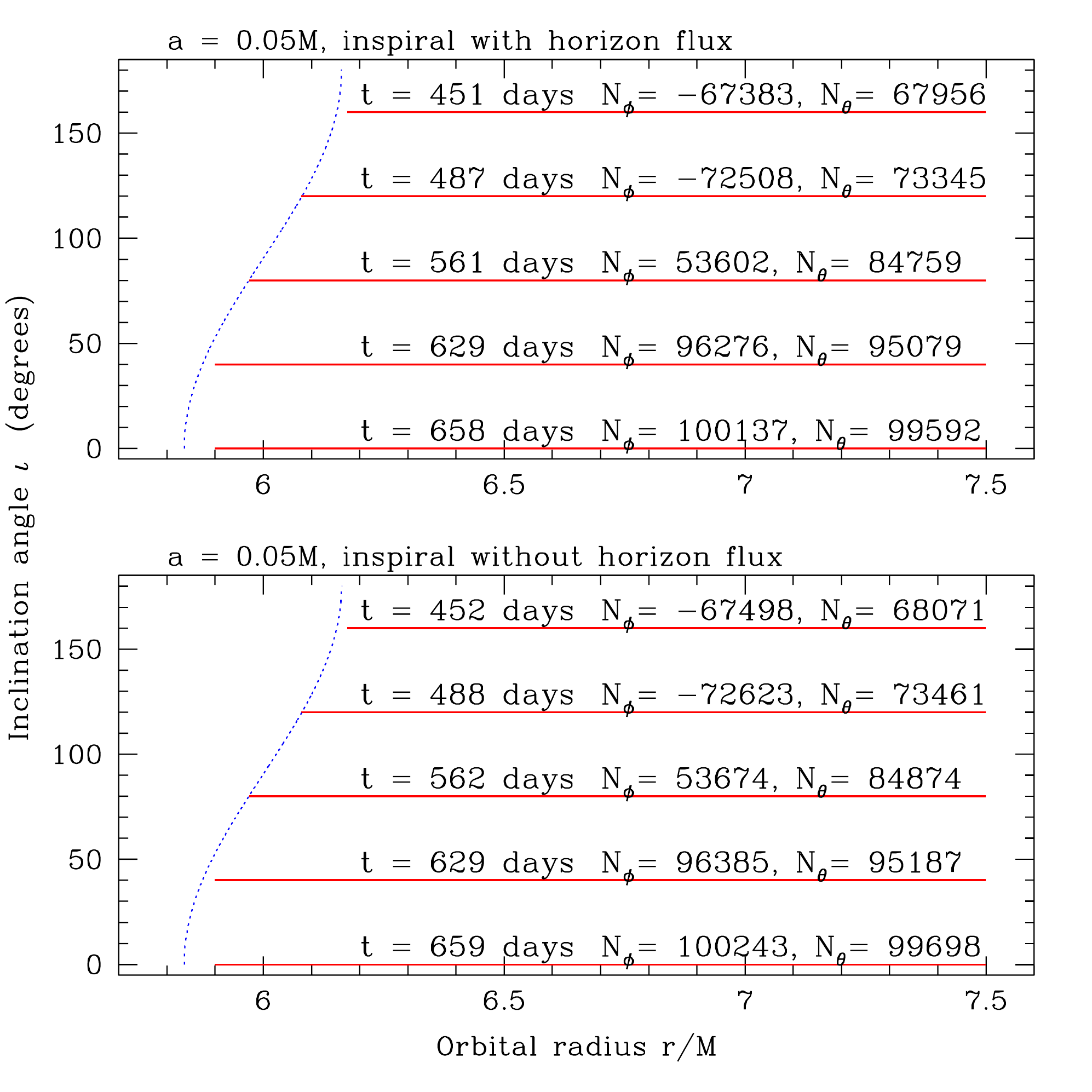}
\includegraphics[width=0.48\textwidth]{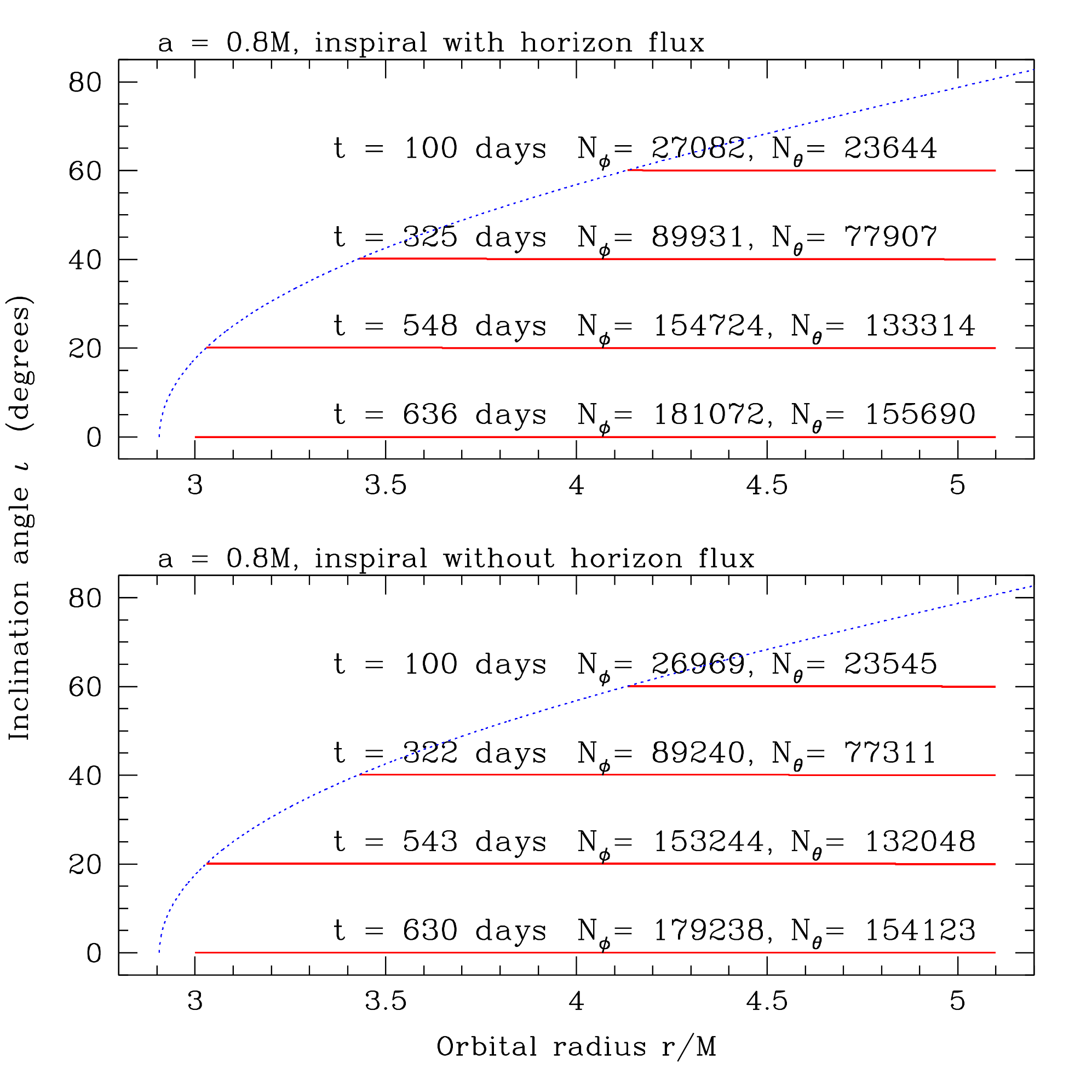}
\caption{\label{fig:heatingEMRIs}
Inspiral trajectories in the strong field of a Kerr BH with $\chi=0.05$ (left) and $\chi=0.8$ (right) in the 
inclination-orbital  radius plane for circular orbits.  The top (bottom) panel includes (exclude) the effect of tidal 
heating, i.e. energy absorption at the horizon. Notice that tidal heating depends strongly on the spin and on the 
orbit. Adapted from Ref.~\cite{Hughes:2001jr}, courtesy of Scott Hughes.
}
\end{center}
\end{figure}

Absence of tidal heating leaves also a detectable imprint in EMRIs~\cite{Hughes:2001jr,Datta:2019euh}. In that case the 
point-particle motion is almost geodesic, with orbital parameters evolving adiabatically because the system loses energy and angular momentum in GWs both at infinity and at the horizon. Energy loss at the horizon is subleading but its putative absence impact the phase 
of the orbits (and hence the GW signal) in a detectable way, especially if the central object is highly 
spinning~\cite{Hughes:2001jr}. In Fig.~\ref{fig:heatingEMRIs} we show a comparison between the inspiral trajectories 
with and without the tidal-heating term.

The effect is clearly important, but the known multiple systematics involved (e.g., due 
to waveform modeling and to parameter estimation in a signal-driven detector like LISA) still need to be quantified. 
Finally, the ability of tidal heating in constraining the closeness parameter $\epsilon$ (or the blueshift 
of photons in Table \ref{tab:constraints} below)
for EMRIs is yet to be understood, both because of the above systematics and also because the absence of tidal heating 
might be directly mapped into a bound on $\epsilon$, since it depends mostly on 
the object interior rather than on the location of the surface (see, however, discussion at the end of 
Sec.~\ref{Sec:tides}).

\subsection{Tidal deformability} \label{sec:test_TLNs}
As discussed in Sec.~\ref{sec:TLNs}, the TLNs of a BH are identically zero, whereas those of an ECO are not.
Although this correction enters at $5$PN order in the waveform, the tidal deformability of an object with radius $r_0$ 
is proportional to $(r_0/M)^5$, so its effect in the GW phase is magnified for less compact objects. 
This effect has been recently explored for boson-star binaries, by investigating the 
distinguishability of binary boson stars from both binary 
BHs~\cite{Cardoso:2017cfl,Sennett:2017etc,Wade:2013hoa,Johnson-McDaniel:2018uvs} and binary 
neutron stars~\cite{Sennett:2017etc}. Second-generation GW detectors at design sensitivity should be able to 
distinguish boson-stars models with no self-potential and with a quartic self-potential (cf. Table~\ref{tab:BSs}) from 
BHs, whereas 3G (resp., LISA) is necessary to distinguish the most compact solitonic boson stars from stellar-mass 
(resp., supermassive) BHs~\cite{Cardoso:2017cfl}. As a rule of thumb, the stronger the boson self-interaction the more 
compact are stable boson-star equilibrium configurations, and hence the smaller the tidal deformability and the chances 
of detectability.
Fits for the TLNs of various boson-star models are provided in Ref.~\cite{Sennett:2017etc}; codes to 
compute these quantities are publicly available~\cite{rdweb}.

For ECOs inspired by Planckian corrections at the horizon scale, the TLNs scale as $k\sim 1/|\log\epsilon|$ for a 
variety of models (see Sec.~\ref{sec:TLNs} and Table~I in Ref.~\cite{Cardoso:2017cfl}). Due to this scaling, in these 
models the TLNs are only roughly 4 
orders of magnitude smaller than for an ordinary neutron star. Nonetheless, measuring such small TLN is probably out of 
reach even with 3G and would require LISA golden binaries~\cite{Maselli:2017cmm} (see right panel of 
Fig.~\ref{fig:heating}).
Due to the logarithmic scaling, in these models the statistical errors on $\epsilon$ would depend 
exponentially on the TLNs and reaching a Planckian requires a very accurate measurement of $k$~\cite{Addazi:2018uhd}. 
Nonetheless, this does not prevent to perform ECO model selection (see Fig.~\ref{fig:lambdaM}).

Finally, in the extreme mass-ratio limit the GW phase~\eqref{tidalphase} \emph{grows linearly} with the mass ratio 
$q=m_1/m_2\gg1$ 
and is proportional to the TLN of the central object, $\psi_\tn{TD}(f)\approx -0.004k_1 q$~\cite{LoveExtreme}. In this 
case the relative measurements errors on $k_1$ scale as $1/\sqrt{q}$ at large SNR. Provided one can overcome the 
systematics on EMRI modeling, this effect might allow to measure TLNs as small as $k_1\approx 10^{-4}$ for EMRI 
with $q=10^6$ detectable by LISA~\cite{LoveExtreme}.
Assuming models for which $k_1\sim 1/\log\epsilon$ (see Eq.~\eqref{Robin}), we can derive the impressive bound
\begin{eqnarray}
\epsilon &\lesssim& \exp\left(-\frac{10^{4}}{\zeta}\right)\,,\label{boundepsEMRIs2}
\end{eqnarray}
where now we defined $\zeta\equiv \frac{k_1}{10^{-4}}$. Note that the above bound is roughly as stringent as that 
in Eq.~\eqref{boundepsEMRIs1} for $k_1\approx \delta {\cal M}_2/M^3$. In both cases the dependence on the departures 
from the BH case is exponential, so the final bound is particularly sensitive also to the prefactors in 
Eqs.~\eqref{Robin} and \eqref{conjectureM}. Also in this case, for models in which $k_2\sim \epsilon^n$ the bound 
on $\epsilon$ would be much less stringent.

\subsection{Resonance excitation}

The contribution of the multipolar structure, tidal heating, and tidal deformability on the gravitational waveform 
is perturbative and produces small corrections relative
to the idealized point-particle waveform. However, there are nonperturbative effects that can be triggered during inspiral,
namely the excitation of the vibration modes of the inspiralling objects. In particular, if the QNMs are of 
sufficiently low frequency, 
they can be excited during inspiral~\cite{Pani:2010em,Macedo:2013qea,Macedo:2013jja,Macedo:2018yoi,VegaInPrep}. This 
case is realized for certain models of ECOs (e.g. ultracompact gravastars and boson stars) and generically for Kerr-like 
ECOs in the $\epsilon\to0$, see Eq.~\eqref{omegaRecho_spin}. In addition to spacetime modes, also model-dependent fluid 
modes might also be excited~\cite{Yunes:2016jcc}. Due to redshift effects, these will presumably play a subdominant role 
in the GW signal.

\subsection{QNM tests}
One of the simplest and most elegant tools to test the BH nature of central objects, and GR itself, 
is to use the uniqueness properties of the Kerr family of BHs: vacuum BHs in GR are fully specified by mass and angular momentum,
and so are their vibration frequencies~\cite{Berti:2005ys,Berti:2009kk}. Thus, detection of one mode (i.e., ringing 
frequency and damping time) allows for an estimate of the mass and angular momentum of the object (assumed to be a GR 
BH). The detection of two or more modes allows to test GR and/or the BH nature of the 
object~\cite{Dreyer:2003bv,Berti:2005ys,Berti:2006qt,Meidam:2014jpa,Berti:2016lat}.

Current detectors can only extract one mode for massive BH mergers, and hence one can estimate the mass and spin of the 
final object, assumed to be a BH~\cite{TheLIGOScientific:2016src}. Future detectors will be able to detect more than one 
mode and perform ``ECO spectroscopy''~\cite{Dreyer:2003bv,Berti:2005ys,Berti:2006qt,Meidam:2014jpa,Berti:2016lat}.

To exclude ECO models, one needs calculations of their vibration spectra. These are available for a wide class of 
objects, including boson stars~\cite{Berti:2006qt,Macedo:2013qea,Macedo:2013jja}, 
gravastars~\cite{Pani:2009ss,Mazur:2015kia,Chirenti:2016hzd}, 
wormholes~\cite{Konoplya:2016hmd,Nandi:2016uzg}, or other quantum-corrected 
objects~\cite{Barcelo:2017lnx,Brustein:2017koc}.
A major challenge in these tests is how to model spin effects properly, since few spinning ECO models are available and 
the study of their perturbations is much more involved than for Kerr BHs.
In general, the post-merger signal from a distorted ECO might be qualitatively similar to that of a neutron-star 
merger, with several long-lived modes excited~\cite{Kokkotas:1999bd} and a waveform that is more involved than a simple 
superposition of damped sinusoids as in the case of BH QNMs.

As discussed previously in Sec.~\ref{sec:echoes} and in Sec.~\ref{sec:testechoes} below, all these extra features are 
expected to become negligible in the $\epsilon\to0$ limit: the \emph{prompt ringdown} of an ultracompact ECOs should 
become indistinguishable from that of a BH in this limit, jeopardizing standard QNM tests.

\subsection{Inspiral-merger-ringdown consistency}
The full nonlinear structure of GR is encoded in the complete waveform from the inspiral and merger of compact objects.
Thus, while isolated tests on separate dynamical stages are important, the ultimate test is that of consistency with the 
full GR prediction: is the full inspiral-merger-ringdown waveform compatible with that of a binary BH coalescence?
Even when the SNR of a given detection is low, such tests can be performed, with some accuracy.
Unfortunately, predictions for the coalescence in theories other than GR and for objects other that BHs
are practically unknown. The exceptions concern evolutions of neutron stars, boson stars, composite fluid 
systems, and axion 
stars~\cite{Liebling:2012fv,Cardoso:2016oxy,Bezares:2017mzk,Bezares:2018qwa,Bezares:2019jcb,Helfer:2016ljl,
Widdicombe:2018oeo,Clough:2018exo} (see Sec.~\ref{sec:formation}), and recent progress in BH mergers in modified 
gravity~\cite{Okounkova:2017yby,Hirschmann:2017psw,Witek:2018dmd,Okounkova:2018pql}.

A model-independent constraint comes from the high merger frequency of 
GW150914~\cite{TheLIGOScientific:2016src}, 
which was measured to be $\nu_{\rm GW}\approx 150\,{\rm Hz}$. The total mass of this system is roughly 
$m_1+m_2\approx 66.2\,M_\odot$. By assuming that the merger frequency corresponds to the Keplerian frequency at 
contact, when the binary is at orbital distance $r=2m(1+\epsilon)$, we obtain the upper bound
\begin{equation}
 \epsilon<0.74\,. 
\end{equation}

Agreement between the mass and spin of the final object as predicted from the inspiral stage
and from a ringdown analysis can be used as a consistency check of GR~\cite{TheLIGOScientific:2016src,Cabero:2017avf}.
For compact boson star mergers, it is possible to find configurations for which either the inspiral 
phase or the ringdown phase match approximately that of a BH coalescence, but not both~\cite{Bezares:2017mzk}. This 
suggests that inspiral-merger-ringdown consistency tests can be very useful to distinguish such binaries.
Thus, although the measurement errors on the mass and spin of the final remnant are currently large, the 
consistency of the ringdown waveform with the full inspiral-merger-ringdown template suggests that the remnant should 
at least be a ClePhO, i.e. places the bound $\epsilon\lesssim {\cal O}(0.01)$, the exact number requires a detailed, 
model-dependent analysis. 
%

\subsection{Tests with GW echoes}\label{sec:testechoes}

For binaries composed of ClePhOs, the GW signal generated during inspiral and merger is expected 
to be very similar to that by a BH binary with the same mass and spin. Indeed,
the multipole moments of very compact objects approach those of Kerr when $\epsilon\to 0$,
and so do the TLNs, etc. Constraining $\epsilon$ (or quantifying up to which point
the vacuum Kerr is a description of the spacetime) is then a question of having sensitive detectors
that can probe minute changes in waveforms. This would also require having sufficiently accurate waveform models to 
avoid systematics. However, there is a clear distinctive feature of horizonless objects: the appearance of late-time 
echoes in the waveforms (see Section~\ref{sec:echoes}). There has been some progress in modeling the echo waveform and data analysis 
strategies are in place to look for such late-time features; some strategies have been also implemented using real data 
\cite{Abedi:2016hgu,Abedi:2017isz,Conklin:2017lwb,Westerweck:2017hus,Tsang:2018uie,Nielsen:2018lkf,Lo:2018sep,
Wang:2019rcf,Uchikata:2019frs}.

\begin{figure}[th]
\begin{center}
\includegraphics[height=0.35\textwidth]{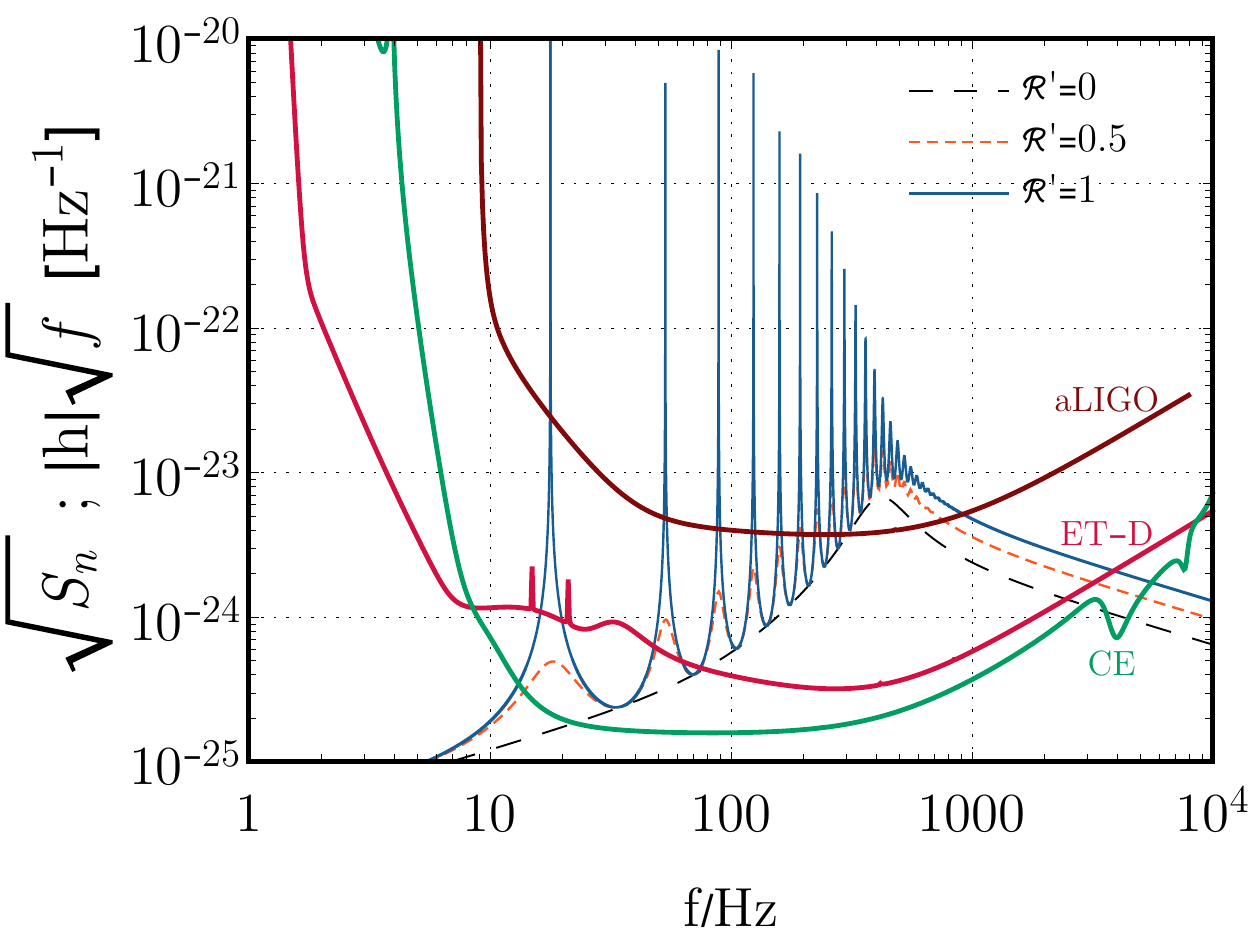} 
\includegraphics[height=0.352\textwidth]{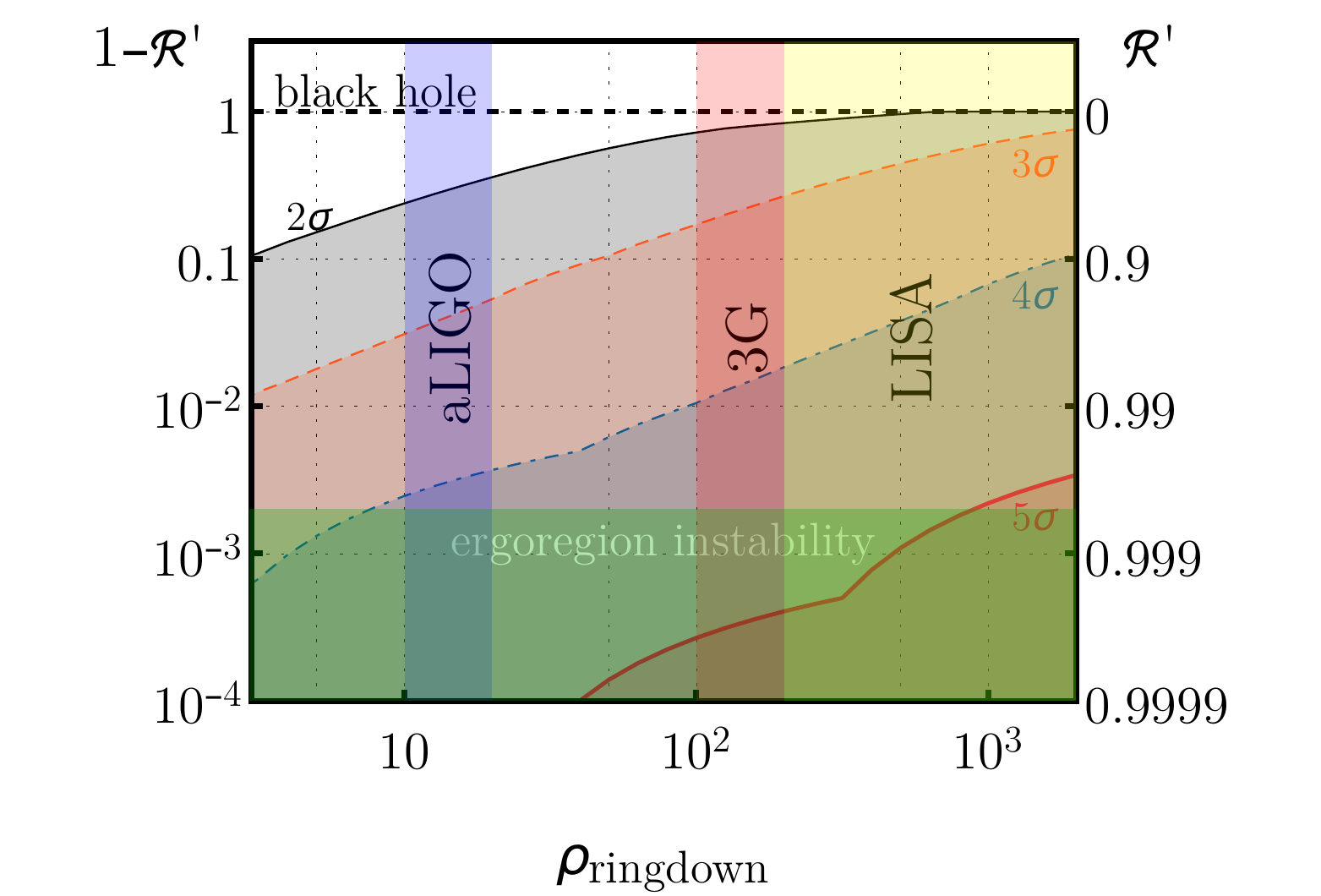}
\caption{\emph{Left:} Representative example of ringdown$+$echo template (Eq.~\eqref{FINALTEMPLATEGEN}) 
compared to the
power spectral densities of various ground-based 
interferometers~\cite{aLIGOzerodet,Evans:2016mbw,Essick:2017wyl,Hild:2010id} as functions 
of the GW frequency $f$. We considered an object with $M=30M_\odot$, at a 
distance of $400\,{\rm Mpc}$, with closeness parameter $\epsilon=10^{-11}$, and various values 
of the reflectivity coefficient ${\cal R}'$ at the surface (see Eq.~\eqref{Rprime}). The case ${\cal R}'=0$ corresponds 
to the pure BH ringdown template.
\emph{Right:} Projected exclusion plot for the ECO reflectivity ${\cal R}'$ as a function of the SNR in the ringdown 
phase and at different $\sigma$ confidence levels, assuming the ringdown template~\eqref{FINALTEMPLATEGEN} based on the 
transfer-function representation and assuming a source near the ECO surface.
Shaded areas represent regions that can be excluded at a given confidence level. Vertical bands are typical SNR 
achievable by aLIGO/Virgo, 3G, and LISA in the ringdown phase, whereas the horizontal band is the 
region excluded by the ergoregion instability, see Sec.~\ref{sec:ERinstability}. 
Adapted from Ref.~\cite{Testa:2018bzd}. 
}
\label{fig:echoes}
\end{center}
\end{figure}

The ability to detect such signals depends on how much energy is converted from the main 
burst into echoes (i.e., on the relative amplitude between the first echo and the prompt 
ringdown signal in Fig.~\ref{fig:ringdown}). Depending on the reflectivity of the ECO, the energy contained in the echoes can exceed that of the standard ringdown alone~\cite{Mark:2017dnq,Testa:2018bzd}, see left panel of Fig.~\ref{fig:echoes}. This suggests 
that it is possible to detect or constrain echoes even when the ringdown is marginally detectable or below 
threshold, as in the case of EMRIs or for comparable-mass coalescences at small SNR.

Searches for echo signals in the detectors based on reliable templates can be used to find new physics, or to set 
very stringent constraints on several models using real data.
Different groups with independent search techniques {\it have found} structure in many of the GW events, compatible with postmerger echoes~\cite{Abedi:2016hgu,Ashton:2016xff,Abedi:2017isz,Conklin:2017lwb,Westerweck:2017hus}.
However, the statistical significance of such events has been put into question~\cite{Westerweck:2017hus,Abedi:2018pst}. 
For GW150914, Refs.~\cite{Abedi:2016hgu,Ashton:2016xff,Conklin:2017lwb}-- using independent search techniques -- report 
evidence for the existence of postmerger echoes in the data. However, Ref.~\cite{Nielsen:2018lkf} finds a lower 
significance and a Bayes factor indicating preference for noise over the echo hypothesis. For other GW events, there is 
agreement between different groups on the existence of postmerger features in the signal, found using echo waveforms. 
The interpretation of these features is under debate.
An independent search in the LIGO-Virgo Catalog 
GWTC-1 found no statistical evidence for the presence of echoes within $0.1\,{\rm s}$ of the main 
burst~\cite{Uchikata:2019frs}.

Any realistic search is controlled by $\eta$ (cf. Eq.~\eqref{templateAbedi}) {\it and} the time delay between main burst 
and echoes~\cite{Abedi:2016hgu,Abedi:2017isz,Conklin:2017lwb,Westerweck:2017hus,Tsang:2018uie,Nielsen:2018lkf,Lo:2018sep}.
Since the SNR of the postmerger signal is controlled by $\eta$ on a integration timescale controlled by $\tau$, even negative searches
can be used to place strong constraints on $\epsilon$~\cite{Westerweck:2017hus,Nielsen:2018lkf}.

Constraints on $\epsilon$ are currently limited by the low SNR. These constraints will greatly improve with next-generation GW detectors. A preliminary analysis in this direction~\cite{Testa:2018bzd} (based on the template~\eqref{FINALTEMPLATEGEN} valid only for nonspinning objects) suggests that perfectly-reflecting ECO models can be detected or ruled out at $5\sigma$ confidence level with SNR in the ringdown of $\rho_{\rm ringdown}\approx 10$. Excluding/detecting echoes for models with smaller values of the 
reflectivity will require SNRs in the post-merger phase of ${\cal O}(100)$. This will be achievable only 
with ground-based 3G detectors and the planned space mission LISA~\cite{Audley:2017drz}, see right panel of 
Fig.~\ref{fig:echoes}.
Simple-minded ringdown searches (using as template an exponentially damped sinusoid~\cite{Abbott:2009km}) can be used to 
look for echoes, separately from the main burst. 
For example, if the first echo carries 20\% of the energy of the main ringdown stage, then
it is detectable with a simple ringdown template. 
LISA will see at least one ringdown event per year, even for 
the most pessimistic population synthesis models used to estimate the 
rates~\cite{Berti:2016lat}. The proposed Einstein Telescope~\cite{Punturo:2010zz} or 
Voyager-like~\cite{Voyager} 3G Earth-based detectors will also be able to 
distinguish ClePhOs from BHs with such simple-minded searches.

Overall, in a large region of the parameter space the signal is large enough to produce effects within reach of
near-future GW detectors, even if the corrections occur at the ``Planck scale'' (by which we mean $\epsilon\sim 
10^{-40}$). This is a truly remarkable prospect.  As the sensitivity of GW detectors increases, the absence of echoes 
might be used to {\rm rule out} ECO models, to set ever stringent upper bounds on the level of absorption in the object's interior, and generically to push tests of gravity closer and 
closer to the horizon scale, as now routinely done for other cornerstones of GR, e.g. in 
tests of the equivalence principle~\cite{Will:2014kxa,Berti:2015itd}.
%

\subsection{Stochastic background}
%
\begin{figure*}[t]
\includegraphics[width=0.48\textwidth]{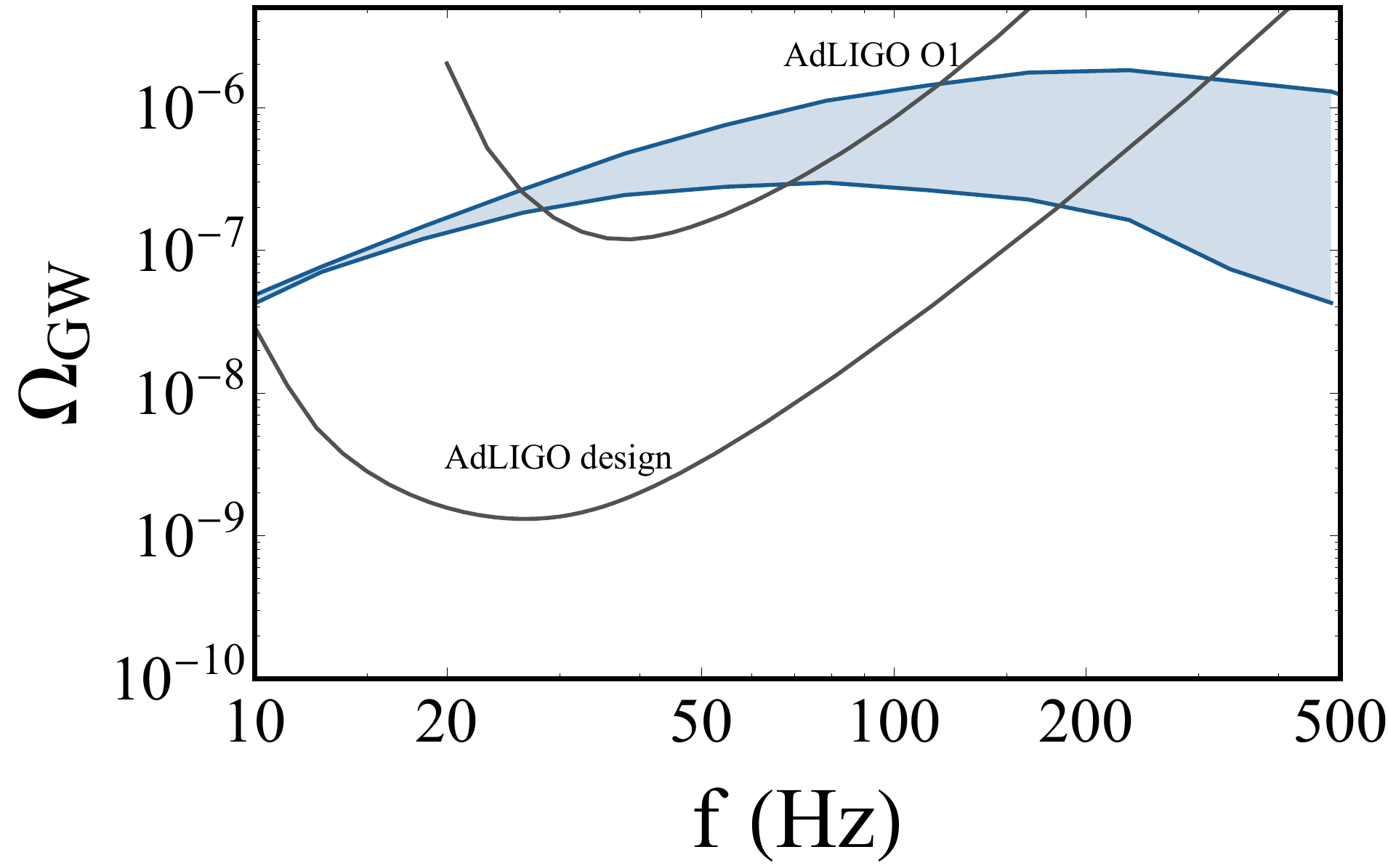}
\includegraphics[width=0.46\textwidth]{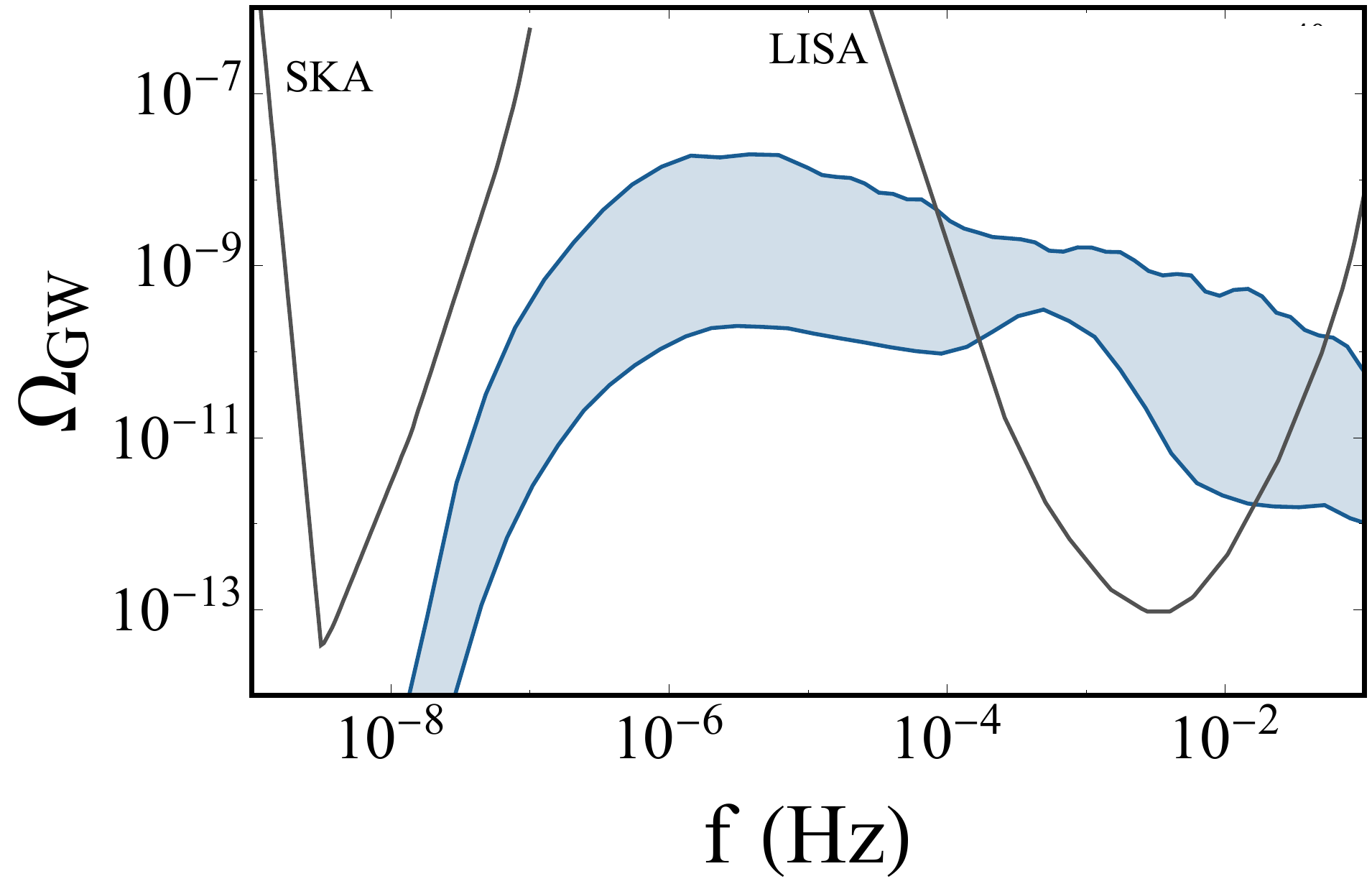}
\caption{Extragalactic stochastic background of GWs in the LIGO/Virgo (left panel), LISA and PTA bands (right panel) 
assuming all BH candidates to be horizonless, described by a Kerr exterior and Dirichlet conditions at the surface 
$r=r_+(1+\epsilon)$, with $\epsilon=10^{-40}$. The bands brackets different population models. The black lines are the 
power-law integrated curves computed using noise power spectral densities for: LISA with one year of observation 
time~\cite{Audley:2017drz}, LIGO's first observing runs (O1), LIGO at design sensitivity, and an SKA-based pulsar timing 
array. Taken from Ref.~\cite{Barausse:2018vdb}. 
\label{fig:background}
}
\end{figure*}

Above in Section~\ref{subsec:regge}, we discussed possible features in the spin distribution of massive compact 
objects. If a large number of massive and dark objects are indeed horizonless and very compact, they will be subjected 
to the ergoregion instability (discussed in Section~\ref{sec:ERinstability}) which drains their rotational energy and 
transfers it to GWs. Thus, the entire universe would be radiating GWs, producing a (potentially) significant amount of 
stochastic GWs~\cite{Barausse:2018vdb,Fan:2017cfw,Du:2018cmp}. Note that such background does {\it not} require 
binaries, isolated ECOs suffice
(isolated compact objects are expected to be $\sim 100$ times more numerous than merging 
binaries~\cite{Dvorkin:2016wac}). 

The background can be characterized by its (dimensionless) energy spectrum
\begin{equation}
\Omega_{\rm GW}=\frac{1}{\rho_c}\frac{d\rho_{\rm gw}}{d\ln f_o}\,, \label{OmegaGW}
\end{equation}
$\rho_{\rm gw}$ being the background's energy density, $f_o$ the frequency measured at the detector
and $\rho_c$ the critical density of the Universe at the present time. Results for a simple ECO, modelled with Kerr 
exterior and Dirichlet conditions at its surface are shown in Fig.~\ref{fig:background}.
The derived constraints assume all BH candidates are horizonless, the bound scales linearly with the fraction of ECOs 
in the population.

\subsection{Motion within ECOs}

In certain models, the ECO interior might be weakly interacting and a further discriminator would be the motion of 
test particles \emph{within} the object. Among other effect, this can produce non-standard signals in EMRIs. As 
discussed in Secs.~\ref{sec:BinaryDM1} and \ref{sec:BinaryDM2}, this motion is driven by the self-gravity of the central 
object, accretion, and dynamical friction.
The study of geodesic motion inside a solitonic boson stars was analyzed in~\cite{Kesden:2004qx}. The effects of 
accretion and drag were included in Refs.~\cite{Macedo:2013qea,Macedo:2013jja,Barausse:2014tra,Barausse:2014pra}.
These effects cannot be directly translated into bounds on $\epsilon$, but would be a smoking-gun signature for 
the existence of structures in supermassive ultracompact objects.


\newpage

\section{Discussion and observational bounds}
The purpose of physics is to describe natural phenomena in the most accurate possible way.
The most outrageous prediction of GR --~that BHs should exist and be always described by the Kerr geometry~-- 
remains poorly quantified. It is a foundational issue, touching on questions such as singularity formation, quantum 
effects in gravity, the behavior of matter at extreme densities, and even DM physics. The quest to quantify the 
evidence for BHs can --~in more than one way~-- be compared with the quest to quantify the equivalence principle, and 
needs to be complemented with tests of the Kerr nature of ultracompact dark objects. Table~\ref{tab:constraints} 
summarizes the observational evidence for BHs.

\begin{table}[h]
\begin{footnotesize}
\begin{tabular}{p{0.2cm}|p{0.8cm}p{1cm}p{8cm}p{3cm}}
\hline \noalign{\smallskip}\hline\noalign{\smallskip}
& \multicolumn{2}{c}{Constraints}  & Source & Reference  \\ 
& $\epsilon (\lesssim)$ & $\frac{\nu}{\nu_\infty}(\gtrsim)$    &  &  \\ 
\noalign{\smallskip}
\hline 
1a.	&${\cal O}(1) $ & ${\cal O}(1)$             &Sgr A$^*$ \& M87               
&\cite{Doeleman:2008qh,Doeleman:2012zc,Johannsen:2015hib,2018A&A...618L..10G,Akiyama:2019cqa}\\
1b.	&$0.74 $ & $1.5$                    &GW150914               
&\cite{Abbott:2016blz}\\
\\
2.	&${\cal O}(0.01)$ & ${\cal O}(10)$					 & GW150914 & \cite{Abbott:2016blz} \\
\\
%
3. &$ 10^{-4.4}$ & $158$    & All with $M>10^{7.5}M_{\odot}$ & \cite{Lu:2017vdx}  \\
\\
4. &$ 10^{-14}$ & $10^7$     & Sgr A$^*$                      &\cite{Lu:2017vdx}  \\
\\
5. &$10^{-40}$ & $10^{20}$   & All with $M<100M_{\odot}$  &\cite{Barausse:2018vdb} \\
\\
6.	&$10^{-47}$ & $10^{23}$  & GW150914 & \cite{Nielsen:2018lkf,Uchikata:2019frs} \\
\\
%
%
7*.	&$e^{-10^4/\zeta}$ & $e^{5000/\zeta}$ & EMRIs & \cite{Babak:2017tow,Barack:2006pq,LoveExtreme} \\
\\
\noalign{\smallskip}\hline \noalign{\smallskip}  \hline
   & \multicolumn{4}{c}{Effect and caveats} \\
   \hline
1a. & \multicolumn{4}{l}{Uses detected orbiting hotspot around Sgr A$^*$ and ``shadow'' of Sgr A$^*$ and M87.}\\
    & \multicolumn{4}{l}{Spin effects are poorly understood; systematic uncertainties not quantified.} \\
1b. & \multicolumn{4}{l}{Merger frequency of GW150914 and measurements of the masses.}\\
    & \multicolumn{4}{l}{Assumes merger frequency equal to Keplerian frequency at contact.} \\
2. & \multicolumn{4}{l}{Consistency of ringdown with BH signal. } \\
   & \multicolumn{4}{l}{Large measurement errors on the QNM frequencies. Precise bounds are model dependent.}\\
   & \multicolumn{4}{l}{Bounds will improve significantly with detailed searches for post-merger echoes.}\\
%
3. & \multicolumn{4}{l}{Lack of optical/UV transients from tidal disruption events.} \\
   & \multicolumn{4}{l}{Assumes: all objects are horizonless, have a hard surface, spherical symmetry, and 
isotropy.}\\
4. & \multicolumn{4}{l}{Uses absence of relative low luminosity from Sgr~A*, compared to disk.} \\
   & \multicolumn{4}{l}{Spin effects and matter-radiation interaction matter poorly understood; assumes spherical symmetry.} \\ 
5. & \multicolumn{4}{l}{Uses absence of GW stochastic background (from ergoregion instability).} \\
   & \multicolumn{4}{l}{Assumes: hard surface (perfect reflection); exterior Kerr; all objects are horizonless.}\\
6. & \multicolumn{4}{l}{Uses absence of GW echoes from post-merger object.} \\
   & \multicolumn{4}{l}{$90\%$ confidence level for $\eta>0.9$, deteriorates for smaller $\eta$. Simplified echo 
template, limited range of priors.}\\
%
%
%
7*. & \multicolumn{4}{l}{Projected EMRI constraints on the spin-induced quadrupole ($\zeta=(\delta 
{\cal M}_2/M^3)/10^{-4}$)  and TLNs ($\zeta=k_/10^{-4}$).}\\
& \multicolumn{4}{l}{Assumes saturation of Eq.~\eqref{conjectureM} (for $\delta {\cal M}_2$) and 
Eq.~\eqref{Robin} (for $k$) and order-unity coefficients in those equations.} \\
   & \multicolumn{4}{l}{Uses PN kludge waveforms, phenomenological deviation for ${\cal M}_2$, and
simplified parameter estimation.}\\
   & \multicolumn{4}{l}{Models for which $\delta {\cal M}_2\sim \epsilon^n$ or $k\sim \epsilon^n$ are much less 
constrained.}\\
\hline
\hline
\end{tabular}
\end{footnotesize}
\vskip 8pt 
\centering 
\caption{\footnotesize
\emph{How well does the BH geometry describe the dark compact objects in our universe?} 
This table quantifies the answer to this question, for selected objects, by {\it excluding} the presence of surfaces
in the spacetime close to the gravitational radius of the object. 
The deviation from the vacuum Kerr geometry, of mass $M$ and angular momentum $J=\chi M^2$, is measured with a 
dimensionless quantity $\epsilon$, such that the structure is localized at a Boyer-Lindquist radius $r_+ (1 + 
\epsilon)$, where $r_+=M(1+\sqrt{1-\chi^2})$. For $\epsilon=0$ the 
spacetime is described by vacuum GR all the way to the horizon. We also express the constraint as measured by the 
blueshift of a radial-directed photon $\nu/\nu_\infty$ (on the equatorial plane, measured by locally non-rotating 
observers) as it travels from large distances to the last point down to which observations are compatible with vacuum.
The constraints come from a variety of observations and tests provided in the references in the last column and 
interpreted as discussed in Sec.~\ref{sec:Tests}.
Alternative quantities that can parametrize the deviation from the vacuum Kerr geometry are the light 
travel time from the light ring to the surface (Eq.~\eqref{tauechospin}) or the proper distance between the light ring 
and the surface. Both quantities depend on $\epsilon$ and on the spin $\chi$ of the object, and scale as 
$\log\epsilon$ as $\epsilon\to0$.
Entries with an asterix refer to projected constraints.
}
\label{tab:constraints}
\end{table}

These bounds can be read in two different ways. On the one hand, they tell us how appropriate the Kerr metric is in 
describing some of the massive and dark objects in our universe. In other words, observations tell us that the Kerr 
description is compatible with observations at least down to $r=r_+(1+\epsilon)$. Alternatively, one can view these 
numbers as constraints on exotic alternatives to BHs. In both cases, the constraint on $\epsilon$ can be translated into 
the ratio of frequencies (or redshift, as measured by locally non-rotating observers~\cite{Bardeen:1972fi}) of a photon 
as it travels from infinity down to the farthest point down to which observations are compatible with vacuum.

Most of the constraints shown in Table~\ref{tab:constraints} are associated with large systematics or modelling 
uncertainties. From a proper understanding of astrophysical environments and their interaction with ultracompact 
objects, the development of a solid theoretical framework, to a proper modeling of the coalescence of such objects 
and data analysis to see such events, the challenges are immense. The pay-off for facing these oustanding issues is to 
be able to quantify the statement that BHs exist in nature.

\clearpage
\newpage

\vskip 2mm
\newpage
\noindent{\bf{\em Acknowledgments.}}
%
We are indebted to Niayesh Afshordi, K. G. Arun, Cosimo Bambi, Carlos Barcel\'o, Ofek Birnholtz, Silke Britzen, Ramy 
Brustein, Collin Capano, Ra\'ul Carballo-Rubio, Ana Carvalho, Miguel Correia, Jan de Boer, Kyriakos Destounis, Valeria 
Ferrari, Valentino Foit, Luis Garay, Steve Giddings, Eric Gourgoulhon, Tomohiro Harada, Carlos Herdeiro, Bob Holdom, Scott Hughes, Bala Iyer, Marios Karouzos, Gaurav Khanna, Joe Keir, Matthew Kleban, Kostas Kokkotas, Pawan Kumar, Claus Laemmerzahl, Jos\'e Lemos, Avi Loeb, Caio Macedo, Andrea Maselli, Samir Mathur, Emil Mottola, Ken-ichi Nakao, Richard Price, Sergey Solodukhin, Nami Uchikata, Chris Van den Broeck, Bert Vercnocke, Frederic Vincent, Sebastian Voelkel, Kent Yagi, and Aaron Zimmerman for providing detailed feedback, useful references, for discussions, or for suggesting corrections to an earlier version of the manuscript.
V.C. acknowledges financial support provided under the European Union's H2020 ERC 
Consolidator Grant ``Matter and strong-field gravity: New frontiers in Einstein's theory'' 
grant agreement no. MaGRaTh--646597.
PP acknowledges financial support provided under the European Union's H2020 ERC, Starting 
Grant agreement no.~DarkGRA--757480 and 
support from the Amaldi Research Center funded by the MIUR program ``Dipartimento di 
Eccellenza'' (CUP: B81I18001170001).
This article is based upon work from COST Action CA16104 ``GWverse'' supported by COST (European Cooperation in Science and Technology).
This work was partially supported by the H2020-MSCA-RISE-2015 Grant No. StronGrHEP-690904 and by FCT Awaken project
PTDC/MAT-APL/30043/2017.
 

\bibliographystyle{utphys}

\bibliography{References}

\end{document}